\documentclass[%
 reprint,
superscriptaddress,
 amsmath,amssymb,
 aps,
prd,
float]{revtex4-2}

\usepackage{graphicx}
\usepackage{dcolumn}
\usepackage{bm}
\usepackage{amsmath}
\usepackage{physics}
\usepackage[usenames,dvipsnames]{color}
\usepackage{xurl}
\usepackage[breaklinks]{hyperref}
\usepackage{hypernat}
\usepackage{breakurl}
\PassOptionsToPackage{hyphens}{url}
\hypersetup{colorlinks, citecolor=Blue, linkcolor=Blue, urlcolor=Blue}

\begin{document}
\title{Scanning X-ray diffraction microscopy for diamond quantum sensing}
\author{Mason C. Marshall}
\affiliation{Department of Electrical and Computer Engineering, University of Maryland, College Park, Maryland, 20742, USA}
\affiliation{Center for Astrophysics $\vert$ Harvard \& Smithsonian, Cambridge, Massachusetts, 02138, USA}
\affiliation{Quantum Technology Center, University of Maryland, College Park, Maryland, 20742, USA}
\author{David F. Phillips}
\affiliation{Center for Astrophysics $\vert$ Harvard \& Smithsonian, Cambridge, Massachusetts, 02138, USA}
\author{Matthew J. Turner}
\affiliation{Quantum Technology Center, University of Maryland, College Park, Maryland, 20742, USA}
\affiliation{Department of Physics, Harvard University, Cambridge, Massachusetts, 02138, USA}
\affiliation{Center for Brain Science, Harvard University, Cambridge, Massachusetts, 02138, USA}
\author{Mark J.H. Ku}
\affiliation{Quantum Technology Center, University of Maryland, College Park, Maryland, 20742, USA}
\affiliation{Department of Physics and Astronomy, University of Delaware, Newark, Delaware, 19716, USA}
\affiliation{Department of Materials Science and Engineering, University of Delaware, Newark, Delaware, 19716, USA}
\author{Tao Zhou}
\affiliation{Center for Nanoscale Materials, Argonne National Laboratory,  Lemont, Illinois, 60439, USA}
\author{Nazar Delegan}
\affiliation{Center for Molecular Engineering, Argonne National Laboratory, Lemont, Illinois, 60439, USA}
\affiliation{Materials Science Division, Argonne National Laboratory,  Lemont, Illinois, 60439, USA}
\author{F. Joseph Heremans}
\affiliation{Center for Molecular Engineering, Argonne National Laboratory, Lemont, Illinois, 60439, USA}
\affiliation{Materials Science Division, Argonne National Laboratory,  Lemont, Illinois, 60439, USA}
\affiliation{Pritzker School of Molecular Engineering, University of Chicago, Chicago, Illinois, 60439,  60637, USA}
\author{Martin V. Holt}
\affiliation{Center for Nanoscale Materials, Argonne National Laboratory,  Lemont, Illinois, 60439, USA}
\author{Ronald L. Walsworth}
\affiliation{Center for Astrophysics $\vert$ Harvard \& Smithsonian, Cambridge, Massachusetts, 02138, USA}
\affiliation{Quantum Technology Center, University of Maryland, College Park, Maryland, 20742, USA}
\affiliation{Department of Electrical Engineering and Computer Science, University of Maryland, College Park, Maryland, 20742, USA}
\affiliation{Department of Physics, University of Maryland, College Park, Maryland, 20742, USA}

\date{Draft: \today}

\begin{abstract}
Understanding nano- and micro-scale crystal strain in CVD diamond is crucial to the advancement of diamond quantum technologies. In particular, the presence of such strain and its characterization present a challenge to diamond-based quantum sensing and information applications -- as well as for future dark matter detectors where directional information of incoming particles is encoded in crystal strain.  Here, we exploit nanofocused scanning X-ray diffraction microscopy to quantitatively measure crystal deformation from defects in diamond with high spatial and strain resolution.  Combining information from multiple Bragg angles allows stereoscopic three-dimensional modeling of strain feature geometry; the diffraction results are validated via comparison to optical measurements of the strain tensor based on spin-state-dependent spectroscopy of ensembles of nitrogen vacancy (NV) centers in the diamond.  Our results demonstrate both strain and spatial resolution sufficient for directional detection of dark matter via X-ray measurement of crystal strain, and provide a promising tool for diamond growth analysis and improvement of defect-based sensing.
\end{abstract}

\maketitle
\newcommand{\w}{3.5in}

\newcommand{\nanoprobeschematic}{
\begin{figure}[htbp!]
\begin{center}
\includegraphics*[width=\columnwidth]{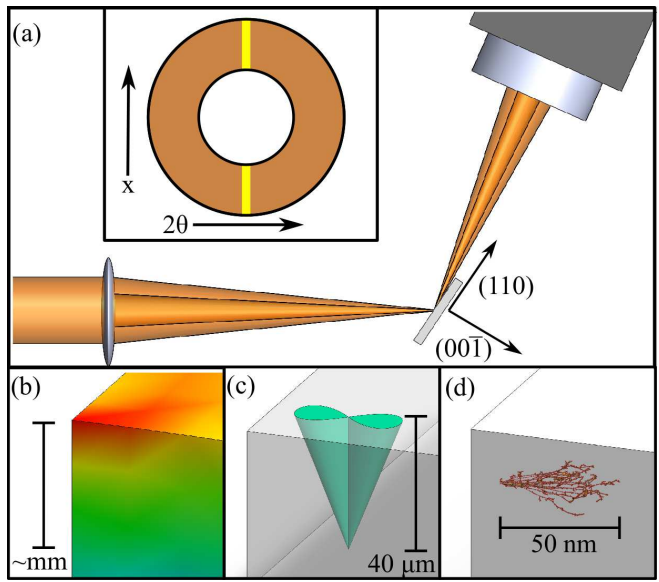}
\caption{\textbf{(a)} Illustration of scanning X-ray diffraction microscopy (SXDM).  A monochromatic beam of hard X-rays (here, 11.2 keV) is focused to a 25-nm spot at the diamond sample surface.  X-rays undergo Bragg diffraction and are collected by a pixelated detector.  The beam is scanned across the sample surface, and a diffraction pattern is collected at each position.  \textbf{Inset}: Simplified schematic of the beam profile (orange ring).  The yellow bar illustrates a simple diffraction pattern, as would be imaged from a perfect bulk crystal.  Crystal strain shifts the Bragg angle $\theta$, moving the diffraction pattern along the detector 2$\theta$ axis.  \textbf{(b)-(d)} Illustrations of different crystal strain features discussed in this work.  \textbf{(b)} Macroscopic strain gradients arise across the diamond, for example due to cleavage of a diamond into segments. \textbf{(c)} Dislocation features incorporated during CVD diamond sample growth may create regions of strained crystal.  \textbf{(d)} Microscopic damage to the diamond crystal - such as nuclear recoils following a WIMP or neutrino impact - leave characteristic strain signatures associated with displaced carbon nuclei in the lattice.}
\label{fig:nanoprobeschematic}
\end{center}
\end{figure}
}

\newcommand{\sage}{
\begin{figure}[htbp!]
\begin{center}
\includegraphics*[width=\columnwidth]{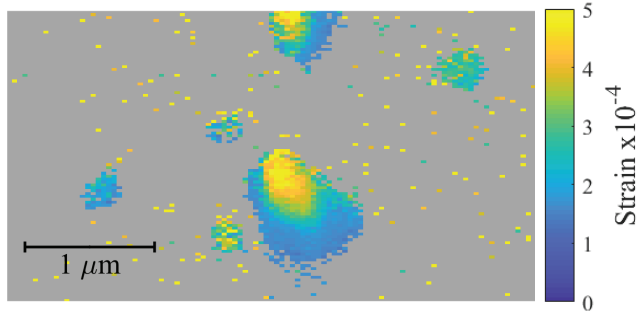}
\caption{Local strain features measured using SXDM on an HPHT diamond sample with a 22 nm scan step. Features are observed on length scales similar to the predicted strain from particle-induced crystal damage \cite{proposal,InvitedPaper} (see Sec.~\ref{sec:darkmatter}).  This data was acquired via ($\bar{1}13$) diffraction; see below for details of the analysis method.}
\label{fig:sage}
\end{center}
\end{figure}
}

\newcommand{\DMdir}{
\begin{figure}[htbp!]
\begin{center}
\includegraphics*[width=\columnwidth]{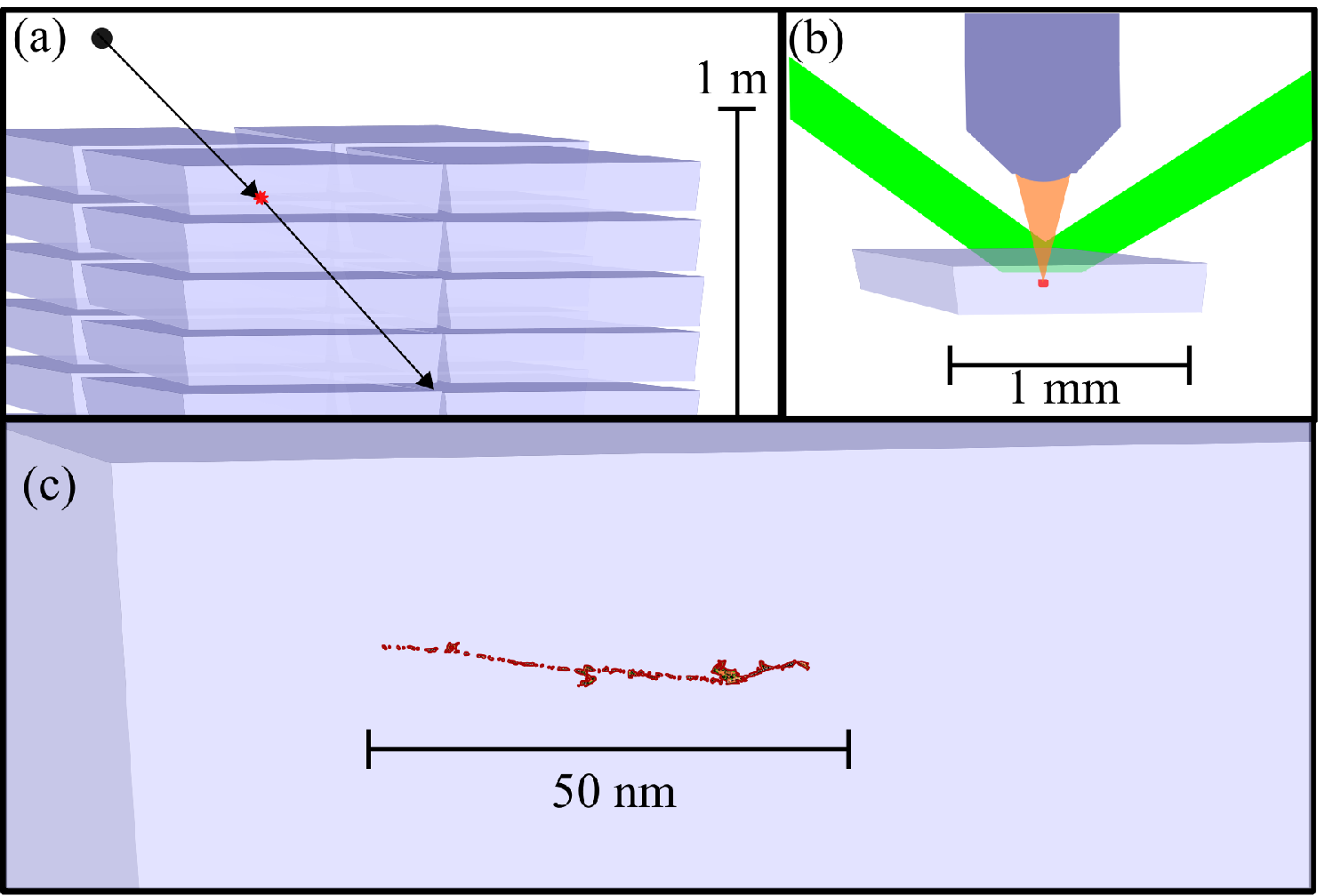}
\caption{Application of spatially resolved nanoscale strain measurement to dark matter detection.  \textbf{(a)} WIMP dark matter interacts with a carbon nucleus in a sectioned diamond target, triggering detector elements. \textbf{(b)} Triggered section is removed from target and imaged or scanned; crystal damage due to nuclear recoils is localized to a $\mu$m-cubed voxel.  \textbf{(c)} Within the diamond section, the orientation and asymmetry of the damage track and resulting nanoscale strain preserve information about the incident WIMP's direction.  This information can be read out using the scanning X-ray nanodiffraction techniques presented here.  The pictured feature is crystal lattice damage from a nuclear recoil track following a 10 keV impact between a WIMP and a carbon nucleus, simulated using the SRIM software \cite{proposal,SRIM}.}
\label{fig:DMdir}
\end{center}
\end{figure}
}

\newcommand{\diamondfeaturestwo}{
\begin{figure}[htbp!]
\begin{center}
\includegraphics*[width=\columnwidth]{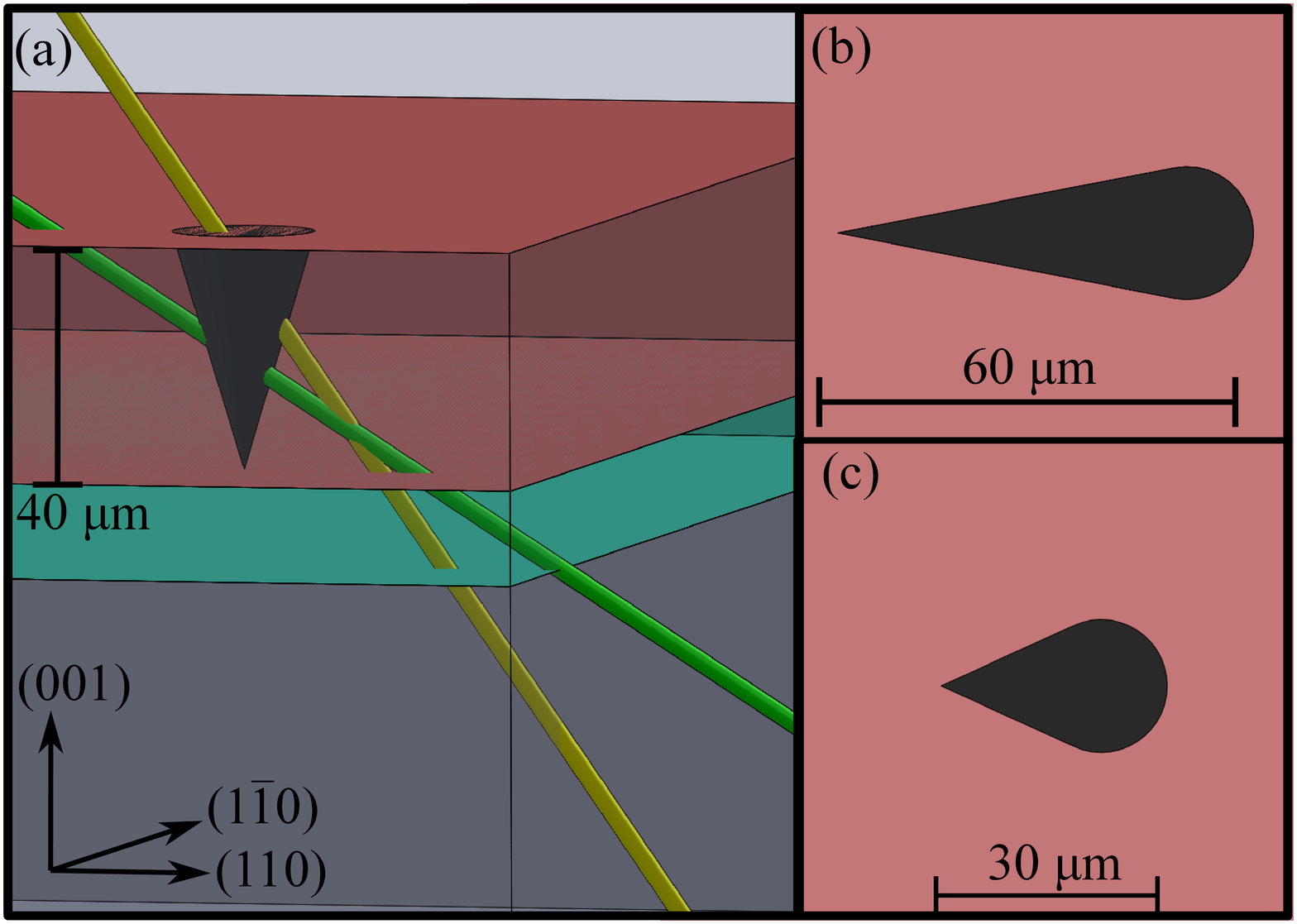}
\caption{\textbf{(a)} Schematic for hard X-ray nanoprobe (HXN) interrogation of the (113) and ($\bm\bar{1}$13) crystal planes on the CVD quantum sensing diamond.  The green beam addresses the (113) plane, while the yellow beam addresses ($\bm\bar{1}$13).  Layers with different lattice spacings and orientations diffract X-rays within the beam differently, creating the diffraction pattern of Fig.\ \ref{fig:sidepeak}a.  Schematically represented diamond components are: high-purity CVD substrate (gray), polishing damage layer at substrate surface (green), NV-doped CVD overgrowth layer (red), which contains strained growth defect features (black).
Right side: Schematic of strained growth defect feature as projected onto the surface by scans addressing the \textbf{(b)} (113) and \textbf{(c)} ($\bm\bar{1}$13) crystal planes.}
\label{fig:diamondfeaturestwo}
\end{center}
\end{figure}
}

\newcommand{\sidepeak}{
\begin{figure}[htbp!]
\begin{center}
\includegraphics*[width=\columnwidth]{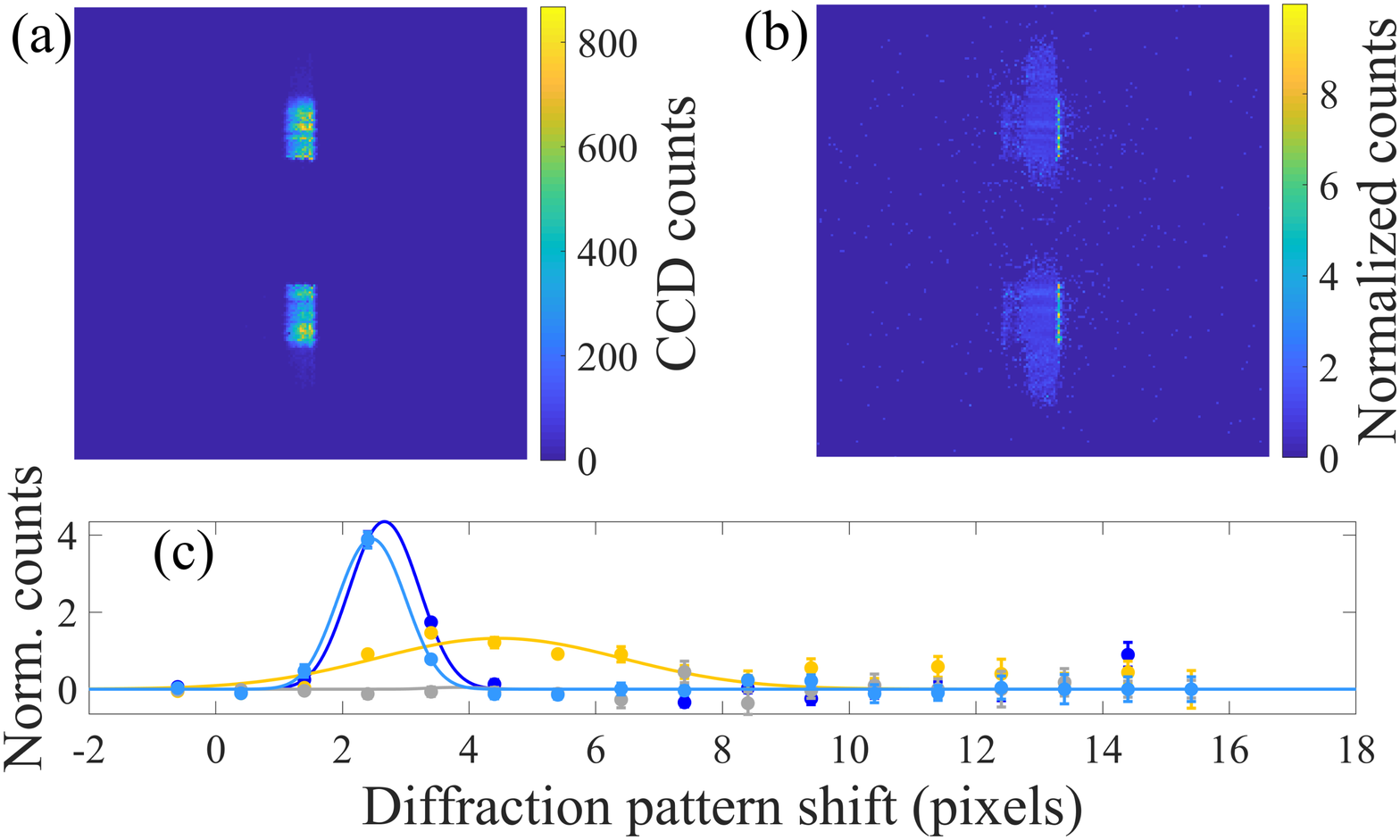}
\caption{\textbf{(a)} Diffraction pattern in the CVD diamond sample measured at a single X-ray beam position, which includes contributions from multiple layers of the diamond as sketched in Fig.~\ref{fig:diamondfeaturestwo}. \textbf{(b)} Diffraction pattern of (a) after dividing by the local host-crystal diffraction profile (see Supplemental Material \cite{SM} Sec.~C for details).  This ratio emphasizes the contribution from sharp, small-length-scale strain features like crystal growth defects or recoil-induced damage. Each column of pixels in (b) is summed, yielding a 1-dimensional diffraction curve, four of which are shown in  \textbf{(c)}.  Each diffraction curve is fit to a Gaussian lineshape to extract a centroid and linewidth.  The centroid position gives the mean strain in the feature, while the linewidth depends on the feature's thickness along the beam path.  Line colors are chosen to approximately match the strain scale of Fig.~\ref{fig:QuantStrain}.}
\label{fig:sidepeak}
\end{center}
\end{figure}
}

\newcommand{\QuantStrain}{
\begin{figure*}[htbp!]
\begin{center}
\includegraphics*[width=\textwidth]{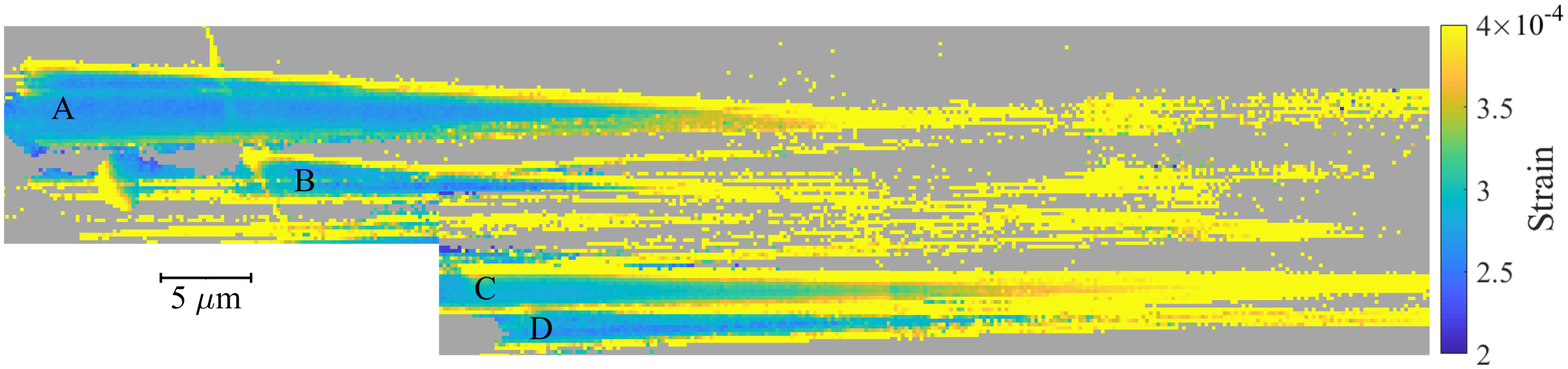}
\caption{Spatial map of strain due to crystal dislocation features in one region of the CVD quantum sensing diamond, as measured with (113) diffraction.  Grey indicates X-ray beam positions that do not intersect a strain feature, as identified via a threshold on the amplitude of the background-subtracted diffraction curve, as illustrated in Figure \ref{fig:sidepeak}c.  The four most prominent features are labeled A-D to simplify discussion in the main text.  The same region and features were also measured with ($\bar{1}13$) diffraction; see Supplemental Material \cite{SM} sec.~G for results.  The color scale limits are chosen for legibility of internal strain within features, rather than enforced by the measurement. Note that the spatial resolution of this scan is set by the scanning step size of 200 nm, chosen to enable measurement of deep strain features over a relatively large field of view; the strain map shown here represents approximately 13 hours of continuous data acquisition.}
\label{fig:QuantStrain}
\end{center}
\end{figure*}
}

\newcommand{\threedmodel}{
\begin{figure}[htbp!]
\begin{center}
\includegraphics*[width=\columnwidth]{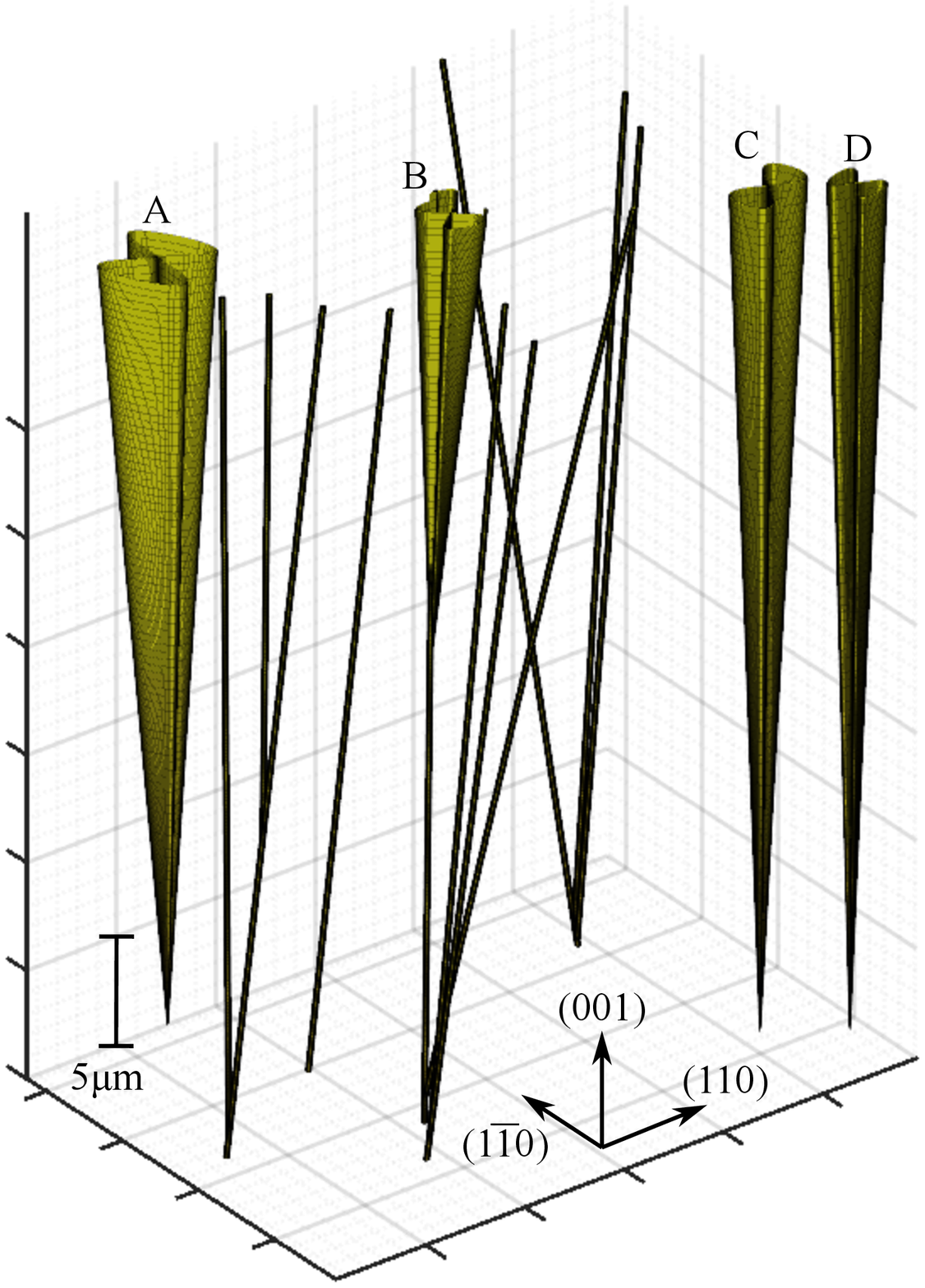}
\caption{Three-dimensional geometric model of compressively strained volumes in the region shown in Fig.~\ref{fig:QuantStrain}, determined from two x-ray projections and constrained by crystal growth assumptions and an associated model of strain features (see text).  Volumes indicated have strain above a threshold of $2\times10^{-4}$.  Labeled regions correspond with labels in Fig.~\ref{fig:QuantStrain}.  For a schematic illustration of the projection of similar features to create data such as that presented in Fig.~\ref{fig:QuantStrain}, see Sec.~I of the Supplemental Material \cite{SM}.  For further discussion of the creation of the three-dimensional model from projected stereoscopic data, see Sec.~K of the supplemental material.}
\label{fig:threedmodel}
\end{center}
\end{figure}
}

\newcommand{\qdmoverlay}{
\begin{figure}[htbp!]
\begin{center}
\includegraphics*[width=\columnwidth]{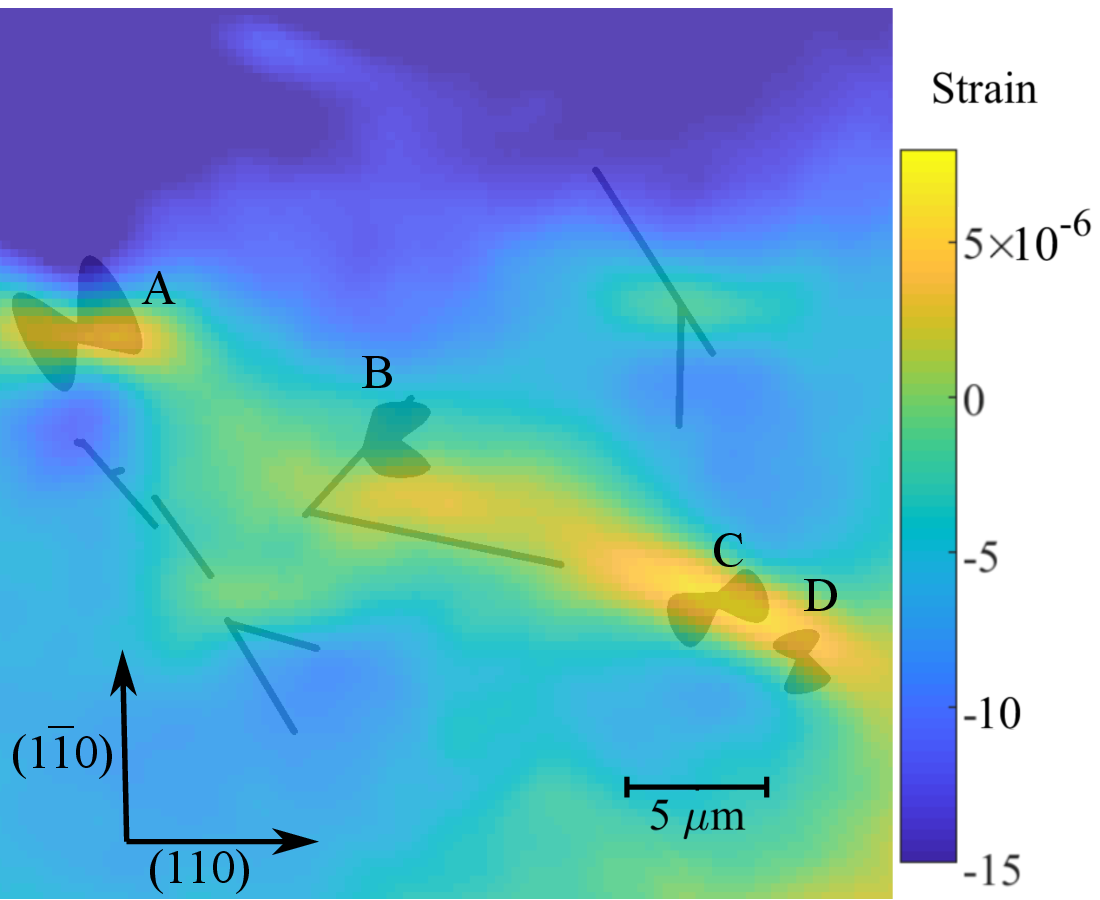}
\caption{Sum of strain tensor elements measured using a quantum diamond microscope (QDM), overlain with vertical projection of a three-dimensional geometric model of strain features derived from SXDM  measurements.  Colormap and corresponding scale bar represent strain measured with QDM, while wireframe objects are the projected model.  Labeled features correspond with labels in Figs.~\ref{fig:QuantStrain} and \ref{fig:threedmodel}.}
\label{fig:qdmoverlay}
\end{center}
\end{figure}
}

\newcommand{\qdmcomparisonfour}{
\begin{figure}[htbp!]
\begin{center}
\includegraphics*[width=\columnwidth]{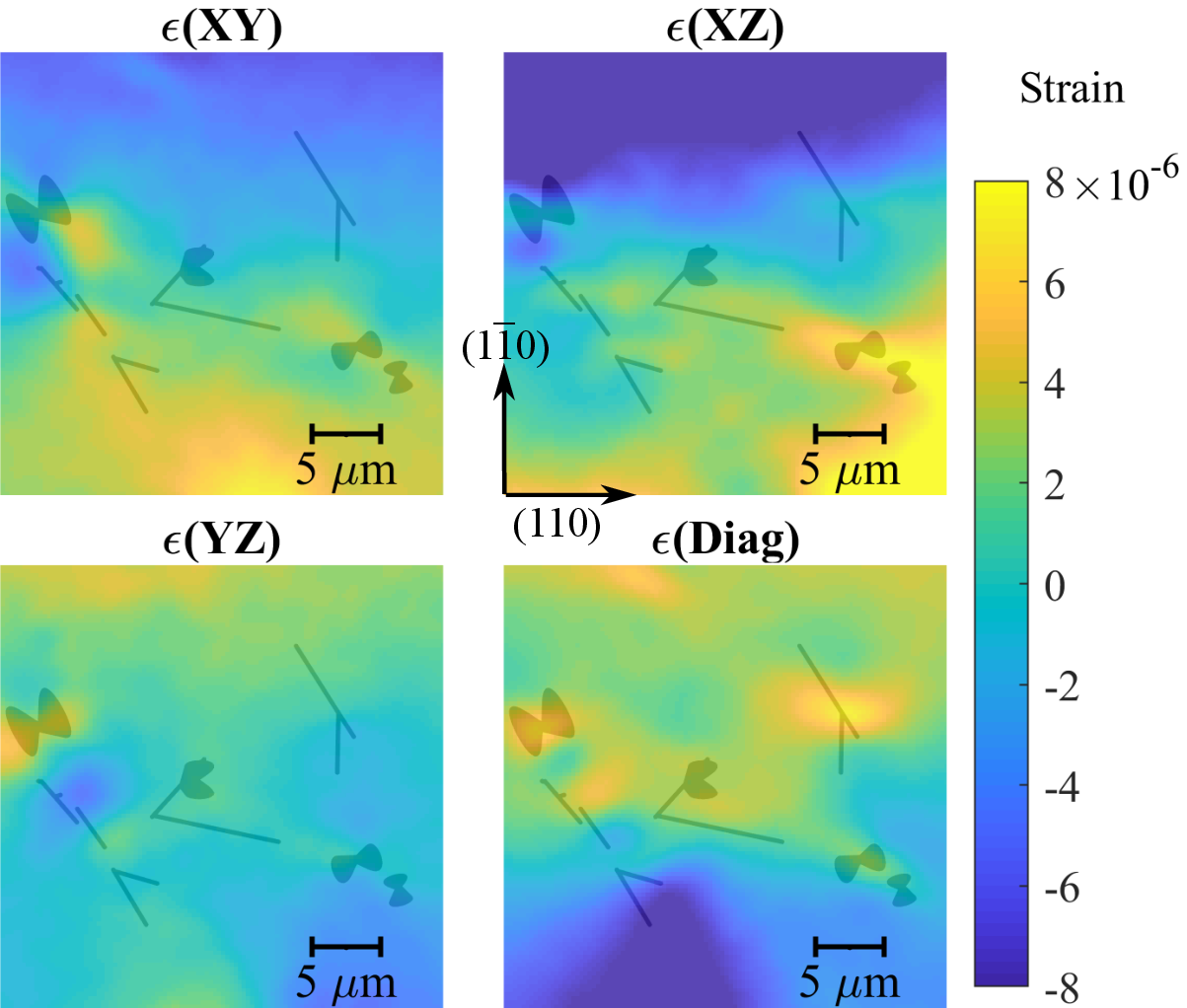}
\caption{Comparison between QDM strain tensor measurements and three-dimensional geometric model of growth defect regions from SXDM measurements.  Strain tensor elements $\epsilon$(ij) give strain on the i crystal plane in the j direction; $\epsilon$(Diag) is the normal strain, while the other tensor elements give shear strain.  Constant offsets have been subtracted from QDM strain tensor measurements to simplify comparison on common color scales.}
\label{fig:qdmcomparisonfour}
\end{center}
\end{figure}
}

\newcommand{\backgroundtwo}{
\begin{figure}[htbp!]
\begin{center}
\includegraphics*[width=\columnwidth]{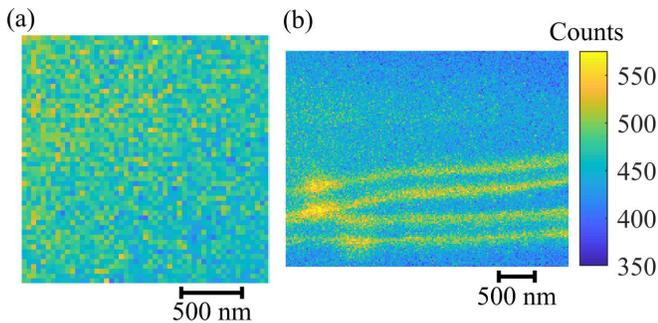}
\caption{SXDM scans of representative ``background'' regions in the CVD quantum sensing diamond, away from large-scale strain features, demonstrate the absence of pre-existing features that could be mistaken for WIMP recoil tracks.  Each plot shows the number of detector counts arising from compressively strained diamond in the CVD overgrowth layer -- see Supplemental Material \cite{SM} Sec.~C for details.  (Note that in these plots we have not applied the background subtraction and strain analysis described in Sec.~\ref{sec:strainmeas}; instead, the total signal from any compressively strained diamond gives a better indication of the presence or absence of small or weak features.)  \textbf{(a)} Despite an overall strain gradient, a 40 nm-increment scan of the nanoprobe did not find features with the $\sim100$ nm length scale expected for WIMP tracks.  \textbf{(b)} A scan of a different background region, this time with 20 nm increments, reveals extended dislocation bundles, which can be easily distinguished from WIMP recoil tracks by their length; again no strain features are observed having $\le100$ nm scale in all three dimensions.}
\label{fig:backgroundtwo}
\end{center}
\end{figure}
}

\newcommand{\bsfit}{
\begin{figure}[htbp!]
\begin{center}
\includegraphics*[width=\columnwidth]{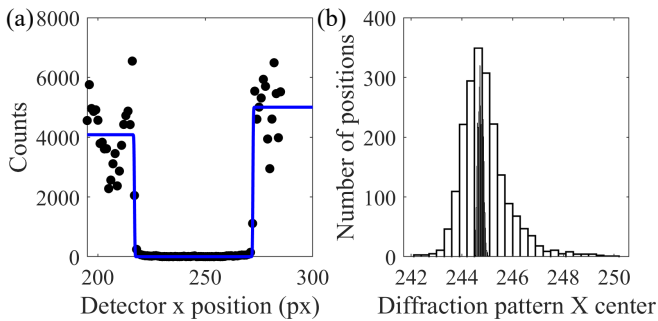}
\caption{\textbf{(a)} Representative diffraction pattern summed over the 2$\theta$ axis, together with a fit to an error function used to determine the center of the diffraction pattern along the X axis.  \textbf{(b)} Histograms of such ``X centers'' for the overall diffraction pattern (black bars) and reduced diffraction pattern (white bars), for X-ray beam positions intersecting growth defect feature A (as labeled in Fig.~4 of the main text).  The white histogram is broader because of the comparatively fewer photons contributing to the reduced lineshape.  The equal centers of these distributions favor compressive strain as the likely origin of the diffraction pattern shifts, as discussed in the text.}
\label{fig:bsfit}
\end{center}
\end{figure}
}

\newcommand{\ROIfig}{
\begin{figure}[htbp!]
\begin{center}
\includegraphics*[width=\columnwidth]{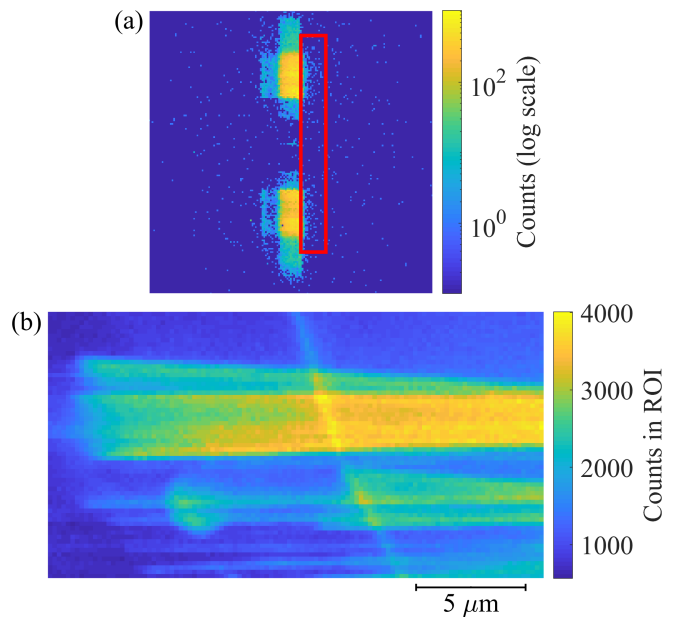}
\caption{\textbf{(a)} Example region of interest (ROI) for initial analysis to classify X-ray beam positions as ``strained'' or ``host crystal;'' the detector ROI for identifying local strain features is marked by the red box.  \textbf{(b)} Total counts in the detector ROI for each beam position within a subset of the region scanned in main text Fig. 5.}
\label{fig:ROIfig}
\end{center}
\end{figure}
}

\newcommand{\calfig}{
\begin{figure}[htbp!]
\begin{center}
\includegraphics*[width=\columnwidth]{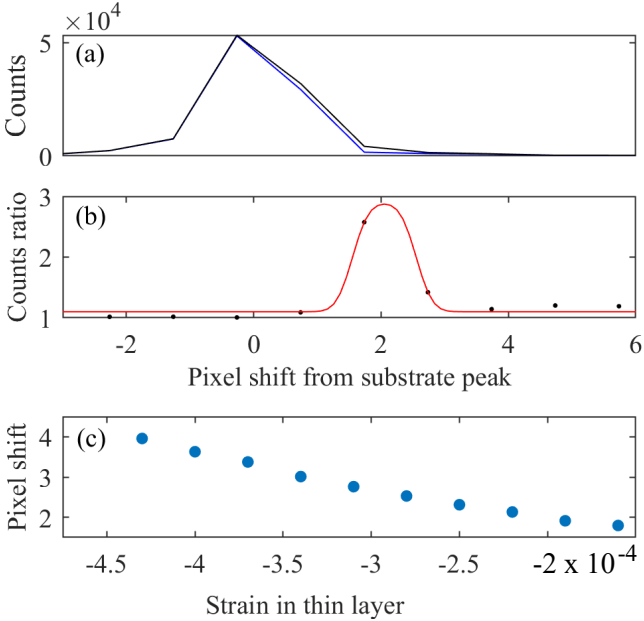}
\caption{Steps for generation of calibration curves to allow conversion between the X-ray diffraction peak position and and short-length-scale strain feature.  \textbf{(a)} A diffraction curve is simulated for an unstrained substrate with (black) and without (blue) a thin layer of strained crystal.  \textbf{(b)} Dividing these two curves gives a simulated reduced diffraction curve; this is fit to obtain its centroid.  \textbf{(c)} This is repeated for different strain values to obtain a calibration curve.}
\label{fig:calfig}
\end{center}
\end{figure}
}

\newcommand{\asymmetriclobes}{
\begin{figure}[htbp!]
\begin{center}
\includegraphics*[width=\columnwidth]{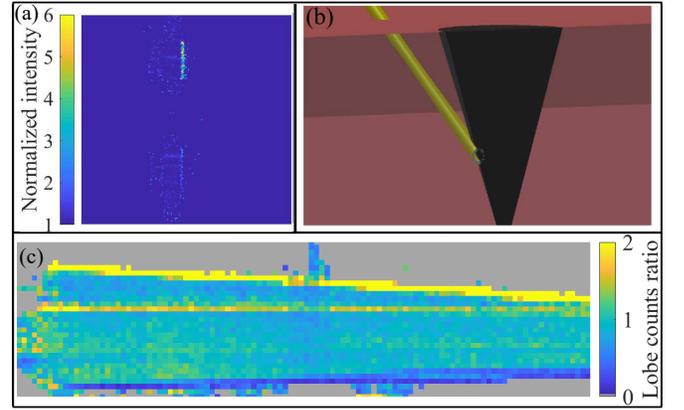}
\caption{Identifying region edges and narrow features using X-ray diffraction-pattern asymmetry.  \textbf{(a)} Reduced lineshape analysis as described in Sec.~2 of the main text, performed on a beam position at the edge of a strained volume.  X-rays are diffracted into the feature peak for only half of the annular beam.   \textbf{(b)} Schematic model of X-ray interactions at edge.  Away from the focal plane, the beam spot size grows; at the sharp edge of a feature, only one-half of the annulus intersects the feature and diffracts X-rays to the detector.  \textbf{(c)} Map of the top-bottom asymmetry of feature peaks, for one ``petal'' feature in the (113) projection shown in main text Fig. 5.  High or low asymmetry values signal the beam is at the edge of a feature.}
\label{fig:asymmetriclobes}
\end{center}
\end{figure}
}

\newcommand{\edgehistogram}{
\begin{figure}[htbp!]
\begin{center}
\includegraphics*[width=\columnwidth]{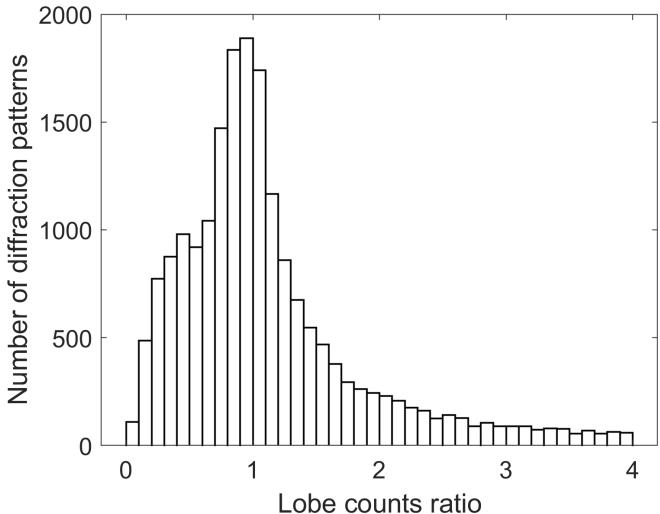}
\caption{Histogram of counts ratios for all X-ray beam positions featured in Fig.~4 of the main text.  We label all features with count ratio less than 0.5 or greater than 2 as ``edge'' points.}
\label{fig:edgehistogram}
\end{center}
\end{figure}
}

\newcommand{\unnormfig}{
\begin{figure}[htbp!]
\begin{center}
\includegraphics*[width=\columnwidth]{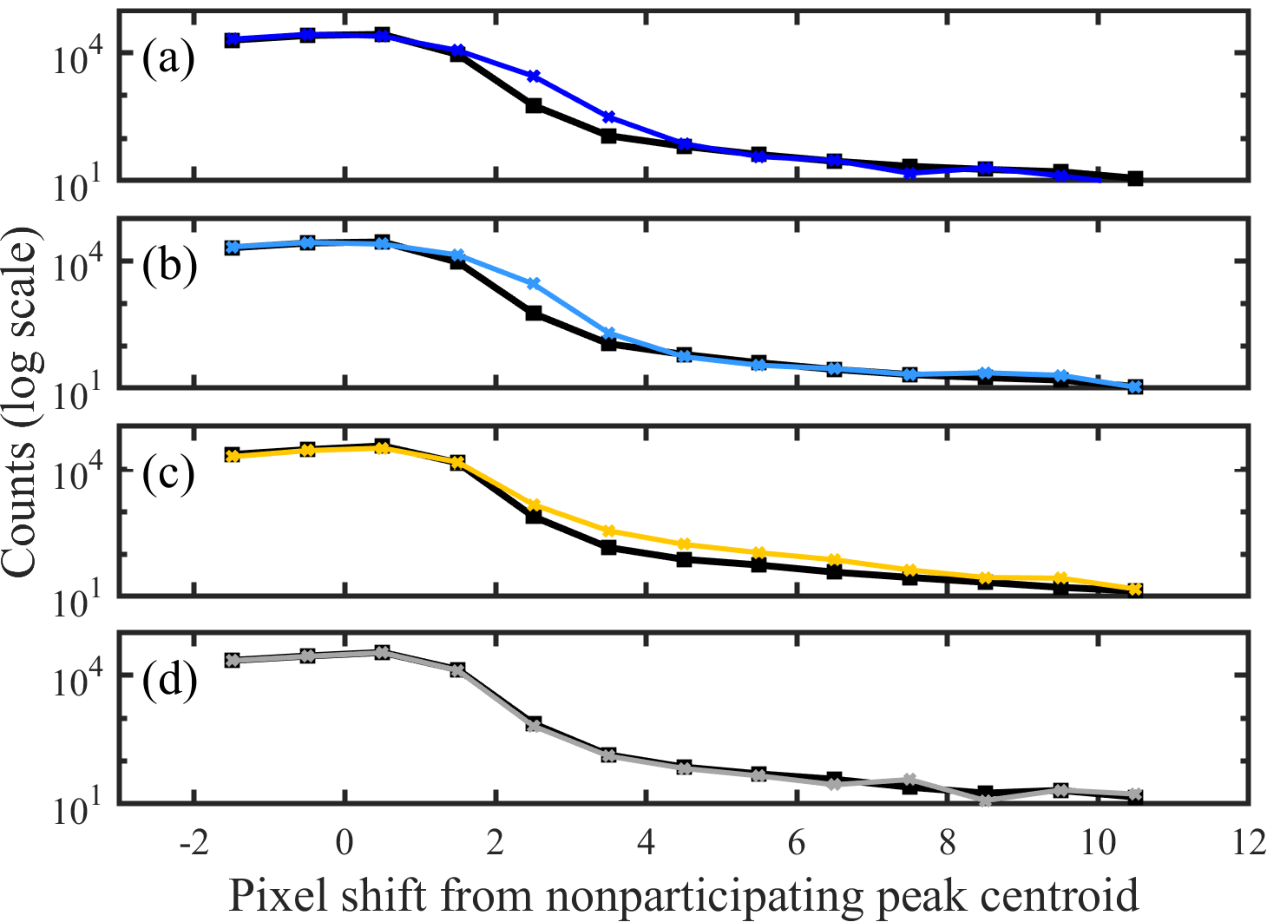}
\caption{Log-scale plot of SXDM diffraction curves without the empirical lineshape analysis (for the same beam positions featured in Fig.~2 of the main text and with the same colors), together with the corresponding local host-crystal diffraction profiles (black points).}
\label{fig:unnormfig}
\end{center}
\end{figure}
}

\newcommand{\otherangle}{
\begin{figure*}[htbp!]
\begin{center}
\includegraphics*[width=\textwidth]{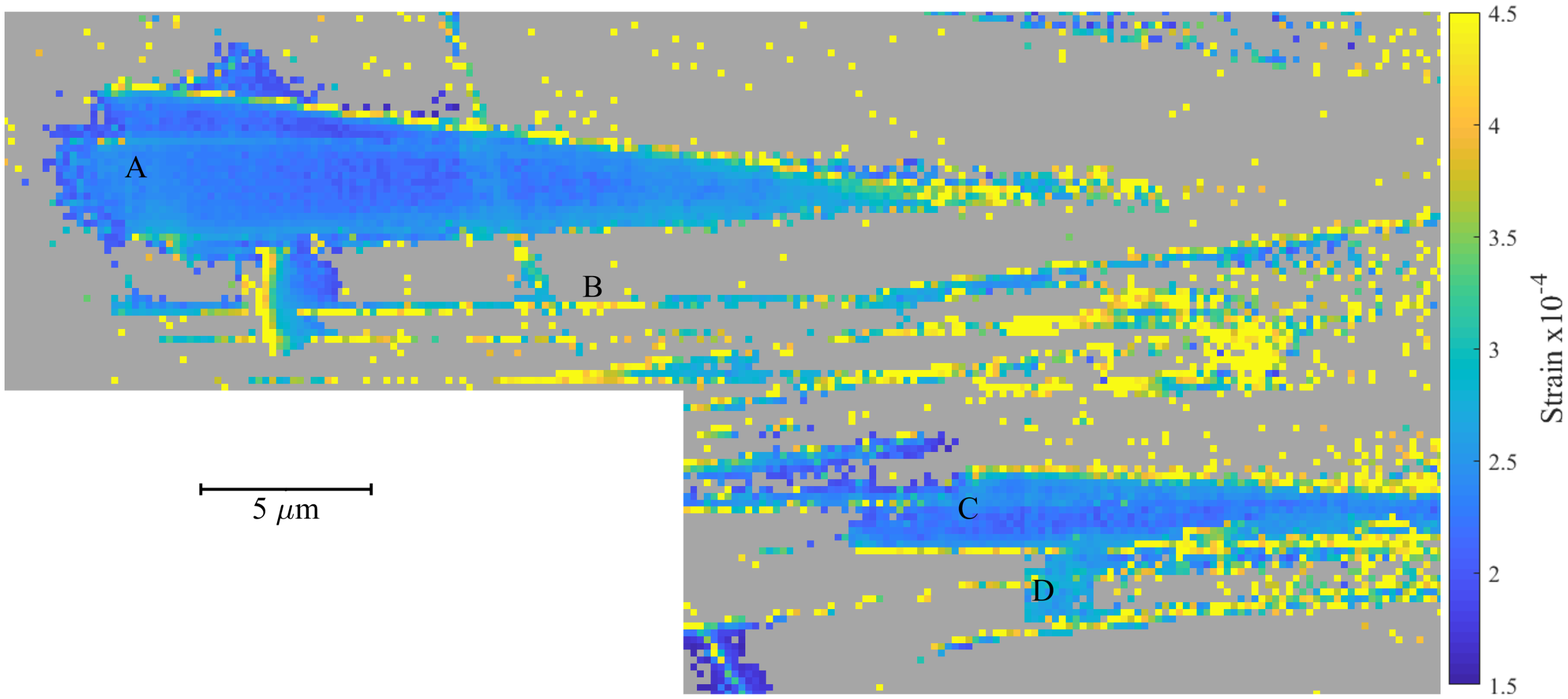}
\caption{SXDM strain map of the region featured in main text Fig.~5, as measured with ($\bar{1}13$) diffraction using a 200 nm scan step size. The data was analyzed using the same background subtraction method detailed in main text Sec.~II.}
\label{fig:otherangle}
\end{center}
\end{figure*}
}

\newcommand{\otheranglesmall}{
\begin{figure}[htbp!]
\begin{center}
\includegraphics*[width=\columnwidth]{otheranglesmall.eps}
\caption{SXDM strain map of the region featured in main text Fig.~5, as measured with ($\bar{1}13$) diffraction. The data was analyzed using the same background subtraction method detailed in main text Sec.~II.}
\label{fig:otherangle}
\end{center}
\end{figure}
}

\newcommand{\constraindata}{
\begin{figure*}[htbp!]
\begin{center}
\includegraphics*[width=\textwidth]{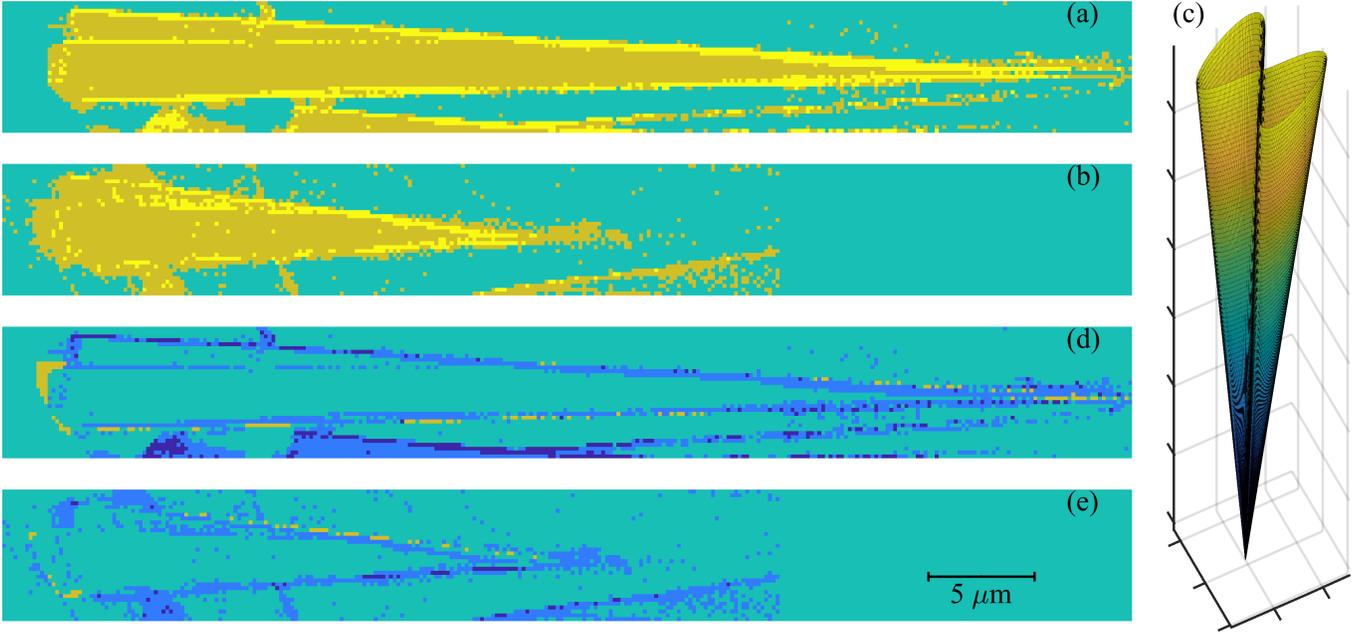}
\caption{Constraining 3-D strain geometric model parameters with SXDM data.  \textbf{(a)-(b)}: Classified data for the upper conical strain feature from the (113) (upper) and ($\bm\bar{1}$13) (lower) projections.  Each beam position is classified based on feature peak fits as either host crystal (teal), strained subvolume (orange), or edge (yellow).  \textbf{(c)}: A three-dimensional model is constructed, which can be projected onto the two measurement planes.  \textbf{(d)-(e)}: Residuals after subtracting each data set from the appropriate model projection.  The sum of squared differences can be fed into a minimizing algorithm to optimize geometric parameters of the model.}
\label{fig:constraindata}
\end{center}
\end{figure*}
}

\newcommand{\fullcomp}{
\begin{figure*}[htbp!]
\begin{center}
\includegraphics*[width=\textwidth]{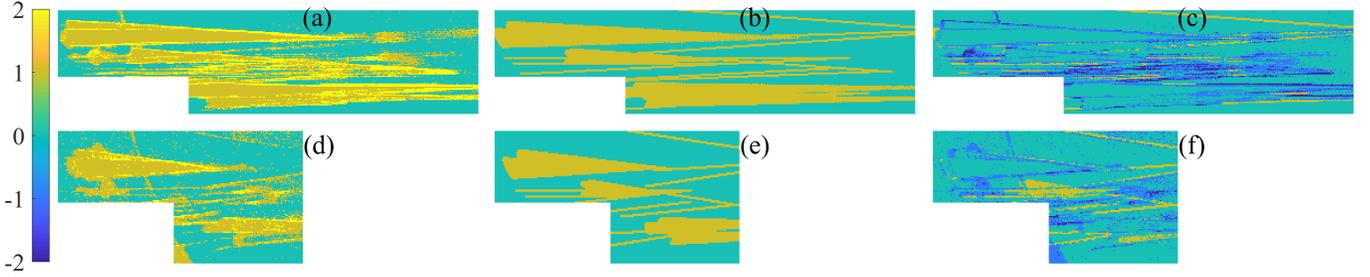}
\caption{Full comparison of strain geometric model with SXDM data.  \textbf{(a)}: Data from the (113) projection, classified in the same way as the example in Fig.~\ref{fig:constraindata}.  \textbf{(b)}: Projection of the geometrical model onto the (113) diffraction plane. \textbf{(c)}: Residuals after subtracting data from projection.  \textbf{(d)-(f)}: Same as above but for the ($\bar{1}13$) projection.}
\label{fig:fullcomp}
\end{center}
\end{figure*}
}

\newcommand{\diamondfeaturesalt}{
\begin{figure}[hbtp!]
\begin{center}
\includegraphics*[width=\columnwidth]{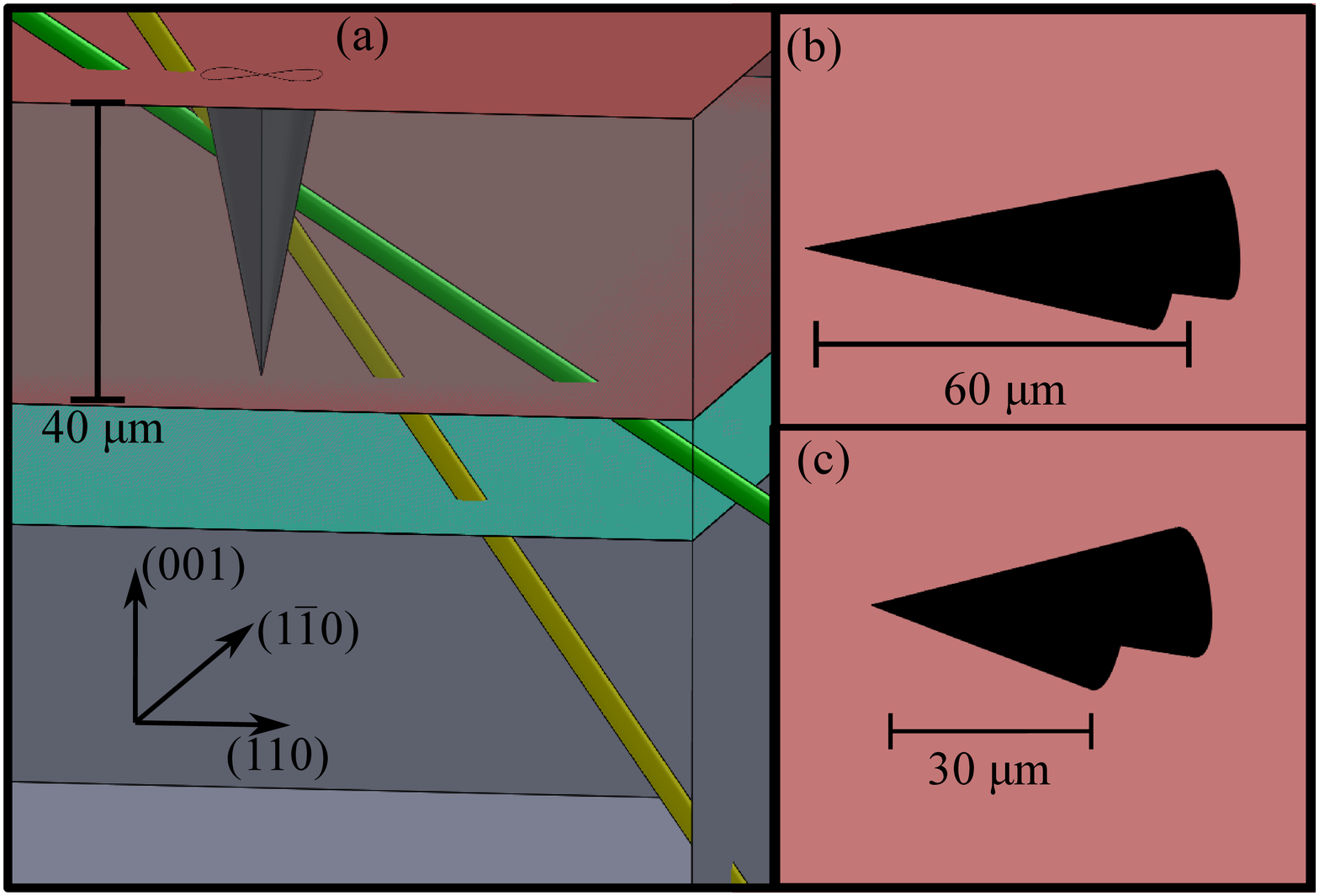}
\caption{Schematic of diamond interrogation with HXN.  This figure demonstrates the interrogation of a two-leaf ``clover'' feature, nucleating at the CVD-substrate interface and propagating along the growth direction, similar to those observed in this study.  As in Fig.~1 of the main text, the green beam addresses the (113) crystal plane, while the yellow beam addresses ($\bm\bar{1}$13).  Panel \textbf{(b)} shows the two-dimensional projection observed by the (113) beam, while \textbf{(c)} shows the ($\bm\bar{1}$13) projection.}
\label{fig:diamondfeaturesalt}
\end{center}
\end{figure}
}

\newcommand{\xrayqdmintro}{
\begin{figure}[hbtp!]
\begin{center}
\includegraphics*[width=\columnwidth]{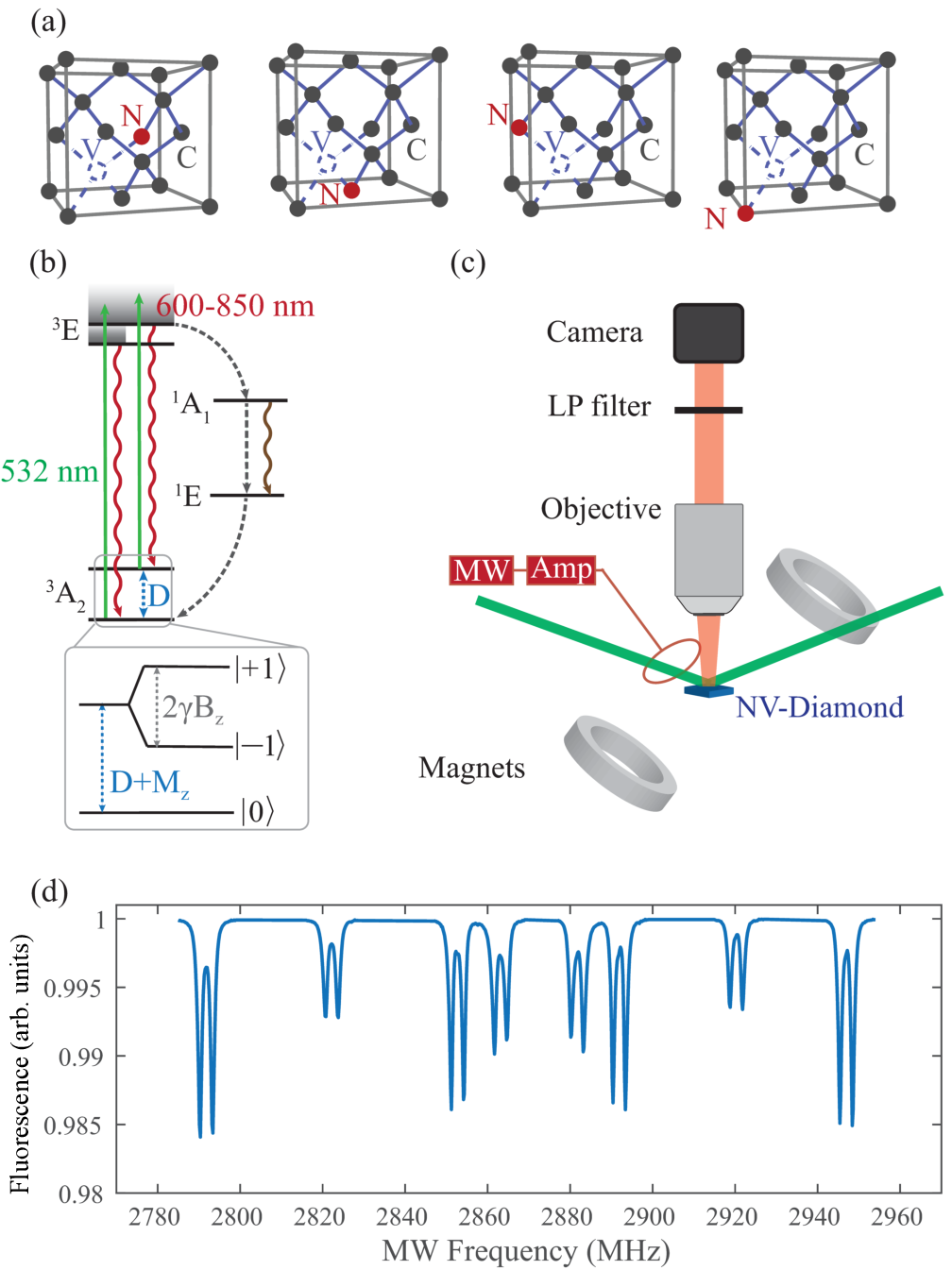}
\caption{Overview of strain imaging with a quantum diamond microscope (QDM).  \textbf{(a)} Crystal structure of the NV center, demonstrating the four possible NV crystallographic axes.    \textbf{(b)} Simplified NV energy level structure.  Inset shows ground  state spin energy levels.  See text for details. \textbf{(b)} Schematic of QDM apparatus.  NVs are excited by green laser light, while red fluorescence is collected by a microscope objective and imaged onto a camera.  A magnetic bias field is provided by permanent magnets, and microwaves are applied using a loop antenna.  \textbf{(d)} Example NV spin spectrum measured by a QDM (see text for details). }
\label{fig:xrayqdmintro}
\end{center}
\end{figure}
}

\section{Introduction}
\label{sec:introduction}
Recently, quantum defects in diamond have emerged as a popular and growing platform for a variety of sensing applications \cite{QDMreview,NVNanoReview2014,NVMRIreview2019,NVMagnetometer2008,NVimaging2008}.  While defect-ensemble based sensing can achieve extremely high spatial resolution and sensitivity, these capabilities are subject to the host diamond's homogeneity and quality \cite{StrainPaper,Friel2009}.  Intrinsic strain in diamond limits many defect-based sensing applications; understanding and optimizing the defect's local crystallographic environment is therefore a major effort in the quantum sensing and diamond growth communities \cite{StrainPaper,PurpleDiamond,Friel2009}.  In particular, strain features with spatial structure near or below the visible-light diffraction limit are difficult to characterize via existing techniques such as optical birefringence \cite{BirefringenceStrain2014} or nitrogen vacancy (NV) center spectroscopy \cite{StrainPaper, AusStrain2019, EnglundStrain2016}.  However, such strain features negatively impact sensing applications by broadening quantum defect spin transition linewidths, thereby limiting their overall sensitivity to electromagnetic fields, temperature, and other phenomena \cite{StrainPaper}.  Additionally, the boundaries of crystallographic strain defects act as charge traps  \cite{SchreckDislocationTrapping2020}, which may have deleterious effects on sensing via reduced spin coherence \cite{ChenDepolarization2015,ChoiDepolarization2017}.  These issues are compounded by the need for fabricated optical microstructures and nanophotonic features for quantum information processing using NV centers and other defects in diamond \cite{NanophotonicPlatform2016,NVNanobeams2013}, which result in large strain over submicron length scales \cite{KnauerFabStrain2020}.  A technique capable of characterizing short-length strain is necessary to better understand the interplay between local strain features and quantum point-defect sensors.

Strain sensing below the optical diffraction limit is also a crucial element of a proposed dark matter (DM) detector in diamond capable of sensing the direction of incident particles \cite{proposal}.  Diamond is a promising target material for future detectors aimed at weakly interacting massive particles (WIMPs) and similar DM candidates \cite{Kurinsky2019}.  Lasting crystal damage from WIMP collisions would act as a memory for the direction of the incoming particle, and would be read out via a multi-step detection process outlined in Sec.~\ref{sec:darkmatter}.  This directional information would enable discrimination from background solar and other neutrinos expected to limit the next generation of WIMP searches \cite{ReadoutDirectional, NeutrinoFloor}.  Demonstrating sensitive, high-spatial-resolution strain measurement in diamond is an important step towards such a detector \cite{InvitedPaper}.

All of the applications highlighted here would also benefit from understanding the three-dimensional topology of strain features. Currently, a variety of techniques are used to produce low strain in quantum defect-hosting diamonds, including substrate surface treatment before CVD deposition and annealing of CVD layers \cite{Friel2009,PurpleDiamond}.  Understanding the three-dimensional topology of remaining strain features, and their evolution under such processes, could help optimize diamond growth and thus improve sensing performance.  Finally, a three-dimensional measurement would also provide a fuller understanding of the strain around and within fabricated microstructures, and allow prediction and mitigation of their effect on nearby defects.

\nanoprobeschematic

In this paper, we demonstrate sub-micron and three-dimensional measurements of diamond strain with scanning X-ray diffraction microscopy (SXDM).  Using the Hard X-ray Nanoprobe (HXN) at the Advanced Photon Source at Argonne National Laboratory \cite{APSNanoprobe2012} --– illustrated schematically in Fig.~\ref{fig:nanoprobeschematic}a --– we measure strain features on two diamond samples with high spatial resolution.  To demonstrate sensitivity to spatially small features, we use a diamond produced by high-pressure, high-temperature (HPHT) synthesis and exhibiting low strain at large length scales.  Measurements on this sample demonstrate the resolution required to characterize local strain environments for single-defect quantum sensing and information applications, as well as to measure crystal damage tracks for a WIMP search (as discussed in Sec.~\ref{sec:darkmatter}).

To demonstrate that SXDM can obtain useful and otherwise inaccessible information in real-world quantum-sensing conditions, without diamonds specifically engineered or selected for the purpose, we use a second diamond sample that is grown by chemical vapor deposition (CVD) for NV quantum sensing (see Supplemental Material Sec. \ref{sec:SuppDiamond} for additional details about both samples).  On this sample, we measure the three-dimensional spatial structure of larger strain features persisting throughout a 40-$\mu$m NV-enriched overgrowth layer; high-resolution measurement of such deep defects is uniquely possible with SXDM, because of the focused X-ray beam's deep penetration into the diamond.  This diamond's multilayered structure, with strain and impurity content variations within and between layers, yields a complex diffraction profile which is not well described by straightforward models.  We therefore demonstrate a background subtraction technique which enables quantitative measurement of small, highly strained volumes occupying a subset of the beam-diamond interaction length.  By combining SXDM measurements at different Bragg conditions and comparing the results to NV spin-spectroscopic strain measurements, we demonstrate that SXDM successfully probes the internal, three-dimensional structure of strain features that limit NV quantum imaging applications \cite{StrainPaper}.   Finally, we characterize both strain sensitivity and the prevalence of pre-existing strain features in the CVD diamond as compared to requirements for the proposed WIMP detector \cite{InvitedPaper}, demonstrating a viable path to directional DM detection.

\section{Scanning X-ray strain measurements in diamond}
\label{sec:strainmeas}

The HXN focuses a bright, monochromatic beam of X-rays to a 10-25 nanometer spot and scans it across a sample held at a Bragg angle \cite{HoltReview2013,MohacsiFresnelOptics2017}.  A two-dimensional photon-counting pixel detector records a diffraction pattern for each scan point.  This pattern encodes the local crystal structure, including its spacing and orientation as well as local impurity content \cite{Warren_1990}.  When the beam passes through regions with differing crystal structure -- such as the CVD diamond's substrate and overgrowth layer -- the resulting diffraction pattern will include contributions from each such region.  Comparing diffraction patterns from nearby beam positions reveals differences in strain over short length scales.  Fig.~\ref{fig:sage} shows a map of local strain features in a region of the HPHT diamond obtained via such a comparison, illustrating the spatial resolution and sensitivity of SXDM on diamond.  The $\sim$100 nanometer length scale of the features observed here is comparable to that expected from a particle-induced damage track, as discussed in Sec.~\ref{sec:darkmatter}.  To extract the strain in these features, we use a background subtraction technique developed to account for the complex diffraction pattern arising from the CVD diamond's multilayered structure (see details below). 

\sage

While the HPHT diamond mapped in Fig.~\ref{fig:sage} has a relatively homogeneous crystal structure, the CVD diamond includes several distinct volumes, schematically illustrated in Fig.~\ref{fig:diamondfeaturestwo}a.  The uppermost, NV-enriched overgrowth layer of the CVD-diamond hosts crystal growth defects, which nucleate at imperfections on the substrate surface and exhibit large strains relative to the surrounding diamond.  The surface at the substrate-layer interface was mechanically polished before CVD overgrowth, resulting in sub-surface damage and creating a broad strain distribution \cite{Friel2009}.  Finally, the substrate is a 500 $\mu$m-thick high-purity diamond plate.

As the beam is scanned in two dimensions across the diamond surface, strain features at different depths in the crystal are therefore projected at an angle determined by the Bragg condition (as illustrated in Figs.~\ref{fig:diamondfeaturestwo}b and \ref{fig:diamondfeaturestwo}c).  We use two distinct Bragg angles to image the CVD diamond, giving diffraction from the crystal planes with Miller indices (113) and ($\bm\bar{1}$13); combining the information from two projections allows us to constrain the three-dimensional structure of features in the sample.  

\diamondfeaturestwo

The lattice spacing and crystal orientation vary through the diamond volume due to defect-induced strain, impurity content, and mechanical processing damage. The diffraction pattern combines the contributions from all of these independent, spatially varying effects.  An example diffraction pattern acquired from the CVD sample is shown in Fig.~\ref{fig:sidepeak}a.  The diffraction pattern from an ideal, unstrained, 40 $\mu$m-thick diamond would appear as a single column of detector pixels in Fig.~\ref{fig:sidepeak}a \cite{YingKinematicCalcs2010, UWDynamicalSims2018}; the broad peak observed here arises from multiple complicating factors, including a variety of strain states among the layers as well as interactions at the layer interfaces.

Creating a full analytical model of the diffraction pattern from such a sample requires detailed microscopic models for vacancy and impurity incorporation during the different CVD growth steps, for polishing damage at the substrate surface, and for crystal structure and X-ray diffraction behavior at the substrate-overgrowth interface.  Developing such models is challenging because of the complexity of the stochastic processes involved, as well as the proprietary nature of diamond growth parameters.  Instead, we analyze local strain features in the overgrowth layer, independent of the complex but slowly spatially-varying overall lineshape of the diffraction pattern associated with the layered diamond structure.  qhw  Two such features are schematically illustrated in Fig.~\ref{fig:nanoprobeschematic}: the clover-shaped dislocation feature of panel (c) and the nuclear recoil track of panel (d).  Conversely, the strain gradient of panel (b) would contribute to the slowly varying overall lineshape of the diffraction pattern, and thus would not be highlighted in our analysis; strain over these larger length scales can be more easily measured by other methods (see Sec.~\ref{sec:comparison}).  While the HPHT diamond mapped in Fig.~\ref{fig:sage} has a more homogeneous crystal structure -- lacking, for example, the NV-enriched overgrowth layer of the CVD diamond -- it still includes large-scale, slowly varying strain effects due to mechanical polishing and a high nitrogen impurity content.  We therefore apply the same analysis to both diamonds.

\sidepeak

\QuantStrain

The slowly varying overall lineshape contains crystal structure and diffraction information about the surrounding diamond; to isolate and quantitatively analyze short-length strain features, we treat this lineshape as a background and subtract it using a technique illustrated in Fig.~\ref{fig:sidepeak}b-c.  First, we construct a local host-crystal diffraction profile, by averaging diffraction patterns from nearby positions where the X-ray beam does not intersect strain features.  (See Supplemental Material Sec. \ref{sec:SuppBkg} for detail on how this profile is constructed.)  We then divide the measured diffraction pattern pixel-by-pixel with the host-crystal diffraction profile.  This de-emphasizes the slowly varying overall lineshape and highlights the diffraction contribution from local strain features, resulting in a ``reduced diffraction pattern'' as shown in Fig.~\ref{fig:sidepeak}b.  Crystal strain changes the Bragg angle $\theta$, which shifts the diffraction pattern along the detector's 2$\theta$ axis (chosen to lie in-plane with the Bragg angle).  We therefore sum each column of the reduced diffraction pattern to create a diffraction curve, as shown in Fig.~\ref{fig:sidepeak}c.  Finally, we fit these curves to a Gaussian lineshape and extract the centroid position.  (We choose a Gaussian because the spread of local impurity concentrations and strains within the beam spot should be approximately normally distributed).  

These measurements are sensitive to changes in both spacing and orientation of the diffracting crystal planes; sensitivity to both of these effects is valuable, as both types of lattice distortion affect defect-based sensing and both are expected to occur in a particle track. Comparing the results obtained from two Bragg angles and incorporating information from the two-dimensional shape of the diffraction pattern constrains, but does not completely determine, the contributions of these effects (see Supplemental Material Sec. \ref{sec:SuppTilt} for this analysis). To evaluate the sensitivity of this measurement and to quantify the amount of lattice distortion observed in our sample, we analyzed the data in the case of uniform lattice orientation, where the signal arises entirely from strain-induced changes in the lattice spacing.  Fig.~\ref{fig:QuantStrain} demonstrates the results of this analysis.  

To convert from the centroid position of the reduced lineshape to strain (under this uniform-orientation condition), we generate calibration curves calculated using kinematic diffraction theory for a multilayered diamond structure \cite{YingKinematicCalcs2010,TilkaOpticalSimulations2016}.  See Supplemental Material Sec. \ref{sec:SuppCal} for details on the generation of these curves.  Our samples are thick enough for dynamical diffraction effects to manifest \cite{BattermanDDtheory1964,UWDynamicalSims2018}, but the resultant uncertainty in measured strain is small compared to other sources - see Supplemental Material Sec. \ref{sec:SuppUnc} for details.

The width of the diffraction peak due to the surrounding crystal and substrate, together with the background subtraction technique, impose a minimum detectable strain threshold on our measurements of local strain features.  Background subtraction allows us to quantitatively investigate strain features in the top CVD layer, despite the presence of a broad and spatially inhomogeneous strain distribution in the surrounding crystal.  However, background subtraction also suppresses features whose diffraction peaks are insufficiently shifted from the host-crystal diffraction profile.  To be detectable with low uncertainty, a feature must have enough strain for its diffraction peak to be shifted out of the slowly varying overall diffraction profile.  The actual minimum threshold depends on the linewidth of the host-crystal diffraction peak; in the CVD sample it is $\sim$1.5 detector pixels.  (See Supplemental Material Sec. \ref{sec:SuppUnnorm} for further details.) This leads to a minimum measurable compressive strain of $\sim1.6\times10^{-4}$.  

Fig.~\ref{fig:QuantStrain} shows a map of strain features in one region of the CVD diamond, as projected onto the (113) crystal axis.  Each X-ray beam position is categorized, based on whether the amplitude of the reduced diffraction curve is greater than a threshold, as either part of a strain feature or as only containing the host crystal.  Points where the beam does not intersect a strain feature are greyed out in Fig.~\ref{fig:QuantStrain}; for strain feature points, the calibrated strain is plotted at each beam position.  The strain magnitude is determined from the centroid of the diffraction curve fit and the calibration curves.  Four extended growth defects are clearly identifiable as blue regions labeled A-D in this map, with internal strains (in the uniformly oriented case) of $2.5-3\times10^{-4}$.  We note that lattice distortion in the features is high at the edges, while feature centers exhibit a smoothly varying structure.  This ability to resolve differences in deformation within a growth defect demonstrates the power of SXDM with appropriate background subtraction to investigate crystal growth and defect incorporation. 

In addition to these four extended defect regions, we also measure several thin, apparently linear high-distortion features; these are most likely the edges of extended defect regions exhibiting strain in a direction to which we are less sensitive.  Bragg diffraction measures changes in the spacing between crystallographic planes -- in other words, the projection of strain onto the axis of diffraction.  For strain within growth features in diamond, the Burgers vector -- the direction of the crystal lattice distortion --  generally points along one of the $\langle$110$\rangle$ family of crystal axes \cite{E6xray2008, Pinto_Jones_2009}.  Unit vectors for axes in this family can have dot products of 0.9, 0.4 or 0 with the (113) and ($\bm\bar{1}$13) mirror planes used in our measurements.  For example, if a feature has an average Burgers vector of (101), a (113) diffraction measurement would yield 0.9 times its nominal strain, while a measurement with ($\bm\bar{1}$13) diffraction would only yield 0.4 times the nominal strain.  Comparing the projection onto different crystal planes thus allows us to constrain the average Burgers vector within particular strain features.

For features in certain projections we may only detect the high-strain edges, while the strain in their central regions would be below this threshold.  For example, features A and C of Fig.~\ref{fig:QuantStrain} have the same strain projection when measured with both $(\bm\bar{1}13)$ and (113) diffraction.  Therefore, assuming uniform lattice orientation as discussed above, their average Burgers vectors are likely along (011).  Conversely, only the edges of features B and D are measurable using $(\bm\bar{1}13)$ diffraction, meaning their average Burgers vectors likely lie along (101) (again, in the uniform-orientation case).  The ``rod-like'' features observed in both diffraction angles are likely of similar origin, representing the high-strain edges of dislocation features whose Burgers vectors lie mostly along the (0$\bm\bar{1}$1), (110) or ($\bm\bar{1}$10) axes, with small projections onto both of our diffracting planes.  The measured strain at the centers of features A-D in Fig.~\ref{fig:QuantStrain} is about $2.5-3\times10^{-4}$.  For features B and D, which we hypothesize have (101) Burgers vectors, the projection onto the ($\bm\bar{1}$13) diffraction axis yields strain of about $1.1-1.3\times10^{-4}$ -- below the minimum measurable strain.  This result is consistent with our observations, where only the edges of features B and D are visible with ($\bm\bar{1}$13) diffraction.

Note that this minimum detectable value applies to compressive strain. A similar analysis applies to tensile strain, which will shift the diffraction peak towards smaller 2$\theta$.  For the CVD diamond sample, tensile strain shifts the strain-feature-induced diffraction peak across the detector region dominated by the slowly varying overall lineshape; the background subtraction technique is therefore less sensitive to small tensile strain.  Tensile strain at the level of $\sim2\times10^{-3}$ would be needed for a feature to be detectable.  In practice, we only observe compressively strained regions of growth defects in the 40-$\mu$m overgrowth layer in this sample.  In general, the method presented here is applicable to both compressive and tensile strains, with the minimum measurable strain depending on the uniformity of the host crystal.  However, a major advantage of our method is the ability to extract information even in the presence of a very inhomogeneous host crystal as found in the CVD quantum sensing diamond.

(In Fig.~\ref{fig:QuantStrain}, we do not measure features with sub-micron spatial scale in all dimensions, similar to those seen in Fig.~\ref{fig:sage}; see discussion in Supplemental Material Sec. \ref{sec:supplack})

\section{Three-dimensional model of strain feature geometry}

High-spatial-resolution, three-dimensional strain mapping within diamond samples is essential to the dark matter detection proposal discussed in section \ref{sec:darkmatter}.  Additionally, three-dimensional, high-resolution strain measurements enhance analysis of CVD diamond growth and optical structure fabrication.  By combining measurements at two diffraction angles with prior information of growth conditions and established knowledge about the structure of growth defects in CVD diamond, we generate a three-dimensional model of strain feature geometry in this 24,000 $\mu$m$^3$ region of the CVD diamond sample, as shown in Fig.~\ref{fig:threedmodel}.  Constructing an assumption-free three-dimensional model using this method would require additional measurements at orthogonal Bragg conditions \cite{HofmannFIBStrain2017,HofmannFullStrainTensor2020}; however, constraints on experimental geometry during the current demonstration limited us to the non-orthogonal (113) and ($\bm\bar{1}$13) mirror planes.  We constrain the model with two assumptions: first, that extended defects in CVD diamond generally nucleate at the interface with the substrate and propagate close to (but not perfectly along) the growth direction \cite{E6xray2008}; and second, that the integrated strain far from a defect must converge to zero \cite{Eshelby1954}.  

Two features types that satisfy these constraints, and have been widely observed in diamonds grown under similar conditions to our sample, are bundles of linear edge or screw dislocations and ``petal'' or ``clover''-shaped extended defects \cite{E6xray2008,StrainPaper,PintoExtendedDefects,RamanStrain2011}.  We therefore build our three-dimensional model from such features, constrained by the diffraction measurements.  See Supplemental Material Sec. \ref{sec:supppetal} for further discussion of these feature types.

\threedmodel

We first identify individual features in the projected strain maps, and classify them either as ``petals'' or ``rods''.  We note that some of the ``rods'' likely represent the high-strain edges of ``petal'' features whose centers have (113) or ($\bm\bar{1}$13) projections below the measurement threshold, as discussed in Sec.~\ref{sec:strainmeas}; measurements at additional projection angles would be required to resolve this ambiguity.  For each identified feature, we construct a corresponding entity in our three-dimensional model, and constrain its geometrical parameters by minimizing the difference between data and model.  (See Supplemental Material Sec. \ref{sec:SuppModel} for further details on constraining parameters of the model.)   

We finally note two nontrivial results obtained from this analysis.  First, the strain features exhibit sharp, distinct edges where they intersect the host crystal matrix; and second, within the features we observe smoothly varying but nonuniform strain.  While constructing a full microscopic model of the strained regions is beyond the scope of this work, our data suggest that the crystal dislocations that comprise such clover-shaped features may be concentrated at or near the ``petal'' boundaries, enclosing a region of strained crystal, rather than evenly spread throughout the strained volume.

\section{Comparison with strain measurements from quantum diamond microscope (QDM)}
\label{sec:comparison}

Optical methods for strain measurements in diamond are limited in spatial resolution by diffraction to a few hundred nm and are not in general sensitive to three-dimensional spatial structure.  In particular, ensemble NV spectroscopy measures integrated stress throughout the nitrogen-doped layer \cite{StrainPaper}.  Despite lower spatial resolution and two-dimensional projective imaging, optical strain/stress measurements via NV spin-state spectroscopy offer complementary advantages over SXDM.  Wider fields of view enable fast location of regions of interest for high-resolution diffraction measurements, which is necessary for directional WIMP detection \cite{InvitedPaper}.  Additionally, such optical methods are capable of measuring strain as low as $10^{-6}$, albeit only as an average over an optical pixel.  Finally, from optical measurements it is possible to reconstruct the entire strain tensor \cite{StrainPaper}, which could inform the choice of Bragg angles used for diffraction.

\qdmoverlay

To illustrate the complementary application of X-ray and optical methods, we used a quantum diamond microscope (QDM) \cite{QDMreview} to image the same field of view featured in our nanoprobe diffraction measurements of the CVD quantum sensing diamond.  In a QDM, NV centers in diamond are excited with green laser light, and their spin-state controlled with applied microwaves; the spin state and environment of the NVs can then be determined.  QDMs are commonly used to image DC magnetic fields, but have also been used for sensitive stress and strain measurements.  QDM results reported in this work were performed by determining the stress Hamiltonian from NV spin-state-dependent optical measurements.  For an overview of QDM methods, see Supplemental Material Sec. \ref{sec:SuppQDM}; for full details of the technique and apparatus used, see Ref.~\cite{StrainPaper}.  The stress tensor was converted to strain using the measured elasticity tensor for diamond \cite{CVDYoungsModulus1992}.  

In the present work, QDM images were taken after the SXDM measurements; to find appropriate regions of interest to develop SXDM on diamond, we performed relatively time-consuming wide-area scans with the HXN.  Future development of SXDM for diamond engineering and dark matter detection will use QDM measurements to more extensively inform SXDM parameters. 

\qdmcomparisonfour

Figures \ref{fig:qdmoverlay} and \ref{fig:qdmcomparisonfour} compare the strain feature geometrical model obtained from SXDM data and our methodology with QDM strain tensor measurements.  The QDM measurements demonstrate poorer spatial resolution (1-10 micron), limited by collection of light from the entire axial focal length of the microscope objective used.  Nonetheless, these figures demonstrate good agreement between the two techniques with regards to strain geometry over large lengthscales.  

The QDM measurement distinguishes between tensile and shear strains, while SXDM measures a projection of both onto the diffracting crystal axes.  For the comparison in Fig.~\ref{fig:qdmoverlay}, we therefore sum all strain tensor elements measured using the QDM, to obtain a map of strain feature positions regardless of orientation; we compare that map to the geometry obtained from SXDM data.   In this comparison, the positions of the four ``petal'' features A, B, C, and D correspond to the three strongest strain features identifiable in the QDM data (features C and D are too close together for the QDM to resolve them individually).  This agreement between the SXDM and QDM approaches validates the SXDM measurement of strain features via the reduced diffraction curve, as well as the three-dimensional model.  Additionally, the ``rod'' positions agree relatively well with the remaining strain features in the QDM measurement.  This supports the hypothesis that some or all ``rods'' may actually be high-strain edges of additional ``petal'' regions, at whose centers the projection onto the (113) and ($\bm\bar{1}$13) axes is below the SXDM detection threshold.

Figure \ref{fig:qdmcomparisonfour} compares individual elements of the strain tensor from QDM measurements with the geometrical model from SXDM measurements.  Generically, a strain tensor element $\epsilon($ij$)$ gives the change in the position of the i crystal plane along the j direction.  For i=j, $\epsilon$ is the normal strain on the i planes, while for i$\neq$j $\epsilon$ is the shear strain.  The comparison in Fig.~\ref{fig:qdmcomparisonfour} shows the additional information that can be gained from performing both QDM and X-ray diffraction measurements on the same sample; note that several of the strain features come through much more strongly in some strain tensor elements than in others, especially the lower-left ``rod'' features in $\epsilon$(YZ) and the upper-right ``rod'' feature in $\epsilon$(Diag).  The QDM thus illuminates the direction of shear strains, and gives information about strain on longer length scales, while the nanoprobe measurement gives higher spatial resolution and reveals the three-dimensional structure of the strain features.  

We note that several other methods have been applied to optically measure strain in CVD diamond, although none fills the role of SXDM for high-resolution, three-dimensional strain measurements.  Optical birefringence measures strain integrated through the length of the crystal \cite{BirefringenceStrain2014}.  Bulk X-ray tomography has been applied to similar CVD diamonds: it provides three-dimensional images, but suffers from lower spatial and strain resolution than SXDM \cite{E6xray2008}.  Strain measurements using spectroscopy of single NV centers are perhaps the most comparable to SXDM in terms of obtaining three-dimensional information \cite{EnglundStrain2016}.  However, the need to resolve individual NV centers in the bulk of the diamond limits the spatial resolution to the $\sim$micron level, and limits the technique's application to diamonds with a narrow range of NV densities.

\section{Towards dark matter detection}
\label{sec:darkmatter}

\DMdir

We finish by characterizing the performance of the SXDM strain mapping technique as applied to the CVD quantum sensing diamond in the context of the proposed directional WIMP detector, see Fig.~\ref{fig:DMdir} \cite{proposal,InvitedPaper}.  We can estimate the strain signal from a WIMP collision following \cite{InvitedPaper}, and compare to the sensitivity analysis of Sec.~\ref{sec:strainmeas}.  We scale the strain per crystal lattice vacancy by the number of vacancies created \cite{proposal} and the $r^{-3}$ distance scaling of strain due to point defects \cite{Eshelby1954}.  For a particle scattering event imparting 10 keV to a carbon nucleus --- the low end of a diamond detector's directional sensitivity range \cite{proposal} --- the average strain within 30 nm of the resulting damage track will be $\sim1.8\times10^{-4}$.  This would be detectable with the SXDM technique even with the imperfect, highly inhomogeneous diamond sample used in this demonstration.    

Figure \ref{fig:backgroundtwo} presents results from a background SXDM scan, taken in two arbitrarily chosen regions away from strain features large enough to appear in birefringence or QDM measurements.  Figure \ref{fig:backgroundtwo}b includes a group of extended linear defects whose observed strain signals are approximately 100 nm wide, demonstrating spatial resolution well below the optical diffraction limit.  This matches the expected length scale of WIMP-induced damage tracks for much of the target energy range.  The spatial resolution limit of the HXN is $\sim10$ nm at the minimum focal size \cite{MohacsiFresnelOptics2017}, which matches requirements for resolving WIMP signatures \cite{InvitedPaper}. If required, higher resolution X-ray strain imaging is possible via advanced techniques such as Bragg projection ptychography \cite{Pfeiffer_2018,3DBPP2017,multiangleBPP2018}.

\backgroundtwo

In addition to an instrument and technique capable of measuring nanoscale damage tracks in diamond, the proposed WIMP directional detection method requires production of CVD diamond crystals without pre-existing strain features that could be misidentified as damage tracks.  The data in Fig.~\ref{fig:backgroundtwo} represents an initial step towards demonstrating such samples; in high-resolution SXDM scans, we see no evidence of strain features with submicron extent in all three dimensions, as expected for WIMP-induced damage tracks.  In Fig.~\ref{fig:backgroundtwo}a, despite the presence of a broad overall gradient we see no sharp features; and in Fig.~\ref{fig:backgroundtwo}b, we observe features that are $\sim100$ nm wide but microns in length.  Although further background characterization will be required before production of a directional WIMP detector, these initial results support the expectation that after high-temperature annealing, defects in CVD diamond should either be point-like or extended, without structure at intermediate length scales $\sim$100 nm in all three dimensions \cite{PintoExtendedDefects,Friel2009}.  

\section{conclusion}

We performed scanning X-ray diffraction microscopy (SXDM) with high spatial and strain resolution on two diamond samples -- an HPHT diamond sample with a relatively homogeneous crystal structure, and a CVD diamond sample featuring a 40 micron layer of dense nitrogen vacancy (NV) centers.  Using a background subtraction technique applied to SXDM diffraction curves, we measured the strain in small growth defect volumes within the crystal matrix in both diamonds.  In the HPHT diamond, we measured spatially-small strain features, with length scales of order 100 nm, demonstrating the resolution and feature sensitivity achievable with SXDM.  In the CVD diamond, we measured strain in micron-scale growth defects persisting throughout the 40-micron overgrowth layer, despite a broad, inhomogeneous diffraction signal arising from the sample's layered structure and its high impurity content.  Such measurements will grant valuable insight into defect development and mitigation during diamond growth, with important implications for future defect-based quantum sensing efforts, as well as for the development of diamond structure fabrication.

By combining measurements at two diffraction conditions, we created a geometrical model of several three-dimensional strained volumes in the CVD quantum sensing diamond.  Subsequent stress measurements performed with NV spin-state spectroscopy using a quantum diamond microscope (QDM) showed good agreement with this model, demonstrating the capability of SXDM to measure features that affect NV sensing, as well as the complementary advantages of the two techniques.  Their combination offers a window into the three-dimensional, microscopic structure of strain features in diamond.

Finally, we evaluated the performance of the SXDM techniques for a proposed directional detector of WIMP dark matter.  The HPHT diamond results demonstrate the ability to measure features at the lengthscales expected for WIMP-induced nuclear recoil tracks.  The strain sensitivity and three-dimensional resolution -- characterized in the inhomogeneous, layered CVD diamond -- approach or exceed the requirements for WIMP detection.  In an ideal detector segment -- a homogeneous crystal without a substrate layer or mechanical polishing damage -- the minimum detectable nanoscale strain with the SXDM method should be between two and twenty times smaller \cite{HoltReview2013}.  Additionally, a limited initial survey found no pre-existing backgrounds to impede a dark matter search.  Such a detector technology -- with high target density and directional detection capability -- would enable WIMP searches to push sensitivity below the neutrino floor, opening a new path for future generations of WIMP detectors.

\section{Acknowledgements}
We acknowledge useful discussions and feedback from Connor Hart  and Reza Ebadi. This work was supported by the DOE QuANTISED program under Award No. DE-SC0019396; the Army Research Laboratory MAQP program under Contract No. W911NF-19-2-0181; the DARPA DRINQS program under Grant No. D18AC00033; and the University of Maryland Quantum Technology Center.  This work was performed in part at the Harvard Center for Nanoscale Systems (CNS), a member of the National Nanotechnology Coordinated Infrastructure Network (NNCI), which is  supported by  the  National  Science  Foundation  under NSF award no. 1541959.  Work at Argonne was supported by the Center for Novel Pathways to Quantum Coherence in Materials, an Energy Frontier Research Center funded by the U.S. Department of Energy, Office of Science, Basic Energy Sciences in collaboration with the U.S. Department of Energy, Office of Science, National Quantum Information Science Research Centers.  The SXDM measurements were performed at the Hard X-ray Nanoprobe Beamline operated by the Center for Nanoscale Materials and the Advanced Photon Source, both Office of Science user facilities supported by the U.S. Department of Energy, Office of Science, Office of Basic Energy Sciences, under Contract No. DE-AC02-06CH11357.

\bibliography{MasonNVbib.bib}

\begin{thebibliography}{50}%
\makeatletter
\providecommand \@ifxundefined [1]{%
 \@ifx{#1\undefined}
}%
\providecommand \@ifnum [1]{%
 \ifnum #1\expandafter \@firstoftwo
 \else \expandafter \@secondoftwo
 \fi
}%
\providecommand \@ifx [1]{%
 \ifx #1\expandafter \@firstoftwo
 \else \expandafter \@secondoftwo
 \fi
}%
\providecommand \natexlab [1]{#1}%
\providecommand \enquote  [1]{``#1''}%
\providecommand \bibnamefont  [1]{#1}%
\providecommand \bibfnamefont [1]{#1}%
\providecommand \citenamefont [1]{#1}%
\providecommand \href@noop [0]{\@secondoftwo}%
\providecommand \href [0]{\begingroup \@sanitize@url \@href}%
\providecommand \@href[1]{\@@startlink{#1}\@@href}%
\providecommand \@@href[1]{\endgroup#1\@@endlink}%
\providecommand \@sanitize@url [0]{\catcode `\\12\catcode `\$12\catcode
  `\&12\catcode `\#12\catcode `\^12\catcode `\_12\catcode `\%12\relax}%
\providecommand \@@startlink[1]{}%
\providecommand \@@endlink[0]{}%
\providecommand \url  [0]{\begingroup\@sanitize@url \@url }%
\providecommand \@url [1]{\endgroup\@href {#1}{\urlprefix }}%
\providecommand \urlprefix  [0]{URL }%
\providecommand \Eprint [0]{\href }%
\providecommand \doibase [0]{https://doi.org/}%
\providecommand \selectlanguage [0]{\@gobble}%
\providecommand \bibinfo  [0]{\@secondoftwo}%
\providecommand \bibfield  [0]{\@secondoftwo}%
\providecommand \translation [1]{[#1]}%
\providecommand \BibitemOpen [0]{}%
\providecommand \bibitemStop [0]{}%
\providecommand \bibitemNoStop [0]{.\EOS\space}%
\providecommand \EOS [0]{\spacefactor3000\relax}%
\providecommand \BibitemShut  [1]{\csname bibitem#1\endcsname}%
\let\auto@bib@innerbib\@empty
\bibitem [{\citenamefont {Levine}\ \emph {et~al.}(2019)\citenamefont {Levine},
  \citenamefont {Turner}, \citenamefont {Kehayias}, \citenamefont {Hart},
  \citenamefont {Langellier}, \citenamefont {Trubko}, \citenamefont {Glenn},
  \citenamefont {Fu},\ and\ \citenamefont {Walsworth}}]{QDMreview}%
  \BibitemOpen
  \bibfield  {author} {\bibinfo {author} {\bibfnamefont {E.~V.}\ \bibnamefont
  {Levine}}, \bibinfo {author} {\bibfnamefont {M.~J.}\ \bibnamefont {Turner}},
  \bibinfo {author} {\bibfnamefont {P.}~\bibnamefont {Kehayias}}, \bibinfo
  {author} {\bibfnamefont {C.~A.}\ \bibnamefont {Hart}}, \bibinfo {author}
  {\bibfnamefont {N.}~\bibnamefont {Langellier}}, \bibinfo {author}
  {\bibfnamefont {R.}~\bibnamefont {Trubko}}, \bibinfo {author} {\bibfnamefont
  {D.~R.}\ \bibnamefont {Glenn}}, \bibinfo {author} {\bibfnamefont {R.~R.}\
  \bibnamefont {Fu}},\ and\ \bibinfo {author} {\bibfnamefont {R.~L.}\
  \bibnamefont {Walsworth}},\ }\bibfield  {title} {\bibinfo {title} {Principles
  and techniques of the quantum diamond microscope},\ }\href
  {https://doi.org/10.1515/nanoph-2019-0209} {\bibfield  {journal} {\bibinfo
  {journal} {Nanophotonics}\ }\textbf {\bibinfo {volume} {8}},\ \bibinfo
  {pages} {1945–1973} (\bibinfo {year} {2019})}\BibitemShut {NoStop}%
\bibitem [{\citenamefont {Schirhagl}\ \emph {et~al.}(2014)\citenamefont
  {Schirhagl}, \citenamefont {Chang}, \citenamefont {Loretz},\ and\
  \citenamefont {Degen}}]{NVNanoReview2014}%
  \BibitemOpen
  \bibfield  {author} {\bibinfo {author} {\bibfnamefont {R.}~\bibnamefont
  {Schirhagl}}, \bibinfo {author} {\bibfnamefont {K.}~\bibnamefont {Chang}},
  \bibinfo {author} {\bibfnamefont {M.}~\bibnamefont {Loretz}},\ and\ \bibinfo
  {author} {\bibfnamefont {C.~L.}\ \bibnamefont {Degen}},\ }\bibfield  {title}
  {\bibinfo {title} {Nitrogen-vacancy centers in diamond: Nanoscale sensors for
  physics and biology},\ }\href
  {https://doi.org/10.1146/annurev-physchem-040513-103659} {\bibfield
  {journal} {\bibinfo  {journal} {Annual Review of Physical Chemistry}\
  }\textbf {\bibinfo {volume} {65}},\ \bibinfo {pages} {83–105} (\bibinfo
  {year} {2014})}\BibitemShut {NoStop}%
\bibitem [{\citenamefont {Boretti}\ \emph {et~al.}(2019)\citenamefont
  {Boretti}, \citenamefont {Rosa}, \citenamefont {Blackledge},\ and\
  \citenamefont {Castelletto}}]{NVMRIreview2019}%
  \BibitemOpen
  \bibfield  {author} {\bibinfo {author} {\bibfnamefont {A.}~\bibnamefont
  {Boretti}}, \bibinfo {author} {\bibfnamefont {L.}~\bibnamefont {Rosa}},
  \bibinfo {author} {\bibfnamefont {J.}~\bibnamefont {Blackledge}},\ and\
  \bibinfo {author} {\bibfnamefont {S.}~\bibnamefont {Castelletto}},\
  }\bibfield  {title} {\bibinfo {title} {Nitrogen-vacancy centers in diamond
  for nanoscale magnetic resonance imaging applications},\ }\href
  {https://doi.org/10.3762/bjnano.10.207} {\bibfield  {journal} {\bibinfo
  {journal} {Beilstein Journal of Nanotechnology}\ }\textbf {\bibinfo {volume}
  {10}},\ \bibinfo {pages} {2128–2151} (\bibinfo {year} {2019})}\BibitemShut
  {NoStop}%
\bibitem [{\citenamefont {Taylor}\ \emph {et~al.}(2008)\citenamefont {Taylor},
  \citenamefont {Cappellaro}, \citenamefont {Childress}, \citenamefont {Jiang},
  \citenamefont {Budker}, \citenamefont {Hemmer}, \citenamefont {Yacoby},
  \citenamefont {Walsworth},\ and\ \citenamefont {Lukin}}]{NVMagnetometer2008}%
  \BibitemOpen
  \bibfield  {author} {\bibinfo {author} {\bibfnamefont {J.~M.}\ \bibnamefont
  {Taylor}}, \bibinfo {author} {\bibfnamefont {P.}~\bibnamefont {Cappellaro}},
  \bibinfo {author} {\bibfnamefont {L.}~\bibnamefont {Childress}}, \bibinfo
  {author} {\bibfnamefont {L.}~\bibnamefont {Jiang}}, \bibinfo {author}
  {\bibfnamefont {D.}~\bibnamefont {Budker}}, \bibinfo {author} {\bibfnamefont
  {P.~R.}\ \bibnamefont {Hemmer}}, \bibinfo {author} {\bibfnamefont
  {A.}~\bibnamefont {Yacoby}}, \bibinfo {author} {\bibfnamefont
  {R.}~\bibnamefont {Walsworth}},\ and\ \bibinfo {author} {\bibfnamefont
  {M.~D.}\ \bibnamefont {Lukin}},\ }\bibfield  {title} {\bibinfo {title}
  {High-sensitivity diamond magnetometer with nanoscale resolution},\ }\href
  {https://doi.org/10.1038/nphys1075} {\bibfield  {journal} {\bibinfo
  {journal} {Nature Physics}\ }\textbf {\bibinfo {volume} {4}},\ \bibinfo
  {pages} {810–816} (\bibinfo {year} {2008})}\BibitemShut {NoStop}%
\bibitem [{\citenamefont {Balasubramanian}\ \emph {et~al.}(2008)\citenamefont
  {Balasubramanian}, \citenamefont {Chan}, \citenamefont {Kolesov},
  \citenamefont {Al-Hmoud}, \citenamefont {Tisler}, \citenamefont {Shin},
  \citenamefont {Kim}, \citenamefont {Wojcik}, \citenamefont {Hemmer},
  \citenamefont {Krueger}, \citenamefont {Hanke}, \citenamefont
  {Leitenstorfer}, \citenamefont {Bratschitsch}, \citenamefont {Jelezko},\ and\
  \citenamefont {Wrachtrup}}]{NVimaging2008}%
  \BibitemOpen
  \bibfield  {author} {\bibinfo {author} {\bibfnamefont {G.}~\bibnamefont
  {Balasubramanian}}, \bibinfo {author} {\bibfnamefont {I.~Y.}\ \bibnamefont
  {Chan}}, \bibinfo {author} {\bibfnamefont {R.}~\bibnamefont {Kolesov}},
  \bibinfo {author} {\bibfnamefont {M.}~\bibnamefont {Al-Hmoud}}, \bibinfo
  {author} {\bibfnamefont {J.}~\bibnamefont {Tisler}}, \bibinfo {author}
  {\bibfnamefont {C.}~\bibnamefont {Shin}}, \bibinfo {author} {\bibfnamefont
  {C.}~\bibnamefont {Kim}}, \bibinfo {author} {\bibfnamefont {A.}~\bibnamefont
  {Wojcik}}, \bibinfo {author} {\bibfnamefont {P.~R.}\ \bibnamefont {Hemmer}},
  \bibinfo {author} {\bibfnamefont {A.}~\bibnamefont {Krueger}}, \bibinfo
  {author} {\bibfnamefont {T.}~\bibnamefont {Hanke}}, \bibinfo {author}
  {\bibfnamefont {A.}~\bibnamefont {Leitenstorfer}}, \bibinfo {author}
  {\bibfnamefont {R.}~\bibnamefont {Bratschitsch}}, \bibinfo {author}
  {\bibfnamefont {F.}~\bibnamefont {Jelezko}},\ and\ \bibinfo {author}
  {\bibfnamefont {J.}~\bibnamefont {Wrachtrup}},\ }\bibfield  {title} {\bibinfo
  {title} {Nanoscale imaging magnetometry with diamond spins under ambient
  conditions},\ }\href {https://doi.org/10.1038/nature07278} {\bibfield
  {journal} {\bibinfo  {journal} {Nature}\ }\textbf {\bibinfo {volume} {455}},\
  \bibinfo {pages} {648–651} (\bibinfo {year} {2008})}\BibitemShut {NoStop}%
\bibitem [{\citenamefont {Kehayias}\ \emph {et~al.}(2019)\citenamefont
  {Kehayias}, \citenamefont {Turner}, \citenamefont {Trubko}, \citenamefont
  {Schloss}, \citenamefont {Hart}, \citenamefont {Wesson}, \citenamefont
  {Glenn},\ and\ \citenamefont {Walsworth}}]{StrainPaper}%
  \BibitemOpen
  \bibfield  {author} {\bibinfo {author} {\bibfnamefont {P.}~\bibnamefont
  {Kehayias}}, \bibinfo {author} {\bibfnamefont {M.~J.}\ \bibnamefont
  {Turner}}, \bibinfo {author} {\bibfnamefont {R.}~\bibnamefont {Trubko}},
  \bibinfo {author} {\bibfnamefont {J.~M.}\ \bibnamefont {Schloss}}, \bibinfo
  {author} {\bibfnamefont {C.~A.}\ \bibnamefont {Hart}}, \bibinfo {author}
  {\bibfnamefont {M.}~\bibnamefont {Wesson}}, \bibinfo {author} {\bibfnamefont
  {D.~R.}\ \bibnamefont {Glenn}},\ and\ \bibinfo {author} {\bibfnamefont
  {R.~L.}\ \bibnamefont {Walsworth}},\ }\bibfield  {title} {\bibinfo {title}
  {Imaging crystal stress in diamond using ensembles of nitrogen-vacancy
  centers},\ }\href {https://doi.org/10.1103/PhysRevB.100.174103} {\bibfield
  {journal} {\bibinfo  {journal} {Physical Review B}\ }\textbf {\bibinfo
  {volume} {100}},\ \bibinfo {pages} {174103} (\bibinfo {year}
  {2019})}\BibitemShut {NoStop}%
\bibitem [{\citenamefont {Friel}\ \emph {et~al.}(2009)\citenamefont {Friel},
  \citenamefont {Clewes}, \citenamefont {Dhillon}, \citenamefont {Perkins},
  \citenamefont {Twitchen},\ and\ \citenamefont {Scarsbrook}}]{Friel2009}%
  \BibitemOpen
  \bibfield  {author} {\bibinfo {author} {\bibfnamefont {I.}~\bibnamefont
  {Friel}}, \bibinfo {author} {\bibfnamefont {S.~L.}\ \bibnamefont {Clewes}},
  \bibinfo {author} {\bibfnamefont {H.~K.}\ \bibnamefont {Dhillon}}, \bibinfo
  {author} {\bibfnamefont {N.}~\bibnamefont {Perkins}}, \bibinfo {author}
  {\bibfnamefont {D.~J.}\ \bibnamefont {Twitchen}},\ and\ \bibinfo {author}
  {\bibfnamefont {G.~A.}\ \bibnamefont {Scarsbrook}},\ }\bibfield  {title}
  {\bibinfo {title} {Control of surface and bulk crystalline quality in single
  crystal diamond grown by chemical vapour deposition},\ }\href
  {https://doi.org/10.1016/j.diamond.2009.01.013} {\bibfield  {journal}
  {\bibinfo  {journal} {Diamond and Related Materials}\ }\bibinfo {series}
  {Proceedings of Diamond 2008, the 19th European Conference on Diamond,
  Diamond-Like Materials, Carbon Nanotubes, Nitrides and Silicon Carbide},\
  \textbf {\bibinfo {volume} {18}},\ \bibinfo {pages} {808–815} (\bibinfo
  {year} {2009})}\BibitemShut {NoStop}%
\bibitem [{\citenamefont {Edmonds}\ \emph {et~al.}(2021)\citenamefont
  {Edmonds}, \citenamefont {Hart}, \citenamefont {Turner}, \citenamefont
  {Colard}, \citenamefont {Schloss}, \citenamefont {Olsson}, \citenamefont
  {Trubko}, \citenamefont {Markham}, \citenamefont {Rathmill}, \citenamefont
  {Horne-Smith}, \citenamefont {Lew}, \citenamefont {Manickam}, \citenamefont
  {Bruce}, \citenamefont {Kaup}, \citenamefont {Russo}, \citenamefont
  {DiMario}, \citenamefont {South}, \citenamefont {Hansen}, \citenamefont
  {Twitchen},\ and\ \citenamefont {Walsworth}}]{PurpleDiamond}%
  \BibitemOpen
  \bibfield  {author} {\bibinfo {author} {\bibfnamefont {A.~M.}\ \bibnamefont
  {Edmonds}}, \bibinfo {author} {\bibfnamefont {C.~A.}\ \bibnamefont {Hart}},
  \bibinfo {author} {\bibfnamefont {M.~J.}\ \bibnamefont {Turner}}, \bibinfo
  {author} {\bibfnamefont {P.-O.}\ \bibnamefont {Colard}}, \bibinfo {author}
  {\bibfnamefont {J.~M.}\ \bibnamefont {Schloss}}, \bibinfo {author}
  {\bibfnamefont {K.}~\bibnamefont {Olsson}}, \bibinfo {author} {\bibfnamefont
  {R.}~\bibnamefont {Trubko}}, \bibinfo {author} {\bibfnamefont {M.~L.}\
  \bibnamefont {Markham}}, \bibinfo {author} {\bibfnamefont {A.}~\bibnamefont
  {Rathmill}}, \bibinfo {author} {\bibfnamefont {B.}~\bibnamefont
  {Horne-Smith}}, \bibinfo {author} {\bibfnamefont {W.}~\bibnamefont {Lew}},
  \bibinfo {author} {\bibfnamefont {A.}~\bibnamefont {Manickam}}, \bibinfo
  {author} {\bibfnamefont {S.}~\bibnamefont {Bruce}}, \bibinfo {author}
  {\bibfnamefont {P.~G.}\ \bibnamefont {Kaup}}, \bibinfo {author}
  {\bibfnamefont {J.~C.}\ \bibnamefont {Russo}}, \bibinfo {author}
  {\bibfnamefont {M.~J.}\ \bibnamefont {DiMario}}, \bibinfo {author}
  {\bibfnamefont {J.~T.}\ \bibnamefont {South}}, \bibinfo {author}
  {\bibfnamefont {J.~T.}\ \bibnamefont {Hansen}}, \bibinfo {author}
  {\bibfnamefont {D.~J.}\ \bibnamefont {Twitchen}},\ and\ \bibinfo {author}
  {\bibfnamefont {R.~L.}\ \bibnamefont {Walsworth}},\ }\bibfield  {title}
  {\bibinfo {title} {Characterisation of cvd diamond with high concentrations
  of nitrogen for magnetic-field sensing applications},\ }\href
  {https://doi.org/10.1088/2633-4356/abd88a} {\bibfield  {journal} {\bibinfo
  {journal} {Materials for Quantum Technology}\ }\textbf {\bibinfo {volume}
  {1}},\ \bibinfo {pages} {025001} (\bibinfo {year} {2021})}\BibitemShut
  {NoStop}%
\bibitem [{\citenamefont {Hoa}\ \emph {et~al.}(2014)\citenamefont {Hoa},
  \citenamefont {Ouisse}, \citenamefont {Chaussende}, \citenamefont {Naamoun},
  \citenamefont {Tallaire},\ and\ \citenamefont
  {Achard}}]{BirefringenceStrain2014}%
  \BibitemOpen
  \bibfield  {author} {\bibinfo {author} {\bibfnamefont {L.~T.~M.}\
  \bibnamefont {Hoa}}, \bibinfo {author} {\bibfnamefont {T.}~\bibnamefont
  {Ouisse}}, \bibinfo {author} {\bibfnamefont {D.}~\bibnamefont {Chaussende}},
  \bibinfo {author} {\bibfnamefont {M.}~\bibnamefont {Naamoun}}, \bibinfo
  {author} {\bibfnamefont {A.}~\bibnamefont {Tallaire}},\ and\ \bibinfo
  {author} {\bibfnamefont {J.}~\bibnamefont {Achard}},\ }\bibfield  {title}
  {\bibinfo {title} {Birefringence microscopy of unit dislocations in
  diamond},\ }\href {https://doi.org/10.1021/cg5010193} {\bibfield  {journal}
  {\bibinfo  {journal} {Crystal Growth \& Design}\ }\textbf {\bibinfo {volume}
  {14}},\ \bibinfo {pages} {5761–5766} (\bibinfo {year} {2014})}\BibitemShut
  {NoStop}%
\bibitem [{\citenamefont {Broadway}\ \emph {et~al.}(2019)\citenamefont
  {Broadway}, \citenamefont {Johnson}, \citenamefont {Barson}, \citenamefont
  {Lillie}, \citenamefont {Dontschuk}, \citenamefont {McCloskey}, \citenamefont
  {Tsai}, \citenamefont {Teraji}, \citenamefont {Simpson}, \citenamefont
  {Stacey}, \citenamefont {McCallum}, \citenamefont {Bradby}, \citenamefont
  {Doherty}, \citenamefont {Hollenberg},\ and\ \citenamefont
  {Tetienne}}]{AusStrain2019}%
  \BibitemOpen
  \bibfield  {author} {\bibinfo {author} {\bibfnamefont {D.}~\bibnamefont
  {Broadway}}, \bibinfo {author} {\bibfnamefont {B.}~\bibnamefont {Johnson}},
  \bibinfo {author} {\bibfnamefont {M.}~\bibnamefont {Barson}}, \bibinfo
  {author} {\bibfnamefont {S.}~\bibnamefont {Lillie}}, \bibinfo {author}
  {\bibfnamefont {N.}~\bibnamefont {Dontschuk}}, \bibinfo {author}
  {\bibfnamefont {D.}~\bibnamefont {McCloskey}}, \bibinfo {author}
  {\bibfnamefont {A.}~\bibnamefont {Tsai}}, \bibinfo {author} {\bibfnamefont
  {T.}~\bibnamefont {Teraji}}, \bibinfo {author} {\bibfnamefont
  {D.}~\bibnamefont {Simpson}}, \bibinfo {author} {\bibfnamefont
  {A.}~\bibnamefont {Stacey}}, \bibinfo {author} {\bibfnamefont
  {J.}~\bibnamefont {McCallum}}, \bibinfo {author} {\bibfnamefont
  {J.}~\bibnamefont {Bradby}}, \bibinfo {author} {\bibfnamefont
  {M.}~\bibnamefont {Doherty}}, \bibinfo {author} {\bibfnamefont
  {L.}~\bibnamefont {Hollenberg}},\ and\ \bibinfo {author} {\bibfnamefont
  {J.-P.}\ \bibnamefont {Tetienne}},\ }\bibfield  {title} {\bibinfo {title}
  {Microscopic imaging of the stress tensor in diamond using in situ quantum
  sensors},\ }\href {https://doi.org/10.1021/acs.nanolett.9b01402} {\bibfield
  {journal} {\bibinfo  {journal} {Nano Letters}\ }\textbf {\bibinfo {volume}
  {19}},\ \bibinfo {pages} {4543–4550} (\bibinfo {year} {2019})}\BibitemShut
  {NoStop}%
\bibitem [{\citenamefont {Trusheim}\ and\ \citenamefont
  {Englund}(2016)}]{EnglundStrain2016}%
  \BibitemOpen
  \bibfield  {author} {\bibinfo {author} {\bibfnamefont {M.~E.}\ \bibnamefont
  {Trusheim}}\ and\ \bibinfo {author} {\bibfnamefont {D.}~\bibnamefont
  {Englund}},\ }\bibfield  {title} {\bibinfo {title} {Wide-field strain imaging
  with preferentially aligned nitrogen-vacancy centers in polycrystalline
  diamond},\ }\href {https://doi.org/10.1088/1367-2630/aa5040} {\bibfield
  {journal} {\bibinfo  {journal} {New Journal of Physics}\ }\textbf {\bibinfo
  {volume} {18}},\ \bibinfo {pages} {123023} (\bibinfo {year}
  {2016})}\BibitemShut {NoStop}%
\bibitem [{\citenamefont {Schreck}\ \emph {et~al.}(2020)\citenamefont
  {Schreck}, \citenamefont {Ščajev}, \citenamefont {Träger}, \citenamefont
  {Mayr}, \citenamefont {Grünwald}, \citenamefont {Fischer},\ and\
  \citenamefont {Gsell}}]{SchreckDislocationTrapping2020}%
  \BibitemOpen
  \bibfield  {author} {\bibinfo {author} {\bibfnamefont {M.}~\bibnamefont
  {Schreck}}, \bibinfo {author} {\bibfnamefont {P.}~\bibnamefont {Ščajev}},
  \bibinfo {author} {\bibfnamefont {M.}~\bibnamefont {Träger}}, \bibinfo
  {author} {\bibfnamefont {M.}~\bibnamefont {Mayr}}, \bibinfo {author}
  {\bibfnamefont {T.}~\bibnamefont {Grünwald}}, \bibinfo {author}
  {\bibfnamefont {M.}~\bibnamefont {Fischer}},\ and\ \bibinfo {author}
  {\bibfnamefont {S.}~\bibnamefont {Gsell}},\ }\bibfield  {title} {\bibinfo
  {title} {Charge carrier trapping by dislocations in single crystal diamond},\
  }\href {https://doi.org/10.1063/1.5140662} {\bibfield  {journal} {\bibinfo
  {journal} {Journal of Applied Physics}\ }\textbf {\bibinfo {volume} {127}},\
  \bibinfo {pages} {125102} (\bibinfo {year} {2020})}\BibitemShut {NoStop}%
\bibitem [{\citenamefont {Chen}\ \emph {et~al.}(2015)\citenamefont {Chen},
  \citenamefont {Zhou}, \citenamefont {Zou}, \citenamefont {Li}, \citenamefont
  {Dong}, \citenamefont {Sun},\ and\ \citenamefont
  {Guo}}]{ChenDepolarization2015}%
  \BibitemOpen
  \bibfield  {author} {\bibinfo {author} {\bibfnamefont {X.-D.}\ \bibnamefont
  {Chen}}, \bibinfo {author} {\bibfnamefont {L.-M.}\ \bibnamefont {Zhou}},
  \bibinfo {author} {\bibfnamefont {C.-L.}\ \bibnamefont {Zou}}, \bibinfo
  {author} {\bibfnamefont {C.-C.}\ \bibnamefont {Li}}, \bibinfo {author}
  {\bibfnamefont {Y.}~\bibnamefont {Dong}}, \bibinfo {author} {\bibfnamefont
  {F.-W.}\ \bibnamefont {Sun}},\ and\ \bibinfo {author} {\bibfnamefont {G.-C.}\
  \bibnamefont {Guo}},\ }\bibfield  {title} {\bibinfo {title} {Spin
  depolarization effect induced by charge state conversion of nitrogen vacancy
  center in diamond},\ }\href {https://doi.org/10.1103/PhysRevB.92.104301}
  {\bibfield  {journal} {\bibinfo  {journal} {Physical Review B}\ }\textbf
  {\bibinfo {volume} {92}},\ \bibinfo {pages} {104301} (\bibinfo {year}
  {2015})}\BibitemShut {NoStop}%
\bibitem [{\citenamefont {Choi}\ \emph {et~al.}(2017)\citenamefont {Choi},
  \citenamefont {Choi}, \citenamefont {Kucsko}, \citenamefont {Maurer},
  \citenamefont {Shields}, \citenamefont {Sumiya}, \citenamefont {Onoda},
  \citenamefont {Isoya}, \citenamefont {Demler}, \citenamefont {Jelezko},
  \citenamefont {Yao},\ and\ \citenamefont {Lukin}}]{ChoiDepolarization2017}%
  \BibitemOpen
  \bibfield  {author} {\bibinfo {author} {\bibfnamefont {J.}~\bibnamefont
  {Choi}}, \bibinfo {author} {\bibfnamefont {S.}~\bibnamefont {Choi}}, \bibinfo
  {author} {\bibfnamefont {G.}~\bibnamefont {Kucsko}}, \bibinfo {author}
  {\bibfnamefont {P.~C.}\ \bibnamefont {Maurer}}, \bibinfo {author}
  {\bibfnamefont {B.~J.}\ \bibnamefont {Shields}}, \bibinfo {author}
  {\bibfnamefont {H.}~\bibnamefont {Sumiya}}, \bibinfo {author} {\bibfnamefont
  {S.}~\bibnamefont {Onoda}}, \bibinfo {author} {\bibfnamefont
  {J.}~\bibnamefont {Isoya}}, \bibinfo {author} {\bibfnamefont
  {E.}~\bibnamefont {Demler}}, \bibinfo {author} {\bibfnamefont
  {F.}~\bibnamefont {Jelezko}}, \bibinfo {author} {\bibfnamefont {N.~Y.}\
  \bibnamefont {Yao}},\ and\ \bibinfo {author} {\bibfnamefont {M.~D.}\
  \bibnamefont {Lukin}},\ }\bibfield  {title} {\bibinfo {title} {Depolarization
  dynamics in a strongly interacting solid-state spin ensemble},\ }\href
  {https://doi.org/10.1103/PhysRevLett.118.093601} {\bibfield  {journal}
  {\bibinfo  {journal} {Physical Review Letters}\ }\textbf {\bibinfo {volume}
  {118}},\ \bibinfo {pages} {093601} (\bibinfo {year} {2017})}\BibitemShut
  {NoStop}%
\bibitem [{\citenamefont {Sipahigil}\ \emph {et~al.}(2016)\citenamefont
  {Sipahigil}, \citenamefont {Evans}, \citenamefont {Sukachev}, \citenamefont
  {Burek}, \citenamefont {Borregaard}, \citenamefont {Bhaskar}, \citenamefont
  {Nguyen}, \citenamefont {Pacheco}, \citenamefont {Atikian}, \citenamefont
  {Meuwly}, \citenamefont {Camacho}, \citenamefont {Jelezko}, \citenamefont
  {Bielejec}, \citenamefont {Park}, \citenamefont {Lončar},\ and\
  \citenamefont {Lukin}}]{NanophotonicPlatform2016}%
  \BibitemOpen
  \bibfield  {author} {\bibinfo {author} {\bibfnamefont {A.}~\bibnamefont
  {Sipahigil}}, \bibinfo {author} {\bibfnamefont {R.~E.}\ \bibnamefont
  {Evans}}, \bibinfo {author} {\bibfnamefont {D.~D.}\ \bibnamefont {Sukachev}},
  \bibinfo {author} {\bibfnamefont {M.~J.}\ \bibnamefont {Burek}}, \bibinfo
  {author} {\bibfnamefont {J.}~\bibnamefont {Borregaard}}, \bibinfo {author}
  {\bibfnamefont {M.~K.}\ \bibnamefont {Bhaskar}}, \bibinfo {author}
  {\bibfnamefont {C.~T.}\ \bibnamefont {Nguyen}}, \bibinfo {author}
  {\bibfnamefont {J.~L.}\ \bibnamefont {Pacheco}}, \bibinfo {author}
  {\bibfnamefont {H.~A.}\ \bibnamefont {Atikian}}, \bibinfo {author}
  {\bibfnamefont {C.}~\bibnamefont {Meuwly}}, \bibinfo {author} {\bibfnamefont
  {R.~M.}\ \bibnamefont {Camacho}}, \bibinfo {author} {\bibfnamefont
  {F.}~\bibnamefont {Jelezko}}, \bibinfo {author} {\bibfnamefont
  {E.}~\bibnamefont {Bielejec}}, \bibinfo {author} {\bibfnamefont
  {H.}~\bibnamefont {Park}}, \bibinfo {author} {\bibfnamefont {M.}~\bibnamefont
  {Lončar}},\ and\ \bibinfo {author} {\bibfnamefont {M.~D.}\ \bibnamefont
  {Lukin}},\ }\bibfield  {title} {\bibinfo {title} {An integrated diamond
  nanophotonics platform for quantum-optical networks},\ }\href
  {https://doi.org/10.1126/science.aah6875} {\bibfield  {journal} {\bibinfo
  {journal} {Science}\ }\textbf {\bibinfo {volume} {354}},\ \bibinfo {pages}
  {847–850} (\bibinfo {year} {2016})}\BibitemShut {NoStop}%
\bibitem [{\citenamefont {Hausmann}\ \emph {et~al.}(2013)\citenamefont
  {Hausmann}, \citenamefont {Shields}, \citenamefont {Quan}, \citenamefont
  {Chu}, \citenamefont {de~Leon}, \citenamefont {Evans}, \citenamefont {Burek},
  \citenamefont {Zibrov}, \citenamefont {Markham}, \citenamefont {Twitchen},
  \citenamefont {Park}, \citenamefont {Lukin},\ and\ \citenamefont
  {Lončar}}]{NVNanobeams2013}%
  \BibitemOpen
  \bibfield  {author} {\bibinfo {author} {\bibfnamefont {B.~J.~M.}\
  \bibnamefont {Hausmann}}, \bibinfo {author} {\bibfnamefont {B.~J.}\
  \bibnamefont {Shields}}, \bibinfo {author} {\bibfnamefont {Q.}~\bibnamefont
  {Quan}}, \bibinfo {author} {\bibfnamefont {Y.}~\bibnamefont {Chu}}, \bibinfo
  {author} {\bibfnamefont {N.~P.}\ \bibnamefont {de~Leon}}, \bibinfo {author}
  {\bibfnamefont {R.}~\bibnamefont {Evans}}, \bibinfo {author} {\bibfnamefont
  {M.~J.}\ \bibnamefont {Burek}}, \bibinfo {author} {\bibfnamefont {A.~S.}\
  \bibnamefont {Zibrov}}, \bibinfo {author} {\bibfnamefont {M.}~\bibnamefont
  {Markham}}, \bibinfo {author} {\bibfnamefont {D.~J.}\ \bibnamefont
  {Twitchen}}, \bibinfo {author} {\bibfnamefont {H.}~\bibnamefont {Park}},
  \bibinfo {author} {\bibfnamefont {M.~D.}\ \bibnamefont {Lukin}},\ and\
  \bibinfo {author} {\bibfnamefont {M.}~\bibnamefont {Lončar}},\ }\bibfield
  {title} {\bibinfo {title} {Coupling of nv centers to photonic crystal
  nanobeams in diamond},\ }\href {https://doi.org/10.1021/nl402174g} {\bibfield
   {journal} {\bibinfo  {journal} {Nano Letters}\ }\textbf {\bibinfo {volume}
  {13}},\ \bibinfo {pages} {5791–5796} (\bibinfo {year} {2013})}\BibitemShut
  {NoStop}%
\bibitem [{\citenamefont {Knauer}\ \emph {et~al.}(2020)\citenamefont {Knauer},
  \citenamefont {Hadden},\ and\ \citenamefont {Rarity}}]{KnauerFabStrain2020}%
  \BibitemOpen
  \bibfield  {author} {\bibinfo {author} {\bibfnamefont {S.}~\bibnamefont
  {Knauer}}, \bibinfo {author} {\bibfnamefont {J.~P.}\ \bibnamefont {Hadden}},\
  and\ \bibinfo {author} {\bibfnamefont {J.~G.}\ \bibnamefont {Rarity}},\
  }\bibfield  {title} {\bibinfo {title} {In-situ measurements of fabrication
  induced strain in diamond photonic-structures using intrinsic colour
  centres},\ }\href {https://doi.org/10.1038/s41534-020-0277-1} {\bibfield
  {journal} {\bibinfo  {journal} {npj Quantum Information}\ }\textbf {\bibinfo
  {volume} {6}},\ \bibinfo {pages} {1–6} (\bibinfo {year}
  {2020})}\BibitemShut {NoStop}%
\bibitem [{\citenamefont {Rajendran}\ \emph {et~al.}(2017)\citenamefont
  {Rajendran}, \citenamefont {Zobrist}, \citenamefont {Sushkov}, \citenamefont
  {Walsworth},\ and\ \citenamefont {Lukin}}]{proposal}%
  \BibitemOpen
  \bibfield  {author} {\bibinfo {author} {\bibfnamefont {S.}~\bibnamefont
  {Rajendran}}, \bibinfo {author} {\bibfnamefont {N.}~\bibnamefont {Zobrist}},
  \bibinfo {author} {\bibfnamefont {A.~O.}\ \bibnamefont {Sushkov}}, \bibinfo
  {author} {\bibfnamefont {R.}~\bibnamefont {Walsworth}},\ and\ \bibinfo
  {author} {\bibfnamefont {M.}~\bibnamefont {Lukin}},\ }\bibfield  {title}
  {\bibinfo {title} {A method for directional detection of dark matter using
  spectroscopy of crystal defects},\ }\href
  {https://doi.org/10.1103/PhysRevD.96.035009} {\bibfield  {journal} {\bibinfo
  {journal} {Physical Review D}\ }\textbf {\bibinfo {volume} {96}},\ \bibinfo
  {pages} {035009} (\bibinfo {year} {2017})}\BibitemShut {NoStop}%
\bibitem [{\citenamefont {Kurinsky}\ \emph {et~al.}(2019)\citenamefont
  {Kurinsky}, \citenamefont {Yu}, \citenamefont {Hochberg},\ and\ \citenamefont
  {Cabrera}}]{Kurinsky2019}%
  \BibitemOpen
  \bibfield  {author} {\bibinfo {author} {\bibfnamefont {N.}~\bibnamefont
  {Kurinsky}}, \bibinfo {author} {\bibfnamefont {T.~C.}\ \bibnamefont {Yu}},
  \bibinfo {author} {\bibfnamefont {Y.}~\bibnamefont {Hochberg}},\ and\
  \bibinfo {author} {\bibfnamefont {B.}~\bibnamefont {Cabrera}},\ }\bibfield
  {title} {\bibinfo {title} {Diamond detectors for direct detection of sub-gev
  dark matter},\ }\href {https://doi.org/10.1103/PhysRevD.99.123005} {\bibfield
   {journal} {\bibinfo  {journal} {Physical Review D}\ }\textbf {\bibinfo
  {volume} {99}},\ \bibinfo {pages} {123005} (\bibinfo {year}
  {2019})}\BibitemShut {NoStop}%
\bibitem [{\citenamefont {O’Hare}\ \emph {et~al.}(2015)\citenamefont
  {O’Hare}, \citenamefont {Green}, \citenamefont {Billard}, \citenamefont
  {Figueroa-Feliciano},\ and\ \citenamefont {Strigari}}]{ReadoutDirectional}%
  \BibitemOpen
  \bibfield  {author} {\bibinfo {author} {\bibfnamefont {C.~A.}\ \bibnamefont
  {O’Hare}}, \bibinfo {author} {\bibfnamefont {A.~M.}\ \bibnamefont {Green}},
  \bibinfo {author} {\bibfnamefont {J.}~\bibnamefont {Billard}}, \bibinfo
  {author} {\bibfnamefont {E.}~\bibnamefont {Figueroa-Feliciano}},\ and\
  \bibinfo {author} {\bibfnamefont {L.~E.}\ \bibnamefont {Strigari}},\
  }\bibfield  {title} {\bibinfo {title} {Readout strategies for directional
  dark matter detection beyond the neutrino background},\ }\href
  {https://doi.org/10.1103/PhysRevD.92.063518} {\bibfield  {journal} {\bibinfo
  {journal} {Physical Review D}\ }\textbf {\bibinfo {volume} {92}},\ \bibinfo
  {pages} {063518} (\bibinfo {year} {2015})}\BibitemShut {NoStop}%
\bibitem [{\citenamefont {Bøehm}\ \emph {et~al.}(2019)\citenamefont {Bøehm},
  \citenamefont {Cerdeño}, \citenamefont {Machado}, \citenamefont {Campo},\
  and\ \citenamefont {Reid}}]{NeutrinoFloor}%
  \BibitemOpen
  \bibfield  {author} {\bibinfo {author} {\bibfnamefont {C.}~\bibnamefont
  {Bøehm}}, \bibinfo {author} {\bibfnamefont {D.~G.}\ \bibnamefont
  {Cerdeño}}, \bibinfo {author} {\bibfnamefont {P.~A.~N.}\ \bibnamefont
  {Machado}}, \bibinfo {author} {\bibfnamefont {A.~O.-D.}\ \bibnamefont
  {Campo}},\ and\ \bibinfo {author} {\bibfnamefont {E.}~\bibnamefont {Reid}},\
  }\bibfield  {title} {\bibinfo {title} {How high is the neutrino floor?},\
  }\href {https://doi.org/10.1088/1475-7516/2019/01/043} {\bibfield  {journal}
  {\bibinfo  {journal} {Journal of Cosmology and Astroparticle Physics}\
  }\textbf {\bibinfo {volume} {2019}}\bibinfo  {number} { (01)},\ \bibinfo
  {pages} {043–043}}\BibitemShut {NoStop}%
\bibitem [{\citenamefont {Marshall}\ \emph {et~al.}(2021)\citenamefont
  {Marshall}, \citenamefont {Turner}, \citenamefont {Ku}, \citenamefont
  {Phillips},\ and\ \citenamefont {Walsworth}}]{InvitedPaper}%
  \BibitemOpen
\bibfield  {number} {  }\bibfield  {author} {\bibinfo {author} {\bibfnamefont
  {M.~C.}\ \bibnamefont {Marshall}}, \bibinfo {author} {\bibfnamefont {M.~J.}\
  \bibnamefont {Turner}}, \bibinfo {author} {\bibfnamefont {M.~J.-H.}\
  \bibnamefont {Ku}}, \bibinfo {author} {\bibfnamefont {D.~F.}\ \bibnamefont
  {Phillips}},\ and\ \bibinfo {author} {\bibfnamefont {R.~L.}\ \bibnamefont
  {Walsworth}},\ }\bibfield  {title} {\bibinfo {title} {Directional detection
  of dark matter with diamond},\ }\href
  {https://doi.org/10.1088/2058-9565/abe5ed} {\bibfield  {journal} {\bibinfo
  {journal} {Quantum Science and Technology}\ }\textbf {\bibinfo {volume}
  {6}},\ \bibinfo {pages} {024011} (\bibinfo {year} {2021})}\BibitemShut
  {NoStop}%
\bibitem [{\citenamefont {Winarski}\ \emph {et~al.}(2012)\citenamefont
  {Winarski}, \citenamefont {Holt}, \citenamefont {Rose}, \citenamefont
  {Fuesz}, \citenamefont {Carbaugh}, \citenamefont {Benson}, \citenamefont
  {Shu}, \citenamefont {Kline}, \citenamefont {Stephenson}, \citenamefont
  {McNulty},\ and\ \citenamefont {Maser}}]{APSNanoprobe2012}%
  \BibitemOpen
  \bibfield  {author} {\bibinfo {author} {\bibfnamefont {R.}~\bibnamefont
  {Winarski}}, \bibinfo {author} {\bibfnamefont {M.}~\bibnamefont {Holt}},
  \bibinfo {author} {\bibfnamefont {V.}~\bibnamefont {Rose}}, \bibinfo {author}
  {\bibfnamefont {P.}~\bibnamefont {Fuesz}}, \bibinfo {author} {\bibfnamefont
  {D.}~\bibnamefont {Carbaugh}}, \bibinfo {author} {\bibfnamefont
  {C.}~\bibnamefont {Benson}}, \bibinfo {author} {\bibfnamefont
  {D.}~\bibnamefont {Shu}}, \bibinfo {author} {\bibfnamefont {D.}~\bibnamefont
  {Kline}}, \bibinfo {author} {\bibfnamefont {G.}~\bibnamefont {Stephenson}},
  \bibinfo {author} {\bibfnamefont {I.}~\bibnamefont {McNulty}},\ and\ \bibinfo
  {author} {\bibfnamefont {J.}~\bibnamefont {Maser}},\ }\bibfield  {title}
  {\bibinfo {title} {A hard x-ray nanoprobe beamline for nanoscale
  microscopy},\ }\href {https://doi.org/10.1107/S0909049512036783} {\bibfield
  {journal} {\bibinfo  {journal} {Journal of Synchrotron Radiation}\ }\textbf
  {\bibinfo {volume} {19}},\ \bibinfo {pages} {1056–1060} (\bibinfo {year}
  {2012})}\BibitemShut {NoStop}%
\bibitem [{\citenamefont {Holt}\ \emph {et~al.}(2013)\citenamefont {Holt},
  \citenamefont {Harder}, \citenamefont {Winarski},\ and\ \citenamefont
  {Rose}}]{HoltReview2013}%
  \BibitemOpen
  \bibfield  {author} {\bibinfo {author} {\bibfnamefont {M.}~\bibnamefont
  {Holt}}, \bibinfo {author} {\bibfnamefont {R.}~\bibnamefont {Harder}},
  \bibinfo {author} {\bibfnamefont {R.}~\bibnamefont {Winarski}},\ and\
  \bibinfo {author} {\bibfnamefont {V.}~\bibnamefont {Rose}},\ }\bibfield
  {title} {\bibinfo {title} {Nanoscale hard x-ray microscopy methods for
  materials studies},\ }\href
  {https://doi.org/10.1146/annurev-matsci-071312-121654} {\bibfield  {journal}
  {\bibinfo  {journal} {Annual Review of Materials Research}\ }\textbf
  {\bibinfo {volume} {43}},\ \bibinfo {pages} {183–211} (\bibinfo {year}
  {2013})}\BibitemShut {NoStop}%
\bibitem [{\citenamefont {Mohacsi}\ \emph {et~al.}(2017)\citenamefont
  {Mohacsi}, \citenamefont {Vartiainen}, \citenamefont {Rösner}, \citenamefont
  {Guizar-Sicairos}, \citenamefont {Guzenko}, \citenamefont {McNulty},
  \citenamefont {Winarski}, \citenamefont {Holt},\ and\ \citenamefont
  {David}}]{MohacsiFresnelOptics2017}%
  \BibitemOpen
  \bibfield  {author} {\bibinfo {author} {\bibfnamefont {I.}~\bibnamefont
  {Mohacsi}}, \bibinfo {author} {\bibfnamefont {I.}~\bibnamefont {Vartiainen}},
  \bibinfo {author} {\bibfnamefont {B.}~\bibnamefont {Rösner}}, \bibinfo
  {author} {\bibfnamefont {M.}~\bibnamefont {Guizar-Sicairos}}, \bibinfo
  {author} {\bibfnamefont {V.~A.}\ \bibnamefont {Guzenko}}, \bibinfo {author}
  {\bibfnamefont {I.}~\bibnamefont {McNulty}}, \bibinfo {author} {\bibfnamefont
  {R.}~\bibnamefont {Winarski}}, \bibinfo {author} {\bibfnamefont {M.~V.}\
  \bibnamefont {Holt}},\ and\ \bibinfo {author} {\bibfnamefont
  {C.}~\bibnamefont {David}},\ }\bibfield  {title} {\bibinfo {title}
  {Interlaced zone plate optics for hard x-ray imaging in the 10 nm range},\
  }\href {https://doi.org/10.1038/srep43624} {\bibfield  {journal} {\bibinfo
  {journal} {Scientific Reports}\ }\textbf {\bibinfo {volume} {7}},\ \bibinfo
  {pages} {43624} (\bibinfo {year} {2017})}\BibitemShut {NoStop}%
\bibitem [{\citenamefont {Warren}(1990)}]{Warren_1990}%
  \BibitemOpen
  \bibfield  {author} {\bibinfo {author} {\bibfnamefont {B.~E.}\ \bibnamefont
  {Warren}},\ }\href@noop {} {\emph {\bibinfo {title} {X-ray Diffraction}}}\
  (\bibinfo  {publisher} {Courier Corporation},\ \bibinfo {year} {1990})\
  \bibinfo {note} {google-Books-ID: wfLBhAbEYAsC}\BibitemShut {NoStop}%
\bibitem [{\citenamefont {Ying}\ \emph {et~al.}(2010)\citenamefont {Ying},
  \citenamefont {Osting}, \citenamefont {Noyan}, \citenamefont {Murray},
  \citenamefont {Holt},\ and\ \citenamefont {Maser}}]{YingKinematicCalcs2010}%
  \BibitemOpen
  \bibfield  {author} {\bibinfo {author} {\bibfnamefont {A.}~\bibnamefont
  {Ying}}, \bibinfo {author} {\bibfnamefont {B.}~\bibnamefont {Osting}},
  \bibinfo {author} {\bibfnamefont {I.~C.}\ \bibnamefont {Noyan}}, \bibinfo
  {author} {\bibfnamefont {C.~E.}\ \bibnamefont {Murray}}, \bibinfo {author}
  {\bibfnamefont {M.}~\bibnamefont {Holt}},\ and\ \bibinfo {author}
  {\bibfnamefont {J.}~\bibnamefont {Maser}},\ }\bibfield  {title} {\bibinfo
  {title} {Modeling of kinematic diffraction from a thin silicon film
  illuminated by a coherent, focused x-ray nanobeam},\ }\href
  {https://doi.org/10.1107/S0021889810008459} {\bibfield  {journal} {\bibinfo
  {journal} {Journal of Applied Crystallography}\ }\textbf {\bibinfo {volume}
  {43}},\ \bibinfo {pages} {587–595} (\bibinfo {year} {2010})}\BibitemShut
  {NoStop}%
\bibitem [{\citenamefont {Pateras}\ \emph {et~al.}(2018)\citenamefont
  {Pateras}, \citenamefont {Park}, \citenamefont {Ahn}, \citenamefont {Tilka},
  \citenamefont {Holt}, \citenamefont {Kim}, \citenamefont {Mawst},\ and\
  \citenamefont {Evans}}]{UWDynamicalSims2018}%
  \BibitemOpen
  \bibfield  {author} {\bibinfo {author} {\bibfnamefont {A.}~\bibnamefont
  {Pateras}}, \bibinfo {author} {\bibfnamefont {J.}~\bibnamefont {Park}},
  \bibinfo {author} {\bibfnamefont {Y.}~\bibnamefont {Ahn}}, \bibinfo {author}
  {\bibfnamefont {J.~A.}\ \bibnamefont {Tilka}}, \bibinfo {author}
  {\bibfnamefont {M.~V.}\ \bibnamefont {Holt}}, \bibinfo {author}
  {\bibfnamefont {H.}~\bibnamefont {Kim}}, \bibinfo {author} {\bibfnamefont
  {L.~J.}\ \bibnamefont {Mawst}},\ and\ \bibinfo {author} {\bibfnamefont
  {P.~G.}\ \bibnamefont {Evans}},\ }\bibfield  {title} {\bibinfo {title}
  {Dynamical scattering in coherent hard x-ray nanobeam bragg diffraction},\
  }\href {https://doi.org/10.1103/PhysRevB.97.235414} {\bibfield  {journal}
  {\bibinfo  {journal} {Physical Review B}\ }\textbf {\bibinfo {volume} {97}},\
  \bibinfo {pages} {235414} (\bibinfo {year} {2018})}\BibitemShut {NoStop}%
\bibitem [{SM()}]{SM}%
  \BibitemOpen
  \href@noop {} {}\bibinfo {note} {See Supplemental Material at [URL inserted
  by publisher], which includes references
  \cite{ICpaper2020,PerneggerChargeCarriers2005,GabryschSecondaryElectrons2008,JayakumarOpticalPatterning2016,BalasubramanianIsotopicPurification2009,UWdynamicalsim,Tsubouchi_TEM_2016}}\BibitemShut
  {NoStop}%
\bibitem [{\citenamefont {Tilka}\ \emph {et~al.}(2016)\citenamefont {Tilka},
  \citenamefont {Park}, \citenamefont {Ahn}, \citenamefont {Pateras},
  \citenamefont {Sampson}, \citenamefont {Savage}, \citenamefont {Prance},
  \citenamefont {Simmons}, \citenamefont {Coppersmith}, \citenamefont
  {Eriksson}, \citenamefont {Lagally}, \citenamefont {Holt},\ and\
  \citenamefont {Evans}}]{TilkaOpticalSimulations2016}%
  \BibitemOpen
  \bibfield  {author} {\bibinfo {author} {\bibfnamefont {J.~A.}\ \bibnamefont
  {Tilka}}, \bibinfo {author} {\bibfnamefont {J.}~\bibnamefont {Park}},
  \bibinfo {author} {\bibfnamefont {Y.}~\bibnamefont {Ahn}}, \bibinfo {author}
  {\bibfnamefont {A.}~\bibnamefont {Pateras}}, \bibinfo {author} {\bibfnamefont
  {K.~C.}\ \bibnamefont {Sampson}}, \bibinfo {author} {\bibfnamefont {D.~E.}\
  \bibnamefont {Savage}}, \bibinfo {author} {\bibfnamefont {J.~R.}\
  \bibnamefont {Prance}}, \bibinfo {author} {\bibfnamefont {C.~B.}\
  \bibnamefont {Simmons}}, \bibinfo {author} {\bibfnamefont {S.~N.}\
  \bibnamefont {Coppersmith}}, \bibinfo {author} {\bibfnamefont {M.~A.}\
  \bibnamefont {Eriksson}}, \bibinfo {author} {\bibfnamefont {M.~G.}\
  \bibnamefont {Lagally}}, \bibinfo {author} {\bibfnamefont {M.~V.}\
  \bibnamefont {Holt}},\ and\ \bibinfo {author} {\bibfnamefont {P.~G.}\
  \bibnamefont {Evans}},\ }\bibfield  {title} {\bibinfo {title} {Combining
  experiment and optical simulation in coherent x-ray nanobeam characterization
  of si/sige semiconductor heterostructures},\ }\href
  {https://doi.org/10.1063/1.4955043} {\bibfield  {journal} {\bibinfo
  {journal} {Journal of Applied Physics}\ }\textbf {\bibinfo {volume} {120}},\
  \bibinfo {pages} {015304} (\bibinfo {year} {2016})}\BibitemShut {NoStop}%
\bibitem [{\citenamefont {Batterman}\ and\ \citenamefont
  {Cole}(1964)}]{BattermanDDtheory1964}%
  \BibitemOpen
  \bibfield  {author} {\bibinfo {author} {\bibfnamefont {B.~W.}\ \bibnamefont
  {Batterman}}\ and\ \bibinfo {author} {\bibfnamefont {H.}~\bibnamefont
  {Cole}},\ }\bibfield  {title} {\bibinfo {title} {Dynamical diffraction of x
  rays by perfect crystals},\ }\href
  {https://doi.org/10.1103/RevModPhys.36.681} {\bibfield  {journal} {\bibinfo
  {journal} {Reviews of Modern Physics}\ }\textbf {\bibinfo {volume} {36}},\
  \bibinfo {pages} {681–717} (\bibinfo {year} {1964})}\BibitemShut {NoStop}%
\bibitem [{\citenamefont {Gaukroger}\ \emph {et~al.}(2008)\citenamefont
  {Gaukroger}, \citenamefont {Martineau}, \citenamefont {Crowder},
  \citenamefont {Friel}, \citenamefont {Williams},\ and\ \citenamefont
  {Twitchen}}]{E6xray2008}%
  \BibitemOpen
  \bibfield  {author} {\bibinfo {author} {\bibfnamefont {M.~P.}\ \bibnamefont
  {Gaukroger}}, \bibinfo {author} {\bibfnamefont {P.~M.}\ \bibnamefont
  {Martineau}}, \bibinfo {author} {\bibfnamefont {M.~J.}\ \bibnamefont
  {Crowder}}, \bibinfo {author} {\bibfnamefont {I.}~\bibnamefont {Friel}},
  \bibinfo {author} {\bibfnamefont {S.~D.}\ \bibnamefont {Williams}},\ and\
  \bibinfo {author} {\bibfnamefont {D.~J.}\ \bibnamefont {Twitchen}},\
  }\bibfield  {title} {\bibinfo {title} {X-ray topography studies of
  dislocations in single crystal cvd diamond},\ }\href
  {https://doi.org/10.1016/j.diamond.2007.12.036} {\bibfield  {journal}
  {\bibinfo  {journal} {Diamond and Related Materials}\ }\textbf {\bibinfo
  {volume} {17}},\ \bibinfo {pages} {262–269} (\bibinfo {year}
  {2008})}\BibitemShut {NoStop}%
\bibitem [{\citenamefont {Pinto}\ and\ \citenamefont
  {Jones}(2009)}]{Pinto_Jones_2009}%
  \BibitemOpen
  \bibfield  {author} {\bibinfo {author} {\bibfnamefont {H.}~\bibnamefont
  {Pinto}}\ and\ \bibinfo {author} {\bibfnamefont {R.}~\bibnamefont {Jones}},\
  }\bibfield  {title} {\bibinfo {title} {Theory of the birefringence due to
  dislocations in single crystal cvd diamond},\ }\href
  {https://doi.org/10.1088/0953-8984/21/36/364220} {\bibfield  {journal}
  {\bibinfo  {journal} {Journal of Physics: Condensed Matter}\ }\textbf
  {\bibinfo {volume} {21}},\ \bibinfo {pages} {364220} (\bibinfo {year}
  {2009})}\BibitemShut {NoStop}%
\bibitem [{\citenamefont {Hofmann}\ \emph {et~al.}(2017)\citenamefont
  {Hofmann}, \citenamefont {Tarleton}, \citenamefont {Harder}, \citenamefont
  {Phillips}, \citenamefont {Ma}, \citenamefont {Clark}, \citenamefont
  {Robinson}, \citenamefont {Abbey}, \citenamefont {Liu},\ and\ \citenamefont
  {Beck}}]{HofmannFIBStrain2017}%
  \BibitemOpen
  \bibfield  {author} {\bibinfo {author} {\bibfnamefont {F.}~\bibnamefont
  {Hofmann}}, \bibinfo {author} {\bibfnamefont {E.}~\bibnamefont {Tarleton}},
  \bibinfo {author} {\bibfnamefont {R.~J.}\ \bibnamefont {Harder}}, \bibinfo
  {author} {\bibfnamefont {N.~W.}\ \bibnamefont {Phillips}}, \bibinfo {author}
  {\bibfnamefont {P.-W.}\ \bibnamefont {Ma}}, \bibinfo {author} {\bibfnamefont
  {J.~N.}\ \bibnamefont {Clark}}, \bibinfo {author} {\bibfnamefont {I.~K.}\
  \bibnamefont {Robinson}}, \bibinfo {author} {\bibfnamefont {B.}~\bibnamefont
  {Abbey}}, \bibinfo {author} {\bibfnamefont {W.}~\bibnamefont {Liu}},\ and\
  \bibinfo {author} {\bibfnamefont {C.~E.}\ \bibnamefont {Beck}},\ }\bibfield
  {title} {\bibinfo {title} {3d lattice distortions and defect structures in
  ion-implanted nano-crystals},\ }\href {https://doi.org/10.1038/srep45993}
  {\bibfield  {journal} {\bibinfo  {journal} {Scientific Reports}\ }\textbf
  {\bibinfo {volume} {7}},\ \bibinfo {pages} {45993} (\bibinfo {year}
  {2017})}\BibitemShut {NoStop}%
\bibitem [{\citenamefont {Hofmann}\ \emph {et~al.}(2020)\citenamefont
  {Hofmann}, \citenamefont {Phillips}, \citenamefont {Das}, \citenamefont
  {Karamched}, \citenamefont {Hughes}, \citenamefont {Douglas}, \citenamefont
  {Cha},\ and\ \citenamefont {Liu}}]{HofmannFullStrainTensor2020}%
  \BibitemOpen
  \bibfield  {author} {\bibinfo {author} {\bibfnamefont {F.}~\bibnamefont
  {Hofmann}}, \bibinfo {author} {\bibfnamefont {N.~W.}\ \bibnamefont
  {Phillips}}, \bibinfo {author} {\bibfnamefont {S.}~\bibnamefont {Das}},
  \bibinfo {author} {\bibfnamefont {P.}~\bibnamefont {Karamched}}, \bibinfo
  {author} {\bibfnamefont {G.~M.}\ \bibnamefont {Hughes}}, \bibinfo {author}
  {\bibfnamefont {J.~O.}\ \bibnamefont {Douglas}}, \bibinfo {author}
  {\bibfnamefont {W.}~\bibnamefont {Cha}},\ and\ \bibinfo {author}
  {\bibfnamefont {W.}~\bibnamefont {Liu}},\ }\bibfield  {title} {\bibinfo
  {title} {Nanoscale imaging of the full strain tensor of specific dislocations
  extracted from a bulk sample},\ }\href
  {https://doi.org/10.1103/PhysRevMaterials.4.013801} {\bibfield  {journal}
  {\bibinfo  {journal} {Physical Review Materials}\ }\textbf {\bibinfo {volume}
  {4}},\ \bibinfo {pages} {013801} (\bibinfo {year} {2020})}\BibitemShut
  {NoStop}%
\bibitem [{\citenamefont {Eshelby}(1954)}]{Eshelby1954}%
  \BibitemOpen
  \bibfield  {author} {\bibinfo {author} {\bibfnamefont {J.~D.}\ \bibnamefont
  {Eshelby}},\ }\bibfield  {title} {\bibinfo {title} {Distortion of a crystal
  by point imperfections},\ }\href {https://doi.org/10.1063/1.1721615}
  {\bibfield  {journal} {\bibinfo  {journal} {Journal of Applied Physics}\
  }\textbf {\bibinfo {volume} {25}},\ \bibinfo {pages} {255–261} (\bibinfo
  {year} {1954})}\BibitemShut {NoStop}%
\bibitem [{\citenamefont {Pinto}\ \emph {et~al.}(2011)\citenamefont {Pinto},
  \citenamefont {Jones}, \citenamefont {Goss},\ and\ \citenamefont
  {Briddon}}]{PintoExtendedDefects}%
  \BibitemOpen
  \bibfield  {author} {\bibinfo {author} {\bibfnamefont {H.}~\bibnamefont
  {Pinto}}, \bibinfo {author} {\bibfnamefont {R.}~\bibnamefont {Jones}},
  \bibinfo {author} {\bibfnamefont {J.~P.}\ \bibnamefont {Goss}},\ and\
  \bibinfo {author} {\bibfnamefont {P.~R.}\ \bibnamefont {Briddon}},\
  }\bibfield  {title} {\bibinfo {title} {Point and extended defects in chemical
  vapour deposited diamond},\ }\href
  {https://doi.org/10.1088/1742-6596/281/1/012023} {\bibfield  {journal}
  {\bibinfo  {journal} {Journal of Physics: Conference Series}\ }\textbf
  {\bibinfo {volume} {281}},\ \bibinfo {pages} {012023} (\bibinfo {year}
  {2011})}\BibitemShut {NoStop}%
\bibitem [{\citenamefont {Crisci}\ \emph {et~al.}(2011)\citenamefont {Crisci},
  \citenamefont {Baillet}, \citenamefont {Mermoux}, \citenamefont {Bogdan},
  \citenamefont {Nesládek},\ and\ \citenamefont {Haenen}}]{RamanStrain2011}%
  \BibitemOpen
  \bibfield  {author} {\bibinfo {author} {\bibfnamefont {A.}~\bibnamefont
  {Crisci}}, \bibinfo {author} {\bibfnamefont {F.}~\bibnamefont {Baillet}},
  \bibinfo {author} {\bibfnamefont {M.}~\bibnamefont {Mermoux}}, \bibinfo
  {author} {\bibfnamefont {G.}~\bibnamefont {Bogdan}}, \bibinfo {author}
  {\bibfnamefont {M.}~\bibnamefont {Nesládek}},\ and\ \bibinfo {author}
  {\bibfnamefont {K.}~\bibnamefont {Haenen}},\ }\bibfield  {title} {\bibinfo
  {title} {Residual strain around grown-in defects in cvd diamond single
  crystals: A 2d and 3d raman imaging study},\ }\href
  {https://doi.org/10.1002/pssa.201100039} {\bibfield  {journal} {\bibinfo
  {journal} {physica status solidi (a)}\ }\textbf {\bibinfo {volume} {208}},\
  \bibinfo {pages} {2038–2044} (\bibinfo {year} {2011})}\BibitemShut
  {NoStop}%
\bibitem [{\citenamefont {Klein}\ and\ \citenamefont
  {Cardinale}(1992)}]{CVDYoungsModulus1992}%
  \BibitemOpen
  \bibfield  {author} {\bibinfo {author} {\bibfnamefont {C.~A.}\ \bibnamefont
  {Klein}}\ and\ \bibinfo {author} {\bibfnamefont {G.~F.}\ \bibnamefont
  {Cardinale}},\ }\bibfield  {title} {\bibinfo {title} {Young’s modulus and
  poisson’s ratio of cvd diamond},\ }in\ \href
  {https://doi.org/10.1117/12.130771} {\emph {\bibinfo {booktitle} {Diamond
  Optics V}}},\ Vol.\ \bibinfo {volume} {1759}\ (\bibinfo  {publisher}
  {International Society for Optics and Photonics},\ \bibinfo {year} {1992})\
  p.\ \bibinfo {pages} {178–193}\BibitemShut {NoStop}%
\bibitem [{\citenamefont {Ziegler}\ \emph {et~al.}(2010)\citenamefont
  {Ziegler}, \citenamefont {Ziegler},\ and\ \citenamefont {Biersack}}]{SRIM}%
  \BibitemOpen
  \bibfield  {author} {\bibinfo {author} {\bibfnamefont {J.~F.}\ \bibnamefont
  {Ziegler}}, \bibinfo {author} {\bibfnamefont {M.~D.}\ \bibnamefont
  {Ziegler}},\ and\ \bibinfo {author} {\bibfnamefont {J.~P.}\ \bibnamefont
  {Biersack}},\ }\bibfield  {title} {\bibinfo {title} {Srim – the stopping
  and range of ions in matter (2010)},\ }\href
  {https://doi.org/10.1016/j.nimb.2010.02.091} {\bibfield  {journal} {\bibinfo
  {journal} {Nuclear Instruments and Methods in Physics Research Section B:
  Beam Interactions with Materials and Atoms}\ }\bibinfo {series} {19th
  International Conference on Ion Beam Analysis},\ \textbf {\bibinfo {volume}
  {268}},\ \bibinfo {pages} {1818–1823} (\bibinfo {year} {2010})}\BibitemShut
  {NoStop}%
\bibitem [{\citenamefont {Pfeiffer}(2018)}]{Pfeiffer_2018}%
  \BibitemOpen
  \bibfield  {author} {\bibinfo {author} {\bibfnamefont {F.}~\bibnamefont
  {Pfeiffer}},\ }\bibfield  {title} {\bibinfo {title} {X-ray ptychography},\
  }\href {https://doi.org/10.1038/s41566-017-0072-5} {\bibfield  {journal}
  {\bibinfo  {journal} {Nature Photonics}\ }\textbf {\bibinfo {volume} {12}},\
  \bibinfo {pages} {9–17} (\bibinfo {year} {2018})}\BibitemShut {NoStop}%
\bibitem [{\citenamefont {Hruszkewycz}\ \emph {et~al.}(2017)\citenamefont
  {Hruszkewycz}, \citenamefont {Allain}, \citenamefont {Holt}, \citenamefont
  {Murray}, \citenamefont {Holt}, \citenamefont {Fuoss},\ and\ \citenamefont
  {Chamard}}]{3DBPP2017}%
  \BibitemOpen
  \bibfield  {author} {\bibinfo {author} {\bibfnamefont {S.~O.}\ \bibnamefont
  {Hruszkewycz}}, \bibinfo {author} {\bibfnamefont {M.}~\bibnamefont {Allain}},
  \bibinfo {author} {\bibfnamefont {M.~V.}\ \bibnamefont {Holt}}, \bibinfo
  {author} {\bibfnamefont {C.~E.}\ \bibnamefont {Murray}}, \bibinfo {author}
  {\bibfnamefont {J.~R.}\ \bibnamefont {Holt}}, \bibinfo {author}
  {\bibfnamefont {P.~H.}\ \bibnamefont {Fuoss}},\ and\ \bibinfo {author}
  {\bibfnamefont {V.}~\bibnamefont {Chamard}},\ }\bibfield  {title} {\bibinfo
  {title} {High-resolution three-dimensional structural microscopy by
  single-angle bragg ptychography},\ }\href {https://doi.org/10.1038/nmat4798}
  {\bibfield  {journal} {\bibinfo  {journal} {Nature Materials}\ }\textbf
  {\bibinfo {volume} {16}},\ \bibinfo {pages} {244–251} (\bibinfo {year}
  {2017})}\BibitemShut {NoStop}%
\bibitem [{\citenamefont {Hill}\ \emph {et~al.}(2018)\citenamefont {Hill},
  \citenamefont {Calvo-Almazan}, \citenamefont {Allain}, \citenamefont {Holt},
  \citenamefont {Ulvestad}, \citenamefont {Treu}, \citenamefont {Koblmüller},
  \citenamefont {Huang}, \citenamefont {Huang}, \citenamefont {Yan},
  \citenamefont {Nazaretski}, \citenamefont {Chu}, \citenamefont {Stephenson},
  \citenamefont {Chamard}, \citenamefont {Lauhon},\ and\ \citenamefont
  {Hruszkewycz}}]{multiangleBPP2018}%
  \BibitemOpen
  \bibfield  {author} {\bibinfo {author} {\bibfnamefont {M.}~\bibnamefont
  {Hill}}, \bibinfo {author} {\bibfnamefont {I.}~\bibnamefont {Calvo-Almazan}},
  \bibinfo {author} {\bibfnamefont {M.}~\bibnamefont {Allain}}, \bibinfo
  {author} {\bibfnamefont {M.}~\bibnamefont {Holt}}, \bibinfo {author}
  {\bibfnamefont {A.}~\bibnamefont {Ulvestad}}, \bibinfo {author}
  {\bibfnamefont {J.}~\bibnamefont {Treu}}, \bibinfo {author} {\bibfnamefont
  {G.}~\bibnamefont {Koblmüller}}, \bibinfo {author} {\bibfnamefont
  {C.}~\bibnamefont {Huang}}, \bibinfo {author} {\bibfnamefont
  {X.}~\bibnamefont {Huang}}, \bibinfo {author} {\bibfnamefont
  {H.}~\bibnamefont {Yan}}, \bibinfo {author} {\bibfnamefont {E.}~\bibnamefont
  {Nazaretski}}, \bibinfo {author} {\bibfnamefont {Y.}~\bibnamefont {Chu}},
  \bibinfo {author} {\bibfnamefont {G.}~\bibnamefont {Stephenson}}, \bibinfo
  {author} {\bibfnamefont {V.}~\bibnamefont {Chamard}}, \bibinfo {author}
  {\bibfnamefont {L.}~\bibnamefont {Lauhon}},\ and\ \bibinfo {author}
  {\bibfnamefont {S.}~\bibnamefont {Hruszkewycz}},\ }\bibfield  {title}
  {\bibinfo {title} {Measuring three-dimensional strain and structural defects
  in a single ingaas nanowire using coherent x-ray multiangle bragg projection
  ptychography},\ }\href {https://doi.org/10.1021/acs.nanolett.7b04024}
  {\bibfield  {journal} {\bibinfo  {journal} {Nano Letters}\ }\textbf {\bibinfo
  {volume} {18}},\ \bibinfo {pages} {811–819} (\bibinfo {year}
  {2018})}\BibitemShut {NoStop}%
\bibitem [{\citenamefont {Turner}\ \emph {et~al.}(2020)\citenamefont {Turner},
  \citenamefont {Langellier}, \citenamefont {Bainbridge}, \citenamefont
  {Walters}, \citenamefont {Meesala}, \citenamefont {Babinec}, \citenamefont
  {Kehayias}, \citenamefont {Yacoby}, \citenamefont {Hu}, \citenamefont
  {Lončar}, \citenamefont {Walsworth},\ and\ \citenamefont
  {Levine}}]{ICpaper2020}%
  \BibitemOpen
  \bibfield  {author} {\bibinfo {author} {\bibfnamefont {M.~J.}\ \bibnamefont
  {Turner}}, \bibinfo {author} {\bibfnamefont {N.}~\bibnamefont {Langellier}},
  \bibinfo {author} {\bibfnamefont {R.}~\bibnamefont {Bainbridge}}, \bibinfo
  {author} {\bibfnamefont {D.}~\bibnamefont {Walters}}, \bibinfo {author}
  {\bibfnamefont {S.}~\bibnamefont {Meesala}}, \bibinfo {author} {\bibfnamefont
  {T.~M.}\ \bibnamefont {Babinec}}, \bibinfo {author} {\bibfnamefont
  {P.}~\bibnamefont {Kehayias}}, \bibinfo {author} {\bibfnamefont
  {A.}~\bibnamefont {Yacoby}}, \bibinfo {author} {\bibfnamefont
  {E.}~\bibnamefont {Hu}}, \bibinfo {author} {\bibfnamefont {M.}~\bibnamefont
  {Lončar}}, \bibinfo {author} {\bibfnamefont {R.~L.}\ \bibnamefont
  {Walsworth}},\ and\ \bibinfo {author} {\bibfnamefont {E.~V.}\ \bibnamefont
  {Levine}},\ }\bibfield  {title} {\bibinfo {title} {Magnetic field
  fingerprinting of integrated-circuit activity with a quantum diamond
  microscope},\ }\href {https://doi.org/10.1103/PhysRevApplied.14.014097}
  {\bibfield  {journal} {\bibinfo  {journal} {Physical Review Applied}\
  }\textbf {\bibinfo {volume} {14}},\ \bibinfo {pages} {014097} (\bibinfo
  {year} {2020})}\BibitemShut {NoStop}%
\bibitem [{\citenamefont {Pernegger}\ \emph {et~al.}(2005)\citenamefont
  {Pernegger}, \citenamefont {Roe}, \citenamefont {Weilhammer}, \citenamefont
  {Eremin}, \citenamefont {Frais-Kölbl}, \citenamefont {Griesmayer},
  \citenamefont {Kagan}, \citenamefont {Schnetzer}, \citenamefont {Stone},
  \citenamefont {Trischuk}, \citenamefont {Twitchen},\ and\ \citenamefont
  {Whitehead}}]{PerneggerChargeCarriers2005}%
  \BibitemOpen
  \bibfield  {author} {\bibinfo {author} {\bibfnamefont {H.}~\bibnamefont
  {Pernegger}}, \bibinfo {author} {\bibfnamefont {S.}~\bibnamefont {Roe}},
  \bibinfo {author} {\bibfnamefont {P.}~\bibnamefont {Weilhammer}}, \bibinfo
  {author} {\bibfnamefont {V.}~\bibnamefont {Eremin}}, \bibinfo {author}
  {\bibfnamefont {H.}~\bibnamefont {Frais-Kölbl}}, \bibinfo {author}
  {\bibfnamefont {E.}~\bibnamefont {Griesmayer}}, \bibinfo {author}
  {\bibfnamefont {H.}~\bibnamefont {Kagan}}, \bibinfo {author} {\bibfnamefont
  {S.}~\bibnamefont {Schnetzer}}, \bibinfo {author} {\bibfnamefont
  {R.}~\bibnamefont {Stone}}, \bibinfo {author} {\bibfnamefont
  {W.}~\bibnamefont {Trischuk}}, \bibinfo {author} {\bibfnamefont
  {D.}~\bibnamefont {Twitchen}},\ and\ \bibinfo {author} {\bibfnamefont
  {A.}~\bibnamefont {Whitehead}},\ }\bibfield  {title} {\bibinfo {title}
  {Charge-carrier properties in synthetic single-crystal diamond measured with
  the transient-current technique},\ }\href {https://doi.org/10.1063/1.1863417}
  {\bibfield  {journal} {\bibinfo  {journal} {Journal of Applied Physics}\
  }\textbf {\bibinfo {volume} {97}},\ \bibinfo {pages} {073704} (\bibinfo
  {year} {2005})}\BibitemShut {NoStop}%
\bibitem [{\citenamefont {Gabrysch}\ \emph {et~al.}(2008)\citenamefont
  {Gabrysch}, \citenamefont {Marklund}, \citenamefont {Hajdu}, \citenamefont
  {Twitchen}, \citenamefont {Rudati}, \citenamefont {Lindenberg}, \citenamefont
  {Caleman}, \citenamefont {Falcone}, \citenamefont {Tschentscher},
  \citenamefont {Moffat}, \citenamefont {Bucksbaum}, \citenamefont
  {Als-Nielsen}, \citenamefont {Nelson}, \citenamefont {Siddons}, \citenamefont
  {Emma}, \citenamefont {Krejcik}, \citenamefont {Schlarb}, \citenamefont
  {Arthur}, \citenamefont {Brennan}, \citenamefont {Hastings},\ and\
  \citenamefont {Isberg}}]{GabryschSecondaryElectrons2008}%
  \BibitemOpen
  \bibfield  {author} {\bibinfo {author} {\bibfnamefont {M.}~\bibnamefont
  {Gabrysch}}, \bibinfo {author} {\bibfnamefont {E.}~\bibnamefont {Marklund}},
  \bibinfo {author} {\bibfnamefont {J.}~\bibnamefont {Hajdu}}, \bibinfo
  {author} {\bibfnamefont {D.~J.}\ \bibnamefont {Twitchen}}, \bibinfo {author}
  {\bibfnamefont {J.}~\bibnamefont {Rudati}}, \bibinfo {author} {\bibfnamefont
  {A.~M.}\ \bibnamefont {Lindenberg}}, \bibinfo {author} {\bibfnamefont
  {C.}~\bibnamefont {Caleman}}, \bibinfo {author} {\bibfnamefont {R.~W.}\
  \bibnamefont {Falcone}}, \bibinfo {author} {\bibfnamefont {T.}~\bibnamefont
  {Tschentscher}}, \bibinfo {author} {\bibfnamefont {K.}~\bibnamefont
  {Moffat}}, \bibinfo {author} {\bibfnamefont {P.~H.}\ \bibnamefont
  {Bucksbaum}}, \bibinfo {author} {\bibfnamefont {J.}~\bibnamefont
  {Als-Nielsen}}, \bibinfo {author} {\bibfnamefont {A.~J.}\ \bibnamefont
  {Nelson}}, \bibinfo {author} {\bibfnamefont {D.~P.}\ \bibnamefont {Siddons}},
  \bibinfo {author} {\bibfnamefont {P.~J.}\ \bibnamefont {Emma}}, \bibinfo
  {author} {\bibfnamefont {P.}~\bibnamefont {Krejcik}}, \bibinfo {author}
  {\bibfnamefont {H.}~\bibnamefont {Schlarb}}, \bibinfo {author} {\bibfnamefont
  {J.}~\bibnamefont {Arthur}}, \bibinfo {author} {\bibfnamefont
  {S.}~\bibnamefont {Brennan}}, \bibinfo {author} {\bibfnamefont
  {J.}~\bibnamefont {Hastings}},\ and\ \bibinfo {author} {\bibfnamefont
  {J.}~\bibnamefont {Isberg}},\ }\bibfield  {title} {\bibinfo {title}
  {Formation of secondary electron cascades in single-crystalline
  plasma-deposited diamond upon exposure to femtosecond x-ray pulses},\ }\href
  {https://doi.org/10.1063/1.2890158} {\bibfield  {journal} {\bibinfo
  {journal} {Journal of Applied Physics}\ }\textbf {\bibinfo {volume} {103}},\
  \bibinfo {pages} {064909} (\bibinfo {year} {2008})}\BibitemShut {NoStop}%
\bibitem [{\citenamefont {Jayakumar}\ \emph {et~al.}(2016)\citenamefont
  {Jayakumar}, \citenamefont {Henshaw}, \citenamefont {Dhomkar}, \citenamefont
  {Pagliero}, \citenamefont {Laraoui}, \citenamefont {Manson}, \citenamefont
  {Albu}, \citenamefont {Doherty},\ and\ \citenamefont
  {Meriles}}]{JayakumarOpticalPatterning2016}%
  \BibitemOpen
  \bibfield  {author} {\bibinfo {author} {\bibfnamefont {H.}~\bibnamefont
  {Jayakumar}}, \bibinfo {author} {\bibfnamefont {J.}~\bibnamefont {Henshaw}},
  \bibinfo {author} {\bibfnamefont {S.}~\bibnamefont {Dhomkar}}, \bibinfo
  {author} {\bibfnamefont {D.}~\bibnamefont {Pagliero}}, \bibinfo {author}
  {\bibfnamefont {A.}~\bibnamefont {Laraoui}}, \bibinfo {author} {\bibfnamefont
  {N.~B.}\ \bibnamefont {Manson}}, \bibinfo {author} {\bibfnamefont
  {R.}~\bibnamefont {Albu}}, \bibinfo {author} {\bibfnamefont {M.~W.}\
  \bibnamefont {Doherty}},\ and\ \bibinfo {author} {\bibfnamefont {C.~A.}\
  \bibnamefont {Meriles}},\ }\bibfield  {title} {\bibinfo {title} {Optical
  patterning of trapped charge in nitrogen-doped diamond},\ }\href
  {https://doi.org/10.1038/ncomms12660} {\bibfield  {journal} {\bibinfo
  {journal} {Nature Communications}\ }\textbf {\bibinfo {volume} {7}},\
  \bibinfo {pages} {12660} (\bibinfo {year} {2016})}\BibitemShut {NoStop}%
\bibitem [{\citenamefont {Balasubramanian}\ \emph {et~al.}(2009)\citenamefont
  {Balasubramanian}, \citenamefont {Neumann}, \citenamefont {Twitchen},
  \citenamefont {Markham}, \citenamefont {Kolesov}, \citenamefont {Mizuochi},
  \citenamefont {Isoya}, \citenamefont {Achard}, \citenamefont {Beck},
  \citenamefont {Tissler}, \citenamefont {Jacques}, \citenamefont {Hemmer},
  \citenamefont {Jelezko},\ and\ \citenamefont
  {Wrachtrup}}]{BalasubramanianIsotopicPurification2009}%
  \BibitemOpen
  \bibfield  {author} {\bibinfo {author} {\bibfnamefont {G.}~\bibnamefont
  {Balasubramanian}}, \bibinfo {author} {\bibfnamefont {P.}~\bibnamefont
  {Neumann}}, \bibinfo {author} {\bibfnamefont {D.}~\bibnamefont {Twitchen}},
  \bibinfo {author} {\bibfnamefont {M.}~\bibnamefont {Markham}}, \bibinfo
  {author} {\bibfnamefont {R.}~\bibnamefont {Kolesov}}, \bibinfo {author}
  {\bibfnamefont {N.}~\bibnamefont {Mizuochi}}, \bibinfo {author}
  {\bibfnamefont {J.}~\bibnamefont {Isoya}}, \bibinfo {author} {\bibfnamefont
  {J.}~\bibnamefont {Achard}}, \bibinfo {author} {\bibfnamefont
  {J.}~\bibnamefont {Beck}}, \bibinfo {author} {\bibfnamefont {J.}~\bibnamefont
  {Tissler}}, \bibinfo {author} {\bibfnamefont {V.}~\bibnamefont {Jacques}},
  \bibinfo {author} {\bibfnamefont {P.~R.}\ \bibnamefont {Hemmer}}, \bibinfo
  {author} {\bibfnamefont {F.}~\bibnamefont {Jelezko}},\ and\ \bibinfo {author}
  {\bibfnamefont {J.}~\bibnamefont {Wrachtrup}},\ }\bibfield  {title} {\bibinfo
  {title} {Ultralong spin coherence time in isotopically engineered diamond},\
  }\href {https://doi.org/10.1038/nmat2420} {\bibfield  {journal} {\bibinfo
  {journal} {Nature Materials}\ }\textbf {\bibinfo {volume} {8}},\ \bibinfo
  {pages} {383–387} (\bibinfo {year} {2009})}\BibitemShut {NoStop}%
\bibitem [{\citenamefont {Gardner}\ \emph {et~al.}()\citenamefont {Gardner},
  \citenamefont {Mueller}, \citenamefont {Tilka},\ and\ \citenamefont
  {Evans}}]{UWdynamicalsim}%
  \BibitemOpen
  \bibfield  {author} {\bibinfo {author} {\bibfnamefont {K.}~\bibnamefont
  {Gardner}}, \bibinfo {author} {\bibfnamefont {E.}~\bibnamefont {Mueller}},
  \bibinfo {author} {\bibfnamefont {J.}~\bibnamefont {Tilka}},\ and\ \bibinfo
  {author} {\bibfnamefont {P.}~\bibnamefont {Evans}},\ }\href@noop {} {\bibinfo
  {title} {X-ray nanobeam dynamical diffraction simulator}},\ \bibinfo
  {howpublished}
  {\url{https://xray.engr.wisc.edu/nanobeamsimulation/dynamicalhtml/index.php}},\
  \bibinfo {note} {accessed: 2020-09-09}\BibitemShut {NoStop}%
\bibitem [{\citenamefont {Tsubouchi}\ and\ \citenamefont
  {Mokuno}(2016)}]{Tsubouchi_TEM_2016}%
  \BibitemOpen
  \bibfield  {author} {\bibinfo {author} {\bibfnamefont {N.}~\bibnamefont
  {Tsubouchi}}\ and\ \bibinfo {author} {\bibfnamefont {Y.}~\bibnamefont
  {Mokuno}},\ }\bibfield  {title} {\bibinfo {title} {Configuration of a single
  grown-in dislocation corresponding to one etch pit formed on the surface of
  cvd homoepitaxial diamond},\ }\href
  {https://doi.org/10.1016/j.jcrysgro.2016.09.030} {\bibfield  {journal}
  {\bibinfo  {journal} {Journal of Crystal Growth}\ }\textbf {\bibinfo {volume}
  {455}},\ \bibinfo {pages} {71–75} (\bibinfo {year} {2016})}\BibitemShut
  {NoStop}%
\end{thebibliography}%

\clearpage

\section{Supplemental Material}

\subsection{Sample information}
\label{sec:SuppDiamond}

The CVD quantum sensing diamond used in this study was produced by Element Six, inc.  It features a 40 $\mu$m thick overgrowth layer, grown via (001) chemical vapor deposition atop an electronic-grade, single-crystal substrate, with less than 5 ppb of both N and NV. Before growth, the substrate was mechanically polished.  The overgrowth layer was grown with isotopically purified carbon, with 99.95\% $^{12}$C, and doped with 10 parts per million (ppm) of $^{15}$N.  After growth, the diamond was electron-irradiated and annealed to form NV centers in the N-doped layer.  The resulting NV$^-$ density is approximately 1 ppm, similar to sister samples used in previous work \cite{StrainPaper, ICpaper2020}.

We note that the electron irradiation also helps to resolve concerns about charge- and dipole-trapping near damage tracks affecting QDM measurements.  In high-NV-density diamonds, which have been irradiated with a high fluence of electrons, a single additional charge will have little to no effect on the electric-field environment.  (However, note that this electron irradiation cannot initiate nuclear recoil chains which would interfere with damage track detection - conservation of momentum dictates each electron will create only one or a few lattice site vacancies.)  Said electron irradiation also does not tend to form large numbers of permanent dipoles, as evidenced by the lack of strong electric-field signals in NV spectroscopy on irradiated diamonds. This may be a result of diamond's semiconductor nature; electron-hole pairs, once induced, will either recombine or drift through the lattice to charge-trapping impurities, rather than form enduring dipoles near the damage track \cite{PerneggerChargeCarriers2005,GabryschSecondaryElectrons2008,JayakumarOpticalPatterning2016}.  The same should be true for ionization induced by WIMP- or particle-induced damage tracks.  (Additionally, see Sec.~\ref{sec:SuppQDM} below for a discussion of potential electric-field contributions to QDM measurements.)

We additionally note that the isotopic purification is significant for many diamond quantum sensing applications, as removal of $^{13}$C nuclear spins enhances quantum defect coherence times and reduces transition linewidths \cite{BalasubramanianIsotopicPurification2009}.  However, the isotopic makeup of the diamond has little effect on the SXDM measurements; and for QDM measurements such as those reported here, isotopic purification is relevant only when the strain-induced frequency shifts to be sensed are small enough to be comparable in magnitude with the linewidth reduction from isotopic purification.

The HPHT diamond featured in main text Fig.~2 was produced by Element Six, inc, using type Ib high-pressure, high-temperature synthesis.  Produced using carbon with natural isotopic abundance together with approximately 100-200 ppm of nitrogen, this HPHT diamond features more homogeneous overall strain than the layered CVD sample, but lacks the dense NV ensemble required for magnetic or strain imaging using NV spin-state spectroscopy.  The HPHT diamond surface contains imperfections or defects, which were identified prior to SXDM scanning via confocal laser microscopy; main text Fig.~2 shows an SXDM scan of one group of these features.  The exact nature of these features is not known, but could include surface imperfections due to microscopic polishing damage; clusters of pointlike defects; or surface contamination from photolithographic processing and metallization of fiducial markers.  Independent of the feature origin, however, the physical scale of the features is comparable to the 100-nm scale expected for WIMP- or particle-induced damage tracks.

\subsection{Constraining contributions from changes in lattice orientation}
\label{sec:SuppTilt}

In this work, crystal lattice strain is measured via shifts in the X-ray Bragg diffraction pattern.  Similar shifts in the diffraction pattern may also result from changes of orientation in the crystal lattice.  However, while strain-induced changes in lattice spacing of the diffracting crystal planes can only shift the diffraction pattern along the detector 2$\theta$ axis, a changing lattice orientation can shift the diffraction pattern along either detector axis.  Characterizing the position of the diffraction pattern along the detector X axis can thus help constrain the contribution of lattice orientation changes to the observed diffraction pattern shifts.  We characterize such shifts by finding the edges of the central low-count region induced by the beam stop \cite{HoltReview2013}, shown in Fig.~\ref{fig:bsfit}a.  We find that the average vertical positions in the strain features and the overall diffraction profile are equal, as shown in Fig.~\ref{fig:bsfit}b.  This disfavors changes in lattice orientation as the main source of the observed diffraction pattern shifts, except for orientations that coincidentally yield little or no X position shift. 

\bsfit

Comparison of the measurements performed using (113) and ($\bm\bar{1}$13) diffraction enable us to further constrain possible contributions from changes in lattice orientation.  Features A and C of Fig.~5 of the main text exhibit a centroid shift in the reduced diffraction curve lineshape by an amount consistent with equal strain along both projections.  We also observe vertical diffraction pattern shifts of one pixel or less in both projections.  This constrains the possible directions of lattice orientation in these strain features to only those that yield nearly zero vertical diffraction shift for both Bragg conditions, as well as a horizontal shift consistent with equal strain.  While such orientation changes cannot be ruled out by our measurements, strain with Burgers vector in the $\langle$110$\rangle$ family of crystal axes is a more plausible explanation, since it both explains the observations and is known to be a dominant class of growth defects in diamond \cite{E6xray2008}.  This contributes to our choice in Sec.~II of the main text to present results in the uniform-orientation scenario.  More generally, while some of the shifts in the diffraction pattern may be caused by orientation changes as well as strain, in this work we demonstrate a sensitivity to strain sufficient, e.g., to detect WIMP dark matter signals.

\subsection{Constructing local host-crystal diffraction profiles}
\label{sec:SuppBkg}
\ROIfig

In order to calculate the reduced diffraction pattern used for analysis of short-length strain features, we first construct local host-crystal diffraction profiles.  This requires an initial classification of each X-ray beam position as ``strained'' or ``host crystal'' depending on whether the beam intersects a strain feature.  We base this initial classification on a ``region of interest'' (ROI) analysis, where we define detector regions that count photons scattered by compressively strained regions of the diamond.  Figure \ref{fig:ROIfig}a shows such an ROI on an example diffraction pattern, while Fig.~\ref{fig:ROIfig}b shows a map of total counts in this ROI for all beam positions in one spatial scan.  Beam positions with more than a threshold number of counts in this ROI are identified as ``strained,'' while positions with less than this threshold are identified as interacting only with the host crystal.  To determine the appropriate threshold, we measure the distribution of counts in a region of ``pristine'' diamond without strain features; the threshold is chosen to be approximately two standard deviations of this distribution above the mode of count numbers in each column of a spatial scan, to account for spatial intensity gradients such as that visible in Fig.~\ref{fig:ROIfig}b.  We check this thresholding procedure by manually defining ``strain'' and ``host crystal'' regions for several spatial scans; the results show good agreement.  We note that this initial classification is used only to generate the host-crystal profile; in the final analysis, each beam position is re-classified based on parameters of the reduced diffraction pattern fit.  The overall analysis procedure is therefore robust to occasional initial misclassifications of specific beam positions.

As can be seen in Fig.~\ref{fig:ROIfig}b, this ROI analysis reveals a smooth horizontal gradient in intensity across the field of view, likely originating from variation in the separation between sample and focusing optics as the X-ray beam is rastered across the sample.  We therefore construct a separate local host-crystal diffraction profile for each column of a spatial scan by averaging all diffraction patterns from ``unstrained'' beam positions within that column.  Dividing by the local host-crystal diffraction profile thus suppresses the inhomogeneous signal from the unstrained diamond, while simultaneously removing the effect of these horizontal gradients.  This procedure yields reduced diffraction patterns such as the one featured in Fig.~4b of the main text.

\subsection{Calibration curves}
\label{sec:SuppCal}
To convert between the position of the diffraction peak on the detector and the strain in the short-length-scale crystal feature, we use a calibration curve.  To generate this curve, we use the kinematic approximation for crystal diffraction \cite{TilkaOpticalSimulations2016,YingKinematicCalcs2010} to simulate diffraction patterns for an unstrained diamond substrate with and without an additional thin strained layer, representing the strain feature.  The results of one such simulation are presented in Fig.~\ref{fig:calfig}a.  We then take the ratio between the two diffraction curves, replicating the process by which we generate the reduced diffraction patterns (as discussed in Sec.~II of the main text).  This gives a simulated reduced diffraction curve which can be fit to extract its centroid, as shown in Fig.~\ref{fig:calfig}b.  Finally, this process is repeated for different strain values in the layer, generating a calibration curve, as shown in Fig.~\ref{fig:calfig}c.  

\calfig

Each point in Fig.~\ref{fig:calfig}a-b represents a column of detector pixels.  In the real detector, these pixels count diffracted X-ray photons integrated over a finite area rather than a single point.  To replicate this in our calibration and avoid possible aliasing effects, we initially oversample the simulated diffraction curve by a factor of 12; we then sum the oversampled calibration curves into bins representing the physical detector pixels.

In order to test the robustness of this calibration and characterize its uncertainty, we generated calibration curves while varying several parameters in the kinematic diffraction simulation.  These include the thickness of the strained layer; the linewidth of the diffraction peak due to the unstrained substrate; the amplitude of diffraction peaks from the substrate and strained layer; the amount of oversampling performed; and the position of bin edges when summing the oversampled points.  Amplitude and oversampling parameters are found not to significantly affect the calibration; uncertainty in the calibration due to strain feature thickness and unstrained diffraction peak linewidth are discussed in supplemental material Sec.~\ref{sec:SuppUnc}.

\subsection{Identifying strain feature edges via vertical asymmetry}
\label{sec:edgeasymmetry}
One additional limitation of the lineshape analysis and calibration procedure relates to strain feature thickness: at positions where the X-ray beam intersects the strained feature for less than $\sim$200 nm, the calibration is less reliable.  The Bragg diffraction peak width from a layer increases as the layer thickness decreases; for very thin layers, some portion of the diffraction peak may overlap with the large, slowly varying diffraction profile and be suppressed by the empirical lineshape reduction.   

\asymmetriclobes

To avoid miscalibrated measurements, we identify the edges of growth defects, where the intersection with the beam may be problematically short. We perform this identification by examining the two-dimensional shape of the reduced diffraction pattern.  The diffraction pattern is divided into two ``lobes'', separated by a low-count region created by the central beam stop of the focusing optic \cite{HoltReview2013} (as is visible in Fig.~4a-b of the main text).  At the sharp edges of a growth defect, we see different diffraction from the two lobes, implying significant strain variation over the tens of nanometers of the beam spot.  Away from the sharp edges, X-rays in each lobe diffract identically, as the separation between lobes is miniscule compared to the length scale of strain features.  Diffraction patterns from the sharp edges of strain features thus show asymmetry between the two lobes, which is emphasized in the reduced diffraction pattern.  Figure \ref{fig:asymmetriclobes}a shows a reduced diffraction pattern for such an edge, with the upper lobe having significantly more intensity after the lineshape reduction. Figure \ref{fig:asymmetriclobes}b shows a schematic example of a beam spot at the edge of a growth defect, with only one half of the beam intersecting the strained diamond, creating such a pattern.
 
\edgehistogram

To quantify this asymmetry, we extract the ratio of counts in the upper and lower lobes for each reduced diffraction pattern.  Fig.~\ref{fig:edgehistogram} shows the distribution of this ratio for beam positions in the region of Fig.~5 of the main text.  If this ratio is less than 0.5 or greater than 2, we classify that beam position as an ``edge'', with additional uncertainty in the calibration between diffraction curve centroid and strain due to the short distance of interaction between the beam and the strained layer.  All such ``edge'' pixels identified in main text Fig.~5 and supplemental material Fig.~\ref{fig:otherangle} feature high measured strain; the color bar upper limit thus represents a lower bound on the strain present in the defect features.  When the layer is thicker than $\sim200$ nm, this calibration uncertainty is limited to 1.5E-5, as presented in table~\ref{table:errorbudget}.

\subsection{Uncertainty in strain measurements}
\label{sec:SuppUnc}
Sec.~II of the main text describes the process of extracting quantitative strain values from SXDM diffraction patterns recorded on the detector.  The most significant systematic uncertainties enter through the calibration curves.  Table \ref{table:errorbudget} presents the leading uncertainties for a typical measurement of a strain feature.

\otherangle

\begin{table}[h]
\begin{center}
\caption{Error budget for a typical SXDM strain feature measurement using methods described in this study.}
\label{table:errorbudget}
\begin{tabular}{||c|c||}
\hline
Error Source & Uncertainty\\
\hline\hline
Statistical & 2.0E-5\\
\hline
Feature thickness & 1.5E-5 \\
\hline
Unstrained peak linewidth & 1.2E-5 \\
\hline
Dynamical diffraction effects & 0.9E-5 \\
\hline \hline
Total & 2.9E-5\\
\hline \hline
\end{tabular}
\end{center}
 \end{table}

The thinner a strain feature is, the wider that feature's diffraction peak. Whatever part of the diffraction peak overlaps the large host-crystal diffraction peak will be suppressed when we divide to emphasize sharply varying features; the wider the true diffraction peak is, the more this suppression will shift the centroid of the reduced peak.  The calibration curve therefore shows some dependence on the thickness of the strained layer, especially for very thin strain features.  However, as discussed in Sec.~\ref{sec:edgeasymmetry}, beam positions that only briefly interact with the edges of strain features can be distinguished by the two-dimensional distribution of counts on the detector.  After removing such positions from our analysis, the remaining effect of feature thickness on the strain calibration is reflected in our error budget above.

Similarly, the linewidth of the unstrained diffraction pattern enters into the denominator of the empirical lineshape reduction analysis.  The error budget reflects the range of host-crystal peak widths measured in this work.

Finally, our calibration curves are constructed using the kinematic approximation for diffraction pattern construction, while a more complete analysis would include dynamical diffraction effects.  To evaluate the potential distortion due to this approximation, we benchmark our kinematic approximation calculations by comparison to publicly available dynamical diffraction code \cite{UWDynamicalSims2018,UWdynamicalsim} for the (004) mirror plane, and include the maximum observed difference as a conservative estimate of the resulting uncertainty.

\subsection{Strain map for ($\bar{1}13$) diffraction}

Figure \ref{fig:otherangle} shows a strain map measured with ($\bar{1}13$) diffraction on the same region of the CVD quantum sensing diamond as featured in Fig.~5 of the main text.  Strain values are extracted using the same analysis procedure described in main text Sec.~II, except with calibration curves generated for this diffraction condition.  Note that ``petal'' features A and C appear in both projections with equal strain (within the measurement uncertainty), while feature B and most of feature D are absent; this is likely due to different average Burgers vectors between features, as discussed in the main text.  As with main text Fig.~5, this scan used a step size of 200 nm, and the color scale is chosen to allow the reader to see the internal strain variation within the ``petals.''

\subsection{Diffraction curves without background subtraction}
\label{sec:SuppUnnorm}

\unnormfig

Figure \ref{fig:unnormfig} shows a log-scale plot of diffraction curves before the background subtraction analysis for the same four beam positions presented in Fig.~3c of the main text, together with the corresponding local host-crystal diffraction profiles.  The ratio of measured diffraction curve to local host-crystal profile gives the reduced diffraction curves of Fig.~3c of the main text.  The minimum detectable strain shift of $\sim1.5$ detector pixels originates from the linewidth of the host-crystal diffraction profile; diffraction signals from features with less than this strain are suppressed in the generation of reduced diffraction curves.

\subsection{Illustration of three-dimensional ``petal'' feature projected onto two-dimensional data}

As discussed in the main text, the largest strain structures observed in this study take the form of ``petal''-shaped growth defect regions, which nucleate at the substrate-CVD interface and propagate upwards.  Only two lobes of each feature are observed here, as opposed to the four lobes in similar features observed in \cite{StrainPaper}, because of the different detection thresholds for compressive and tensile strain discussed in Sec.~II of the main text.  Fig.~\ref{fig:diamondfeaturesalt} shows a schematic illustration of two-dimensional projection of such features by the X-ray beam, which creates (for example) the data patterns of Fig.~5 of the main text.

\diamondfeaturesalt

\subsection{``Petal'' and ``rod'' strain features}
\label{sec:supppetal}

We construct a three-dimensional geometrical model of strain features from two feature types, referred to as ``petals'' and ``rods.'' ``Petal'' features include conical volumes with a droplet-shaped cross-section and a shared core, nucleating at or near the substrate-layer interface and growing upwards close to the CVD growth direction. ``Rod'' features are thin cylinders which also nucleate at or near the substrate-layer interface, but which may branch out as they grow.

Defects with these shapes have been identified in past studies of strain features in CVD diamond. ``Petal'' defects have been observed experimentally \cite{StrainPaper,RamanStrain2011} and modeled theoretically \cite{Pinto_Jones_2009}. This modelling has shown that they comprise bundles of many individual dislocations which together create the observed strain fields. Meanwhile, ``rod'' dislocations could represent either the high-strain edges of otherwise unmeasured ``petals'', or single or bundled linear dislocations. X-ray tomography identified ``rod''-like features as both edge and mixed dislocations in bundles of different densities \cite{E6xray2008} (albeit with insufficient spatial resolution to differentiate ``petal'' features). In both cases, these features propagated at a variety of angles to the growth direction. TEM studies of single dislocations in CVD diamond \cite{Tsubouchi_TEM_2016} have shown that they alternate between screw and edge segments with different propagation directions, giving an average propagation at an angle to the growth axis.

We use observed ``petal'' and ``rod'' features as examples of three-dimensional strained volumes that affect NV sensing, and with which we can demonstrate the capabilities of SXDM.  The observed ``petal'' features have internal strain similar to that expected near a particle-induced crystal damage track; measuring their structure serves as a demonstration of the three-dimensional capability of stereoscopic SXDM over tens of microns of depth, as well as a benchmark of strain sensitivity.  While the measurements presented in main text Fig.~5 and supplemental material Fig.~\ref{fig:otherangle} do not allow conclusive determination of the structure of the ``petals,'' they do reveal certain interesting features about these structures -- especially the increasing strain at their edges.  This may indicate that the density of individual dislocations is higher towards the edges or at the outer surface of the volume.

\constraindata

We note that past X-ray tomography \cite{E6xray2008} has distinguished edge from mixed dislocations in ``rod''-like strain features based on their Burgers vector.  However, to create a high-fidelity three-dimensional model we only analyzed ``rods'' which appear in both (113) and ($\bar{1}13$) projections, leaving their Burgers vectors ambiguous.  We also note that the local strain fields of screw, edge, and mixed dislocations fall off with different powers of distance from the dislocation core; however, to efficiently map the entire volume around features A, B, C, and D, we opted for a scan step size of 200 nm, which gives insufficient spatial resolution to perform such an analysis.

\subsection{Lack of ``small'' strain features in main text Fig.~5}
\label{sec:supplack}
We note the lack of observed strain features $\sim$1 micron or smaller in main text Fig.~5 for the CVD diamond, in comparison to those mapped in the HPHT diamond and shown in main text Fig.~2.  This result is in part due to the larger scan step size used in main text Fig.~5, which would have caused some of the features in main text Fig.~2 to appear as single pixels, if at all.  We note that strain features $\sim100$ nm in at least one dimension are observed when a smaller scan step size is used on this diamond, as seen in main text Fig.~10b. 

Aspects of the manufacturing and history of the CVD diamond also likely reduce the prevalence of small features.  For example, the surface of the NV-enriched overgrowth layer was not mechanically polished after growth, meaning there are no small strain features associated with microscopic mechanical surface.  (While the substrate-overgrowth interface was polished, microscopic defects there would propagate through the overgrowth layer; this is in fact a likely origin for features like A, B, C, and D, as discussed below.)  Additionally, the CVD diamond was high-temperature annealed after electron irradiation to create NV centers.  This likely dissociated or alleviated strain due to vacancies and vacancy clusters incorporated during CVD growth and irradiation, as discussed in main text Sec.~V.  

\subsection{Constraining geometrical parameters of the three-dimensional model}
\label{sec:SuppModel}

\fullcomp

To create a geometric model of strained volumes in the CVD diamond, we first identify ``petal'' and ``rod'' features.  ``Petal'' features are identified by conical shapes containing strained diamond, as features A, B, C, and D in main text Fig.~5.  ``Rod'' features are identified as thin linear high-strain features propagating through the overgrowth layer.  To avoid unconstrained features and overfitting, we only identify ``rod'' features that are at least partially present in both (113) and ($\bar{1}13$) projections.  For each identified feature, we then construct a surface in parametric coordinates.  For ``petals,'' we start with paired conical surfaces with drop-shaped cross-section, defined with respect to parametric coordinates $(R,\Theta)$:

\begin{equation}
    \begin{split}
        X=&R \left(\sin{(\pm \phi_1)} \cos{(\Theta)}\right)\\
        Y=&R \left(a\sin{\left(\Theta\right)}\right) \sin{\left(\frac{\Theta}{2}\right)^l}\\
        Z=&R.
    \end{split}
\end{equation}

We rotate each surface by $\pm\theta_1$ about the Y-axis, leaving the tips of the drop-shaped cross-sections touching at the shared core of the ``petal.''  We then rotate the entire shape by angles $\phi_2$ and $\phi_3$ around the X and Z axes, respectively.  Fig.~\ref{fig:constraindata}c shows a model ``petal'' constructed in this way.  The free parameters for the shape are thus $a$ and $\phi_1$, which determine the two elliptical axes of the cross-section; $l$, which determines the steepness of the drop profile; and $\phi_2$ and $\phi_3$, which determine the overall spatial orientation of the feature; as well as the origin (x,y,z) near the CVD-substrate interface.  We construct ``rods'' similarly, with a circle of radius $a$ replacing the drop-shaped cross-section, and with additional rods branching off along the length of the original rod.  After constructing the volumes, we project each entity onto surfaces at angles corresponding to the (113) and ($\bm\bar{1}$13) Bragg diffraction conditions.  Finally, we compare these projections to categorized SXDM data giving the positions of strain features, and constrain the geometrical free parameters by minimizing residuals of this comparison.

Fig.~\ref{fig:constraindata} gives an example of the optimization procedure for determining the parameters of a single ``petal'' feature.  Fig.~\ref{fig:constraindata}a-b show the categorized data used for this comparison.  We classify each point as either only intersecting the host crystal, part of the strained subvolume, or an edge of the strained subvolume, based on the amplitude of the reduced diffraction curve (see Fig.~3 of the main text) and the vertical asymmetry in the diffraction pattern (see supplemental material Sec.~\ref{sec:edgeasymmetry}).  Fig.~\ref{fig:constraindata}c shows the initial three-dimensional model for a ``petal'' feature.  (Only two of the four lobes of the ``petal'' are present because we only observe the compressively strained parts of the feature, while the diffraction signal from the tensile-strained lobes is concealed by the large host-crystal diffraction profile from the substrate - see Sec.~II of the main text for details.) Fig.~\ref{fig:constraindata}d-e show the comparison between SXDM strain data and the model.  To calculate residuals, we assign a value of 2 to each pixel classified as an ``edge'' and 1 to each classified as part of the strained subvolume; we subtract pixel-by-pixel to obtain the difference maps in Fig.~\ref{fig:constraindata}d-e.  To obtain a quantity that can be minimized, we sum the absolute value of all pixels in the difference maps.  Minimizing these residuals gives the position, shape, size, and angle of the feature for the three-dimensional strain feature model of Fig.~6 of the main text.  Figure \ref{fig:fullcomp} shows the classified data, projected model, and subtracted residuals for both Bragg conditions.

\subsection{Overview of strain measurement methods using a quantum diamond microscope (QDM)} 
\label{sec:SuppQDM}

The quantum diamond microscope (QDM) measurements presented in the main text use nitrogen vacancy (NV) center spin-state spectroscopy for strain imaging of diamond on micron to millimeter scales.  The methods and apparatus used for these QDM measurements are presented in detail in Ref.~\cite{StrainPaper}; in this section we give a brief overview.

\xrayqdmintro

The NV center comprises a substitutional nitrogen impurity adjacent to a vacancy in the carbon lattice.  The NV axis can lie along the four $\langle$111$\rangle$ crystallographic axes; Figure \ref{fig:xrayqdmintro}a sketches the crystal structure for these four possible classes of NV centers.  Collectively, the NV forms a spin-1 electronic system, with a simplified schematic of the NV spin and electronic energy level structure given in Fig.~\ref{fig:xrayqdmintro}b.  The spin energy levels in the electronic ground state are sensitive to environmental parameters, including both electromagnetic fields \cite{QDMreview,NVNanoReview2014,NVMRIreview2019,NVMagnetometer2008,NVimaging2008} and strain \cite{StrainPaper,AusStrain2019,EnglundStrain2016}.  In the presence of a bias magnetic field $|\vec{B}| > 1$ mT (as used in this work), the NV ground-state spin Hamiltonian reduces to \cite{StrainPaper}
\begin{equation}
\label{eqn:NVham}
    H_\kappa=(D+M_{z,\kappa})S_{z,\kappa}^2+\gamma\vec{B}\boldsymbol{\cdot}\vec{S}_{z,\kappa},
\end{equation}
where $\kappa=(1,2,3,4)$ represents the four possible crystallographic orientations of the NV axis within the crystal lattice, $D\approx2870$ MHz is the temperature-dependent zero-field splitting, $M_z$ is the energy level shift induced by stress along the axis between nitrogen and vacancy (defined to be the z axis in the NV coordinate system), and $S_z$ is the spin-1 projection operator along that axis.  Spectroscopic measurements of the spin transition frequencies under this Hamiltonian form the basis of QDM measurements.

NV centers have several attractive properties for quantum sensing.  Specifically, the electronic structure enables both optical initialization and readout of the spin state, allowing fast and accurate microwave spectroscopy of the spin transition frequencies.  In a typical NV measurement, green laser light excites the electronic state into a phonon sideband; the excited state usually decays via spin-state-conserving spontaneous emission of red fluorescence.  A second decay path through a pair of singlet states exhibits increased coupling to the $\ket{\rm{m_s}=\pm 1}$ states over $\ket{\rm{m_s}=0}$, and does not emit into the same wavelength band.  NV centers in the $\ket{\rm{m_s}=0}$ state therefore exhibit enhanced fluorescence, allowing optical spin-state readout.  Further, the alternative decay path does not conserve the spin state, meaning repeated excitation with green fluorescence gives nearly complete spin polarization into $\ket{\rm{m_s}=0}$.  

A QDM performs microwave spectroscopy of the ground-state spin transitions using this optical initialization and readout.  A schematic of a typical QDM apparatus is shown in Fig.~\ref{fig:xrayqdmintro}c: green laser light excites the NV centers for optical initialization and readout, while red fluorescence is collected by a microscope objective and imaged onto a camera.  Permanent magnets provide a bias magnetic field, and microwaves broadcasted to the NVs by a loop antenna drive spin transitions when resonant.  For NV strain imaging, the microwave drive frequency is swept across the resonance profile, and the red fluorescence collected at each frequency step.  When the microwaves are off-resonant, the laser polarizes the NV spins into $\ket{\rm{m_s}=0}$, maximizing the collected fluorescence.  However, when the microwaves are resonant they drive transitions into either the $\ket{\rm{m_s}=+1}$ or $\ket{\rm{m_s}=-1}$ state with enhanced coupling to the alternative decay path, reducing the fluorescence.  

Figure \ref{fig:xrayqdmintro}d shows an example spectrum acquired in this fashion.  The eight visible resonances correspond to the $\ket{\rm{m_s}=0}\rightarrow\ket{\rm{m_s}=+1}$ and $\ket{\rm{m_s}=0}\rightarrow\ket{\rm{m_s}=-1}$ transition frequencies for each of the four NV classes, which are typically equally populated in a CVD-grown ensemble diamond.  The two narrower lines within each resonance represent hyperfine transitions with the spin-1/2 $^{15}$N nucleus.  In this measurement, the magnetic field direction was chosen to have a different projection along each of the four NV axes, giving different values of $\gamma\vec{B}\boldsymbol{\cdot}\vec{S}_{z,\kappa}$ and therefore different spin transition frequencies.  The spectrum can be simultaneously fit to equation \eqref{eqn:NVham} to extract $M_{z,\kappa}$ for all four NV axes.  Measuring both the $\ket{\rm{m_s}=+1}$ and $\ket{\rm{m_s}=-1}$ transitions allows separation of the stress and magnetic field contributions to the Hamiltonian, as the stress term is proportional to $S_z^2$ and therefore shifts both transitions in the same direction, while the magnetic field term is proportional to $S_z$ and therefore shifts the $\ket{\rm{m_s}=+1}$ and $\ket{\rm{m_s}=-1}$ transitions in opposite directions.  

To create a strain image such as those presented in the main text of this work, a camera acquires spatially resolved images of red fluorescence at each frequency point of the microwave sweep; this yields a spectrum such as that shown in Fig.~\ref{fig:xrayqdmintro}d for each camera pixel.  Fitting each of these spectra to equation \eqref{eqn:NVham} determines all four $M_{z,\kappa}$ values at each pixel of the image.  Finally, the proper linear combination of the four $M_{z,\kappa}$ values gives the stress tensor elements, which can be converted to strain via the measured elasticity tensor of diamond \cite{CVDYoungsModulus1992} (as mentioned in the main text).  For details of the conversion between $M_{z,\kappa}$ and stress tensor, see Ref. \cite{StrainPaper}.  

To complement the earlier discussion of charge- and dipole-trapping near damage tracks, we briefly note that while electric fields and strain couple through the same operators in an inseparable way for individual NV centers, in an ensemble of NV centers the effects can be distinguished, because stress is a tensor whereas electric field is a vector quantity.  For each of the four NV classes within the ensemble, the NV axis points along a different crystallographic direction; within each of these classes, there are two possible orientations between the nitrogen and vacancy (N-V and V-N, so to speak).  While NVs pointing along both of these directions see the same stress-induced frequency shift $M_{z,\kappa}$, an electric field will shift their resonance frequencies in opposite directions.  Large electric fields will therefore split each of the eight resonances into two sub-resonances, while small electric fields may broaden the resonance linewidths; but electric fields cannot contribute the resonance frequency shifts from which $M_{z,\kappa}$ is determined.

\end{document}